\documentclass[article,10pt,nofootinbib,notitlepage,twocolumn]{revtex4-1}
\usepackage{graphicx}
\usepackage{amsmath}
\usepackage{amssymb}
\usepackage{dcolumn}
\usepackage{bm}
\usepackage{color}
\usepackage{dsfont}
\usepackage{feynmp}
\usepackage{marvosym}
\usepackage{phaistos}

\newcommand \beq{\begin{eqnarray}}
\newcommand \eeq{\end{eqnarray}}

\begin{document}
\unitlength=1mm
\allowdisplaybreaks


\title{Signatures of the Yang-Mills deconfinement transition\\ from the gluon two-point correlator}

\author{Duifje Maria van Egmond}
\affiliation{Centre de Physique Th\'eorique, CNRS, Ecole polytechnique, IP Paris, F-91128 Palaiseau, France.}

\author{Urko Reinosa}
\affiliation{Centre de Physique Th\'eorique, CNRS, Ecole polytechnique, IP Paris, F-91128 Palaiseau, France.}

\date{\today}

\begin{abstract}
We evaluate the longitudinal or {\it (chromo-)electric} Yang-Mills gluon propagator in the recently proposed center-symmetric Landau gauge at finite temperature \cite{vanEgmond:2021jyx}. To model the effect of the Gribov copies in the infrared, we use the Curci-Ferrari model which, in turn, allows us to rely on perturbative calculations. At one-loop order in the SU(2) case, the so-obtained longitudinal gluon propagator provides a clear signature for $Z_2$ center-symmetry breaking with a singular behavior, characteristic of a continuous phase transition. This is in sharp contrast with what is found within the standard Landau gauge. We also identify various signatures for $Z_3$ center-symmetry breaking in the SU(3) case in the form of genuine order parameters. Among those, we find that the gluon propagator, although degenerate along the diagonal color directions in the confining phase, becomes non-degenerate in the deconfined phase. Our results open new ways of identifying the transition from correlation functions both within continuum approaches and on the lattice. 
\end{abstract}

\maketitle

\section{Introduction}

The infrared behavior of the gluon propagator in SU(N) Yang-Mills (YM) theories has been the subject of intense study in the past few decades, both numerically \cite{Boucaud:2011ug} and analytically \cite{Alkofer:2000wg}. The main motivation for these studies comes from the continuum, where the calculation of gauge-invariant quantities requires  that of correlation functions. Among the most powerful semi-analytical tools to access correlations functions are the Schwinger-Dyson (SD) equations, which provide non-perturbative relations between Green's function \cite{vonSmekal:1997ohs, Alkofer:2000wg, Zwanziger:2001kw, Fischer:2003rp, Bloch:2003yu, Aguilar:2004sw, Boucaud:2006if, Aguilar:2007ie, Aguilar:2008xm, Boucaud:2008ky, Fischer:2008uz, Rodriguez-Quintero:2010qad}. This leads, however, to an infinite set of  coupled equations, which needs to be truncated in practice. The same is true for the infinite set of coupled equations obtained in the framework of the  Non-Perturbative/Functional Renormalization Group (NPRG/FRG) \cite{Wetterich:1992yh, Berges:2000ew, Pawlowski:2003hq, Fischer:2004uk,Pawlowski:2005xe,Cyrol:2016tym,Dupuis:2020fhh}.

Despite being gauge-variant quantities defined only once a gauge has been specified, the correlation functions should arrange within observables to produce gauge-invariant results. This, of course, requires a good handle on the truncations that are considered in the continuum as well as an accurate description of the correlation functions, starting from the gluon propagator.  For this reason, the development of continuum approaches has proceeded alongside (gauge-fixed) lattice simulations of ever increasing precision \cite{Cucchieri:2007rg, Cucchieri:2008fc, Bornyakov:2008yx, Cucchieri:2009zt, Bogolubsky:2009dc, Bornyakov:2009ug, Dudal:2010tf,Duarte:2016iko} that provide, among many other things, a benchmark for continuum calculations. 

It is sometimes speculated that correlation functions could also reveal physical information more directly. There are two distinct ways in which this could happen. First, certain quantities extracted from a correlation function can be gauge invariant and thus directly sensitive to the Physics in any gauge. Well known examples are the poles of the correlation functions in the Minkowskian domain \cite{Kobes:1990dc}. Another possibility which we address in this work is that certain gauge-variant quantities extracted from a correlation function can provide sharp physical signatures of the phenomena of interest within specific gauges. In this work, we would like to investigate this scenario with regard to the YM confinement/deconfinement transition at finite temperature. 

It has been firmly established by lattice simulations that pure SU(N) YM theories present a phase transition at high temperatures \cite{Gavai:1982er,Gavai:1983av,Celik:1983wz,Svetitsky:1985ye}, associated to the breaking of the $Z_N$ center symmetry \cite{Pisarski:2002ji,Greensite:2011zz}. This transition is of second order for the SU(2) group and first order for SU(3). A relevant, gauge-invariant order parameter is the Polyakov loop \cite{Polyakov:1978vu} which aquires a non-zero value when center symmetry is broken above some transition temperature $T_c$. Since the Polyakov loop is inversely proportional to the  exponential of the free energy of a static quark, the latter takes finite values in the broken phase, corresponding to the deconfined phase. Below $T_c$, center symmetry is restored, the Polyakov loop vanishes and the free energy becomes infinite, signalling confinement. 

Clearly, the transition is reflected in the Polyakov loop and also in the correlation functions of the latter, which are all physical quantities. Now, because the Polyakov loop is directly related to the temporal component of the gluon field, it is expected that the transition is equivalently encoded in the thermal average of that gluon component and its fluctuations. The way this happens clearly depends on the gauge. An interesting question is then whether, within an appropriate choice of gauge, the transition can be reflected already in the lowest order correlators like $\langle A_0^a(x)\rangle$ or $\langle A_0^a(x)A_0^b(y)\rangle$. Besides its purely theoretical aspect, this question is of great practical interest as these correlators are the most easily accessible quantities to continuum approaches (as compared to the Polyakov loop itself for instance).

Both (gauge-fixed) lattice simulations as well as various analytical studies have searched for signatures of the phase transition in the Landau gauge {\it (chromo-)electric} or longitudinal gluon propagator, which involves the temporal component of the gluon field. More precisely, they looked at the electric susceptibility, the electric gluon propagator at vanishing frequency and momentum. The picture that arises is not completely clear, however. 

In the SU(2) case, early simulations suggested that the electric susceptibility might be a sensible probe for the confinement-deconfinement transition \cite{Cucchieri:2010lgu,Fischer:2010fx,Maas:2011ez}, but later studies with larger volumes contradicted this, finding no clear signature of the transition \cite{Mendes:2015jea}. Some general expectations have been given in Ref.~\cite{Maas:2011ez} but they rely on the background effective action/potential associated to background field gauges, see below, and do not apply directly to the Landau gauge. The SU(3) case has been considered in Refs.~\cite{Fischer:2010fx,Maas:2011ez,Aouane:2011fv,Silva:2013maa}. There, although some quantities extracted from the propagators (such as masses) react to the transition by displaying distinctive features and some even seem to behave as order parameters, it is not completely clear how this is connected to the breaking of center-symmetry. A very interesting step in this direction has been taken in Ref.~\cite{Silva:2016onh} where a connection between center-symmetry breaking and a ``sectorized" gluon propagator (defined by classifying gluon field configurations according to the domain of the complex plane in which the Polyakov lies) has been established. To date, unfortunately, this is not an easy quantity to evaluate in the continuum. As for continuum approaches, non-perturbative calculations of the electric susceptibility in the Landau gauge mostly show a non-monotonic behavior around the transition temperature \cite{Fister:2011uw, Fischer:2012vc, Fukushima:2013xsa, Huber:2012kd, Quandt:2015aaa}, but no clear sign of the transition, in qualitative agreement with the lattice results on larger volumes.

One of the main points to be defended in this work is that there is a priori no strong reason to believe that the electric susceptibility in the Landau gauge unambiguously reflects the breaking of center symmetry. By this, we do not mean that the propagator does not react at all to the transition, but rather that there are a priori no sharp signatures to the breaking of center symmetry in this case. This relates to the fact that, in the Landau gauge, the thermal average of the temporal component of the gluon field does not qualify as an order parameter for this symmetry, as it does not derive from the minimization of a center-symmetric effective action.\footnote{We refer to \cite{Reinosa:2020mnx,vlongue} for a detailed discussion on these questions.}  In particular, at the continuous SU(2) transition, within a generic gauge, there is no reason for the electric susceptibility (related to the inverse curvature of the effective action at its minimum) to diverge.

On the other hand, in gauges for which the temporal gluon field average is an order parameter, the electric susceptibility should become a good probe of the transition. Examples of such gauges can be constructed using background field gauge (BFG) methods \cite{Abbott:1980hw,Abbott:1981ke}. In particular, the background field generalization of the Landau gauge is the family of Landau-deWitt (LDW) gauges parametrized by a background configuration whose main purpose is to formally restore the gauge symmetry. In its standard implementation at finite temperature \cite{Braun:2007bx, Braun:2010cy, Fister:2013bh}, the background is chosen to coincide with the gauge field average, which we refer to as the {\it self-consistent (background) Landau gauge.} Beyond its obvious simplifying aspect as one needs only to work with one field rather than two, the self-consistent backgrounds, and thus the gauge-field average, can be shown to provide order parameters for the breaking of center symmetry \cite{Braun:2007bx,Reinosa:2020mnx}. Self-consistent backgrounds are obtained from the minimization of the \textit{background effective action} which is explicitly center-symmetric and predicts correctly the order of the transition for both $\smash{N=2}$ and $\smash{N=3}$, while giving good estimates of the respective transition temperatures \cite{Braun:2007bx, Braun:2010cy, Fister:2013bh}. 

Despite these remarkable features, it should be noted that the use of the background effective action is not free of subtleties \cite{Reinosa:2020mnx,vlongue}. Indeed, although it can be shown that minimizing the background effective action is equivalent to minimizing a Legendre transform, this equivalence can be broken in practice due to the use of approximations or of modelling.\footnote{By modelling, we mean for instance the use of {\it Ans\"atze} for some of the vertex functions in the problem, or the use of additional operators meant to capture the gauge-fixing procedure in the infrared (such a mass operator in the Curci-Ferrari model, see below) beyond the standard but incomplete Faddeev-Popov procedure. Another source of difficulty is the use of infrared regulators in intermediated steps, such as those used within the FRG.} This can potentially introduce spurious effects in the predictions obtained from the background effective action. Recently, a different implementation of background field methods has been put forward which avoids these difficulties \cite{vanEgmond:2021jyx}. Rather than considering a self-consistent background, it has been proposed to choose a center-symmetric background at all temperatures, which defines what we refer to as the {\it center-symmetric Landau gauge.} Even though this approach requires working with two fields since the field average does not coincide with the background anymore, it is based on the minimization of a regular Legendre transform, which turns out to be explicitly center-symmetric. In turn, its minima qualify as order parameters for the breaking of center symmetry. Moreover, since the propagator is obtained from the second derivative of this same Legendre transform, the behavior of the order parameter directly impacts that of the propagator in a standard fashion. In particular, the electric susceptibility diverges at the continuous transition of the SU(2) theory. Finally, as argued in detail in Ref.~\cite{vanEgmond:2021jyx}, the center-symmetric Landau gauge should be easily implementable in lattice simulations, which is not the case of the self-consistent Landau gauge, at least not in the deconfined phase.

In this work we would like to confirm these expectations by computing the electric propagator in the center-symmetric Landau gauge. We shall do this within the context of the Curci-Ferrari (CF) model \cite{Tissier:2010ts,Tissier:2011ey}. One of the motivations for the latter is the observation on the lattice that the Landau gauge gluon propagator reaches a strictly positive finite value in the deep infrared for space-time Euclidean dimensions $\smash{d>2}$  \cite{Cucchieri:2007rg, Bogolubsky:2009dc, Dudal:2010tf}. This decoupling behavior can be accounted for by a simple massive extension of the Faddeev-Popov (FP) action for Landau gauge YM theories, which is a particular case of the CF model \cite{Curci:1976bt}. That the FP action is incomplete is a well known fact as it does not properly account for the presence of Gribov copies. In Ref.~\cite{Serreau:2012cg, Serreau:2012uvx}, it was argued that the CF model could be part of a complete gauge-fixing in the Landau gauge, since an effective gluon mass equivalent to the CF mass may arise after the Gribov copies have been taken care of via an averaging procedure.\footnote{Another approach based on the (Refined) Gribov-Zwanziger model \cite{Gribov:1977wm, Zwanziger:1989mf, Zwanziger:1992qr, Dudal:2008sp} is more specifically aimed at restricting the number of gauge transformed configurations, i.e. Gribov copies, that satisfy a  certain gauge condition.} The true form of this statement still needs to be figured out, however.\footnote{We refer to \cite{condensate} for a recently unveiled connection between the Curci-Ferrari model and the dynamical generation of a dimension two gluon condensate.} Consequently, in this paper, the CF model will be used as a phenomenological take on that question.

Among the many properties of the model, we note that it is renormalizable and that it reproduces the FP perturbation theory for large momenta. In the low momentum regime, the Landau pole that usually spoils the FP approach can be avoided by an adequate renormalization scheme. Moreover, the effective gluon mass suppresses higher order corrections to the propagator, possibly opening a perturbative window into the infrared region. This goes in line with lattice simulations of the running YM coupling \cite{Duarte:2016iko}. The latter shows no trace of a Landau pole and even decreases to perturbative values in the deep infrared. Based on these considerations, the perturbative CF model has then been applied quite successfully to evaluate YM correlations functions \cite{Tissier:2011ey,Pelaez:2013cpa,Gracey:2019xom}. Upon some controlled extension beyond perturbation theory, it can also be used within a QCD context where it captures the spontaneous breaking of chiral symmetry \cite{Pelaez:2020ups}. At finite temperature, when extended in the presence of a background along the lines of \cite{Braun:2007bx, Braun:2010cy, Fister:2013bh}, it was shown to provide a good qualitative and even quantitative account of the deconfinement transition of YM theories \cite{Reinosa:2014ooa,Reinosa:2014zta,Reinosa:2015gxn}. We refer to \cite{Pelaez:2021tpq} for a recent review of the applications of the CF model to QCD.

Finite temperature propagators have also been investigated within the CF model both in the standard and in the self-consistent Landau gauge. In the former case, one obtains qualitatively similar results as with non-perturbative approaches with little sensitivity of the electric propagator to the transition region \cite{Reinosa:2013twa}. In the latter case, and even though the phase transition leaves an imprint on the electric propagator, it is not as sharp as expected \cite{Maas:2011ez}. In particular, the SU(2) electric susceptibility does not diverge at the transition \cite{Reinosa:2016iml}. As discussed in Ref.~\cite{vanEgmond:2021jyx}, and as recalled above, this relates to the fact that the equivalence between the minimization of the background effective action and of a Legendre transform can be broken when approximations and/or modelling are involved. More recently, the electric susceptibility has been studied within the center-symmetric Landau gauge \cite{vanEgmond:2021jyx}. In this case, despite the use of approximations and modelling of the gauge-fixing in the infrared, the SU(2) electric susceptibility  diverges at the transition. For SU(3), one finds instead a sharp peak, with a slight discontinuity at the transition, in line with the fact that the transition is first order. Let us also note that, in both cases, the approach based on the center-symmetric Landau gauge predicts values for $T_c$ in better agreement with the lattice results than those obtained within the self-consistent Landau gauge. 

In this work, we pursue our study of the center-symmetric Landau gauge by computing the corresponding gluon two-point correlator. The paper is organized as follows. In Sec.~\ref{sec:center}, we recall the definition of the center-symmetric Landau gauge and its implementation within the CF model. In Sec.~\ref{sec:rules}, we derive the corresponding Feynman rules with special emphasis on the benefits of using Cartan-Weyl color bases that allow one to keep track of color conservation constraints. Section \ref{sec:prop} is devoted to the evaluation of the gluon propagator, and more precisely on dealing with the various contractions that occur in the Feynman integrals. In Sec.~\ref{sec:long}, we restrict to the case of the electric propagator along neutral color modes, at vanishing frequency since this is expected to be the most sensitive component of the propagator to the transition. As a cross-check of our calculation, we verify that the zero-momentum limit gives back the electric susceptibility as obtained in Ref.~\cite{vanEgmond:2021jyx}. Sections \ref{sec:mom} and \ref{sec:sums} provide details on the practical evaluation of the sum-integrals. Finally, Sec.~\ref{sec:res} presents our results for both the SU(2) and SU(3) groups, together with a discussion comparing to results in other gauges. The conclusion is followed by a series of Appendices which gather further computational details.

\section{The center-symmetric\\ Landau gauge}\label{sec:center}
We consider the gauge-fixing condition
\beq
\bar D_\mu(A^a_\mu-\bar A^a_\mu)=0\,,\label{eq:cond}
\eeq
where 
\beq
\bar D_\mu\varphi^a\equiv\partial_\mu\varphi^a+gf^{abc}\bar A^b_\mu\varphi^c\label{eq:cDer}
\eeq
is the (adjoint) covariant derivative in the presence of a background field configuration $\bar A$ and where $f^{abc}$ denote the SU(N) structure constants. Rather than specifying a particular gauge, (\ref{eq:cond}) should be seen as defining a class of gauges parametrized by the background $\bar A$ which plays then the role of an infinite collection of gauge-fixing parameters. Each choice of $\bar A$ defines a specific gauge within the class (\ref{eq:cond}). 

The associated FP gauge-fixed action  writes
 \beq\label{eq:S0} 
S_{\bar A}[A]=\!\!\int_x\!\left\{\frac{1}{4}F_{\mu\nu}^aF_{\mu\nu}^a\!+\!\bar D_\mu\bar c^a
D_\mu c^a\!+\!ih^a\bar D_\mu(A_\mu^a-\bar A_\mu^a)\right\},\nonumber\\
\eeq
where $c^a$ and $\bar c^a$ denote the FP ghost and antighost fields and $h^a$ the Nakanishi-Lautrup field. By construction, the action (\ref{eq:S0}) satisfies [it is implicitly understood that the ghost fields and the Nakanishi-Lautrup field are gauge rotated accordingly]
\beq\label{eq:cov}
S_{\bar A^U}[A^U]=S_{\bar A}[A]\,,
\eeq
where $U$ denotes an arbitrary gauge transformation:
\beq
A^U_\mu\equiv U A_\mu U^\dagger-\frac{i}{g}U\partial_\mu U^\dagger\,.
\eeq
  
\subsection{Center symmetry}
 The identity (\ref{eq:cov}) is particularly useful at finite temperature for it encodes center symmetry. The latter symmetry relates to transformations $U$ that do not change the periodic boundary conditions of the gauge field at finite temperature [$\smash{\beta\equiv 1/T}$ denotes the inverse temperature]:
\beq
A_\mu^a(\tau+\beta,\vec{x})=A_\mu^a(\tau,\vec{x})\,.
\eeq
These transformations, which form a group denoted ${\cal G}$ in what follows, are themselves periodic up to an element of the center of the underlying group. For SU(N), they are then such that
\beq
U(\tau+\beta,\vec{x})=e^{i2\pi k/N}U(\tau,\vec{x})\,, \,\, k=0,\dots, N-1\,.\label{eq:boundary}
\eeq
Their relevance is that they act non-trivially on the {\it Polyakov loop}
\beq
\ell\equiv \frac{1}{N}\left\langle {\cal P}\exp\left\{ig\int_0^\beta d\tau\,A_0^a(\tau,\vec{x})t^a\right\}\right\rangle,\label{eq:loop}
\eeq
an observable that measures the ability of a thermal bath of gluons for accepting static colored sources.\footnote{More precisely, static colored sources in representations of non-vanishing $N$-ality.} In fact, the Polyakov loop transforms as $\ell\to e^{i2\pi k/N}\ell$. Thus, when the center symmetry is not broken, the Polyakov loop vanishes which, in turn, signals a color confining phase. 

The benefit of Eq.~(\ref{eq:cov}) is that it makes center symmetry explicit within a gauge-fixed setting. One could legitimately wonder why it is so important to have the center symmetry explicit in the formalism. After all, the gauge fixing deals with an unphysical symmetry and should not, in principle, impact the consequences of physical symmetries such as center symmetry. Although strictly correct at an exact level of treatment, this picture is not always easily met in practice since one needs to resort to approximations and/or modelling. In those cases, center symmetry might not be explicit in the formalism. We refer to Ref.~\cite{Reinosa:2020mnx,vlongue} for a thorough discussion, in particular regarding the differences between a continuum setting and a lattice setting.

In those situations, it is convenient to find appropriate gauges where center symmetry remains explicit despite the considered approximations or modelling.  Now, coming back to the property (\ref{eq:cov}), it can be shown that it is inherited by the effective action in the case where $U$ is taken within ${\cal G}$:
\beq
\forall U\in {\cal G}\,,\,\,\,\Gamma_{\bar A^U}[A^U]=\Gamma_{\bar A}[A]\,.
\eeq
This certainly makes center-symmetry explicit within the class of background Landau gauges. However, this is only to the prize of connecting two different gauges, corresponding to $\bar A$ and $\bar A^U$ respectively. As we argue in Refs.~\cite{Reinosa:2020mnx,vlongue}, this makes it cumbersome to track down center symmetry within a given choice of background gauge. In particular, an absolute minimum $A_{\rm min}[\bar A]$ of $\Gamma_{\bar A}[A]$ is transformed under $U\in{\cal G}$ into an absolute minimum $A_{\rm min}[\bar A^U]$ of the effective action $\Gamma_{\bar A^U}[A]$. In turn the absolute minima of $\Gamma_{\bar A}[A]$ do not necessarily qualify as order parameters for the breaking of center symmetry.

As we now review, there is one particular choice of background, however, where the symmetry becomes explicit in that gauge and the minima of the effective action turn into order parameters.

\subsection{Center-symmetric Landau gauge}
One way to ensure that the formalism remains explicitly center-symmetric within a specific gauge is to work within a background gauge corresponding to $\smash{\bar A=\bar A_c}$ where $\bar A_c$ is a {\it center-symmetric background configuration,} that is a background configuration invariant under center transformations.

To fully grasp the nature of the center-symmetric backgrounds, it should be stressed that the center-symmetry group is actually not ${\cal G}$ but rather ${\cal G}/{\cal G}_0$ where ${\cal G}_0$ denotes the subgroup of periodic transformations, corresponding to $k=0$ in Eq.~(\ref{eq:boundary}). The subgroup ${\cal G}_0$ should be seen as the group of ``true'' gauge transformations that do not change the state of the system while ${\cal G}/{\cal G}_0$ corresponds to physical transformations that change at least one observable, the Polyakov loop (\ref{eq:loop}). It follows that a center-symmetric background $\bar A_c$ is invariant under ${\cal G}$ only modulo ${\cal G}_0$:\footnote{As a matter of fact, there is no background configuration invariant under the whole ${\cal G}$.}\footnote{A simpler example allowing to understand this condition can be borrowed from standard electromagnetism. Consider the case of a homogeneous magnetic field $\vec{B}$. This physical field describes obviously a translation invariant situation. Yet, its potential $\vec{A}(\vec{x})=\vec{B}\times\vec{x}/2$ is not tranlation invariant in the usual sense, but translation invariant modulo a gauge transformation $$\vec{A}(\vec{x}+\vec{u})=\vec{A}(\vec{x})+\vec{B}\times\vec{u}/2=\vec{A}(\vec{x})+\vec{\nabla}(\vec{x}\cdot(\vec{B}\times\vec{u})/2)$$ In other words, it is invariant under the combined application of a translation and a gauge transformation: $\vec{A}^\Lambda(\vec{x}+\vec{u})=\vec{A}(\vec{x})$, with $\Lambda(\vec{x})=-\vec{x}\cdot(\vec{B}\times\vec{u})/2$.}
\beq
\forall U\in {\cal G}\,,\,\, \exists U_0\in {\cal G}_0\,,\,\, \bar A_c^{U_0U}=\bar A_c\,.
\eeq
 Examples of center-symmetric backgrounds are given below for the SU(2) and SU(3) cases. For such backgrounds, Eq.~(\ref{eq:cov}) becomes
\beq
\forall U\in {\cal G}\,,\,\, \exists U_0\in {\cal G}_0\,, \,\, S_{\bar A_c}[A^{U_0U}]=S_{\bar A_c}[A]\,,\label{eq:constraint}
\eeq
an identity which is inherited by the effective action:
\beq
\forall U\in {\cal G}\,,\,\, \exists U_0\in {\cal G}_0\,, \,\, \Gamma_{\bar A_c}[A^{U_0U}]=\Gamma_{\bar A_c}[A]\,.\label{eq:constraint}
\eeq
This is now a center symmetry constraint within a specific gauge, corresponding to the particular choice of background $\bar A_c$. In particular, the absolute minima of $\Gamma_{\bar A_c}[A]$ become order parameters for the breaking of center symmetry \cite{vanEgmond:2021jyx,vlongue}. This is the main advantage of working within the center-symmetric Landau gauge. Also, as compared to the background effective action mentioned in the Introduction and defined as $\smash{\tilde\Gamma[\bar A]\equiv\Gamma_{\bar A}[A=\bar A]}$, the effective action $\Gamma_{\bar A_c}[A]$ is a regular Legendre transform.  Therefore, the interpretation of its minimum as the gauge-field average in the limit of zero sources is a direct one.

We should stress that, as long as one works at an exact level of treatment, it should be entirely equivalent to work with $\Gamma_{\bar A}[A]$ or $\Gamma_{\bar A_c}[A]$. In particular, the minimum of these functionals gives the free-energy of the system which should not depend on the choice of $\bar A$ and should therefore reflect the transition irrespectively of the fact that the center-symmetry is explicit or not at the level of the effective action. As already discussed above, the previous statement is not necessarily true in practice, due to approximations or modelling, and it is then more convenient to work within a setting where the center symmetry is as explicit as possible. This is provided by the effective action $\Gamma_{\bar A_c}[A]$ in the center-symmetric Landau gauge.

 \subsection{Curci-Ferrari model}
It is well known that the action (\ref{eq:S0}) is at best an approximation since, at low energies, the gauge fixing is hindered by the Gribov ambiguity and the Faddeev-Popov procedure needs to be revisited. As already mentioned in the Introduction, in this work we shall rely on a phenomenological approach to the gauge fixing in the infrared which has shown a surprising efficiency in describing many aspects of YM theories at low energies \cite{Pelaez:2021tpq}. It amounts to adding a gluonic mass term to the action (\ref{eq:S0}):
\beq
\delta S_{\bar A}[A]=\!\!\int_x \frac{m^2}{2}(A_\mu^a-\bar A_\mu^a)(A_\mu^a-\bar A_\mu^a)\,.
\eeq
This type of model was originally introduced within the standard Landau gauge (in the absence of background). In that case, the motivation for introducing a mass term is the fact that Landau gauge YM correlators as computed on the lattice comply with the {\it decoupling} scenario, with, in particular, a saturation of the gluon propagator at low momenta. The precise form of the mass term in the presence of $\bar A$ ensures that Eq.~(\ref{eq:cov}) and thus Eq.~(\ref{eq:constraint}) remain true.

\section{Feynman rules}\label{sec:rules}

We now derive the Feynman rules for the model $S_{\bar A_c}[A]+\delta S_{\bar A_c}[A]$. In fact, it is convenient to derive them more generally within the class of gauges corresponding to the choice
\beq
\bar A_\mu=\delta_{\mu0}\frac{T}{g} \bar r^jt^j\,,\label{eq:bg_generic}
\eeq
where the $t^j$ span the diagonal or commuting part of the algebra. Center-symmetric backgrounds can then be obtained by adapting the $t^j$ to the group under consideration and choosing specific values for the $\bar r^j$. In the SU(2) case for instance, a center-symmetric background is
\beq
\bar A_{c,\mu}=\delta_{\mu0}\frac{T}{g} \pi\frac{\sigma^3}{2}\,,
\eeq
with $\sigma_3$ the third Pauli matrix. In the SU(3) case, we can take
\beq
\bar A_{c,\mu}=\delta_{\mu0}\frac{T}{g} \frac{4\pi}{3}\frac{\lambda^3}{2}\,,
\eeq
with $\lambda_3$ the third Gell-Mann matrix. The benefit of working with the generic form (\ref{eq:bg_generic}) is that it allows one to perform the computations in the general SU(N) case while keeping track of color conservation constraints.

 \subsection{Color constraints} 
 The presence of a preferred color direction as dictated by (\ref{eq:bg_generic}) makes the color structure of $\smash{S_{\bar A}[A]+\delta S_{\bar A}[A]}$ more intricate than in the absence of background. In order to simplify the discussion, it is useful to identify and exploit as many symmetries as possible.

One such symmetry is provided by the color rotations $\smash{U_\theta\equiv\exp\{i\theta^jt^j\}}$ whose generators $t^j$ lie in the commuting part of the algebra. Under such rotations, the background (\ref{eq:bg_generic}) is trivially invariant and Eq.~(\ref{eq:cov}) turns into the symmetry constraint
\beq
S_{\bar A}[A^{U_\theta}]=S_{\bar A}[A]\,,\label{eq:theta}
\eeq
$\forall \theta$. The conserved Noether charges associated to this continuous symmetry are nothing but the generators $t^j$ in the appropriate representation. For the fields present in the action (\ref{eq:S0}), these are the adjoint charges $[t^j,\,\,]$. By expanding the fields along a basis of eigenstates of these charges, one unveils the conservation rules at the level of the action which will later facilitate the derivation and interpretation of the Feynman rules. Such a basis is known as a {\it Cartan-Weyl basis,} the definition of which we now recall.

\begin{figure}[t]
\begin{center}
\includegraphics[height=0.2\textheight]{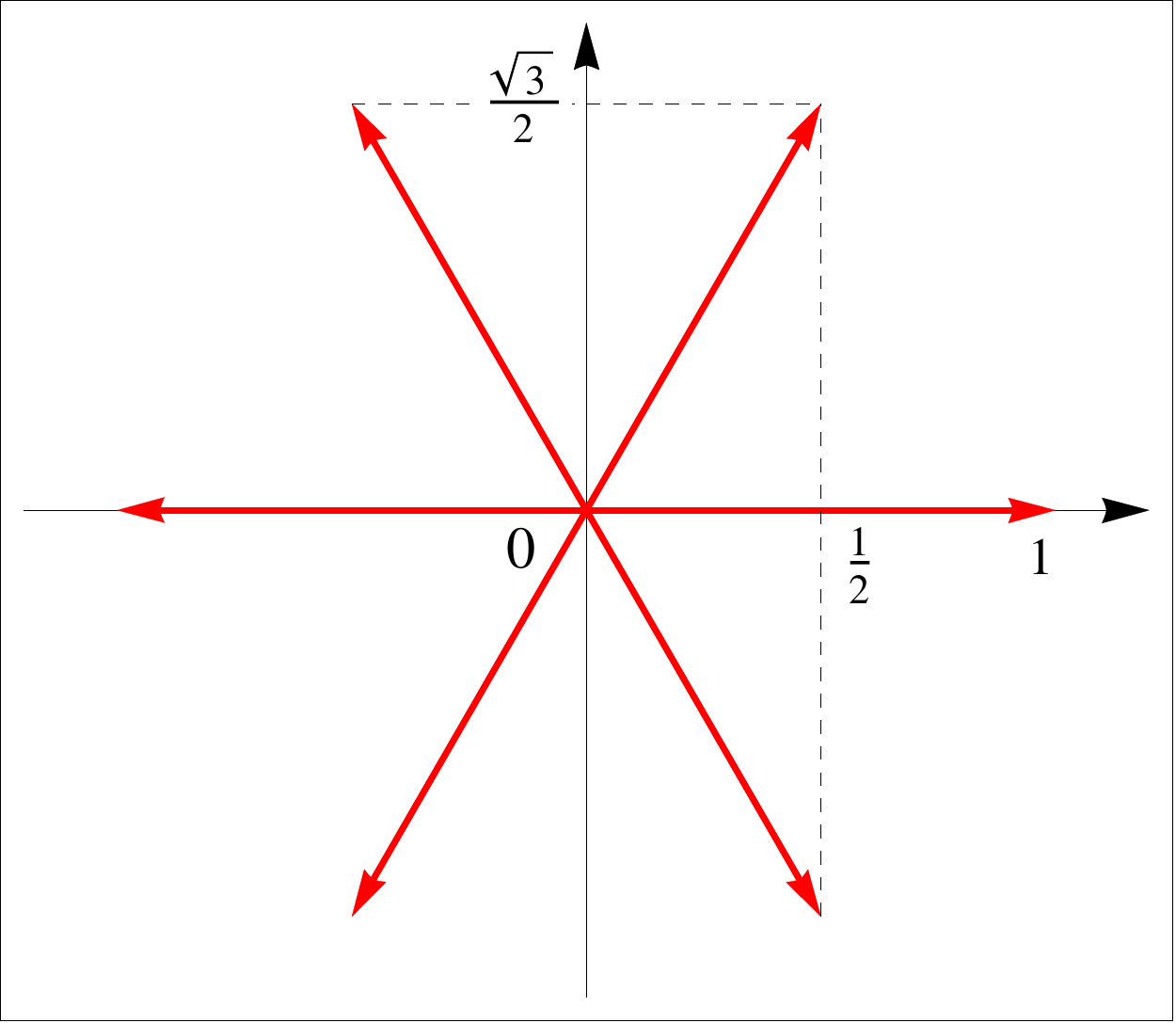}
\caption{Root diagram of the su($3$) algebra.}
\label{fig:roots}
\end{center}
\end{figure}

\subsection{Cartan-Weyl bases}
 A Cartan-Weyl basis, denoted $\{t^\kappa\}$ in what follows, is defined by the conditions
\beq
[t^j,t^\kappa]=\kappa^j t^\kappa\,,\label{eq:comm}
\eeq 
where $\kappa^j$ is the (eigen)value taken by the charge $[t^j,\,\,]$ on the color (eigen)state $t^\kappa$. The labels $\kappa$ are to be seen as real-valued vectors in a space isomorphic to the commuting subalgebra. There are of two types. 

First, any $t^j$ qualifies itself as a $t^\kappa$ with $\smash{\kappa=0}$. To distinguish between the various $t^j$, we shall write instead $\smash{\kappa=0^{(j)}}$, which we refer to as {\it zeros}. This notational trick will only be needed when $\kappa$ refers to the label of a generator. In equations where it does not play the role of a label, $\kappa$ is a mere vector and, therefore, $0^{(j)}$ can be replaced by $0$. This will always be clear from the context. The other possible values of $\kappa$ are (non-degenerate) non-zero vectors known as {\it roots} and denoted $\alpha$, $\beta$, \dots The set of roots characterizes the algebra under consideration. In the SU(2) case for instance, there is one zero and two (one-dimensional) roots, $\pm 1$. In the SU(3) case, there are two zeros and six (two-dimensional) roots, $\pm(1,0)$, $\pm(1/2,\sqrt{3}/2)$ and $\pm(1/2,-\sqrt{3}/2)$, see Fig.~\ref{fig:roots}. In what follows, we shall also refer to the zeros and the roots as the {\it neutral modes} and the {\it charged modes,} respectively.

The properties of Cartan-Weyl bases are numerous. As we have already mentioned, they make the color constraints explicit. This is in particular the case at the level of the vertex functions. Indeed, under the action of $U_\theta$, a field $\varphi$ in the adjoint representation transforms as $\smash{U_\theta \varphi U_\theta^\dagger}$. If we decompose $\varphi$ along a Cartan-Weyl basis, owing to Eq.~(\ref{eq:comm}), this becomes 
\beq
\varphi^\kappa e^{i\theta^j[t^j,\,\,]}t^\kappa=e^{i\theta\cdot\kappa}\varphi^\kappa t^\kappa\,,
\eeq 
with $\smash{\theta\cdot\kappa\equiv \theta^j\kappa^j}$. The transformation is thus very simple in this basis, $\varphi^\kappa\to e^{i\theta\cdot\kappa}\varphi^\kappa$. In particular, any correlation function should satisfy
\beq
\left\langle\varphi_1^{\kappa_1}\cdots\varphi_n^{\kappa_n}\right\rangle=e^{i\theta\cdot(\kappa_1+\dots+\kappa_n)}\left\langle\varphi_1^{\kappa_1}\cdots\varphi_n^{\kappa_n}\right\rangle
\eeq
and thus $\smash{\left\langle\varphi_1^{\kappa_1}\cdots\varphi_n^{\kappa_n}\right\rangle=0}$ unless $\smash{\kappa_1+\cdots+\kappa_n=0}$, that is unless the color labels $\kappa_1,\dots,\kappa_n$ conserve color. For instance, one-point functions $\langle\varphi^\kappa\rangle$ vanish unless $\kappa$ is a zero, and two-point functions $\langle\varphi_1^{\kappa_1}\varphi_2^{\kappa_2}\rangle$ do not connect zeros and roots, $\langle \varphi_1^{0^{(j)}}\varphi_2^\alpha\rangle=\langle\varphi_1^\alpha\varphi_2^{0^{(j)}}\rangle=0$, and have a diagonal structure in the root sector, $\langle\varphi_1^{\alpha_1}\varphi_2^{\alpha_2}\rangle\sim \delta^{\alpha_1,-\alpha_2}$. As for the zero sector, there is no obvious reason why the two-point functions should be diagonal and couplings could appear between distinct zeros, $0^{(j_1)}$ and $0^{(j_2)}$ with $j_1\neq j_2$. In the case of pure YM theory, this situation is excluded for the SU(2) gauge group (trivially) and also for the SU(3) gauge group (due to charge conjugation invariance, see Ref.~\cite{Reinosa:2016iml}). We shall restrict to these cases in what follows. In particular, the gluon propagator ${\cal G}_{\mu\nu}^{\kappa\lambda}$ to be defined below can always be taken of the form $\delta^{\kappa,-\lambda}{\cal G}^\lambda_{\mu\nu}$.

Within a Cartan-Weyl basis, color conservation is also directly explicit at the level of the action. In particular, the usual structure constants $f^{abc}$ are replaced by $f^{\kappa\lambda\tau}$ which vanish whenever the associated charges $\kappa$, $\lambda$ and $\tau$ do not add up to zero. In this way, the interactions explicitly and locally conserve color.\footnote{Not all the color charges are conserved, only the diagonal or commuting ones.} We refer to \cite{Reinosa:2015gxn} for more details and only note here for later purpose that
\beq\label{eq:fs}
f^{(-\kappa)(-\lambda)(-\tau)}=-(f^{\kappa\lambda\tau})^*\,.
\eeq
Another very useful feature of Cartan-Weyl bases is that they allow one to diagonalize the action of the background covariant derivative (\ref{eq:cDer}) as
\beq
\bar D^\kappa_\mu\varphi & = & \partial_\mu\varphi^\kappa-ig\bar A^j_\mu\kappa^j\nonumber\\
& = & \partial_\mu\varphi^\kappa-i\delta_{\mu0}T\bar r^j_\mu\kappa^j\,.
\eeq
In Fourier space, this becomes $-\bar Q_\mu^\kappa\varphi^\kappa$, where we have defined the generalized momentum
\beq
\bar Q^\kappa_\mu\equiv Q_\mu+\delta_{\mu0} T\bar r\cdot\kappa\,.\label{eq:gen_mom}
\eeq
Here, $Q$ corresponds to $(\omega_q,q)$ where the first entry $\smash{\omega_q=2\pi qT}$ denotes a bosonic Matsubara frequency ($q\in\mathds{Z}$) and the second entry denotes the $3$-momentum vector.\footnote{For convenience, we use the same letter $q$ to denote the integer that labels the Matsubara frequency $\omega_q$ and the $3$-momentum $q$. No confusion is possible as long as the integer $q$ is attached to the frequency $\omega_q$.}

Let us mention finally that, within the gauge specified by the choice $\bar A$, the Feynman rules are derived after expanding the gauge field under the functional integral around its expectation value, $A\to A+\tilde a$,\footnote{As it is customary, we denote the dynamical gluon field and its expectation value by the same letter.} which is not necessarily $\bar A$. For backgrounds of the form (\ref{eq:bg_generic}), and as a direct consequence of Eq.~(\ref{eq:theta}), the expectation value of the gauge field is also of the form
 \beq
A_\mu=\delta_{\mu0}\frac{T}{g} r^jt^j\,.\label{eq:1pt}
\eeq
This expectation value plays effectively the role of a second background whose associated generalized momentum
\beq
Q^\kappa_\mu\equiv Q_\mu+\delta_{\mu0} T r\cdot\kappa
\eeq
also enters the Feynman rules, as we now discuss in more detail.

\subsection{Free propagators}
Let us first derive the expressions for the free propagators since this requires only the quadratic part of the action upon expansion of the gluon field. In Fourier space, this quadratic part reads
\beq\label{eq:S2}
& & \sum_\kappa \int_Q^T\, \Bigg\{\bar c^{\kappa}(Q)^*\,Q^\kappa_\mu \bar Q^\kappa_\mu \,c^\kappa(Q)+h^{\kappa}(Q)^* \bar Q_\mu^\kappa \tilde a_\mu^\kappa(Q)\nonumber\\
 & & \hspace{1.5cm}+\,\frac{1}{2}\tilde a_\mu^{\kappa}(Q)^*(Q_\kappa^2P^\perp_{\mu\nu}(Q_\kappa)+m^2\delta_{\mu\nu})\,\tilde a_\nu^\kappa(Q)\Bigg\},\label{eq:quad}\nonumber\\
\eeq
where we defined $\bar c^\kappa(Q)^*\equiv \bar c^{-\kappa}(-Q)$ and where
\beq
P^\perp_{\mu\nu}(Q_\kappa) & \equiv & \delta_{\mu\nu}-\frac{Q^\kappa_\mu Q^\kappa_\nu}{Q^2_\kappa}\,,
\eeq
projects orthogonally to $Q_\kappa$. The notation 
\beq
\int_Q^T f(Q)\equiv T\sum_{q\in\mathds{Z}}\mu^{2\epsilon}\int\frac{d^{d-1}q}{(2\pi)^{d-1}}f(\omega_q,q)
\eeq
stands for a bosonic sum-integral. We define the latter within dimensional regularization with $\smash{d=4-2\epsilon}$ and $\mu$ the associated renormalization scale.

From Eq.~(\ref{eq:S2}), it follows immediately that the tree-level ghost propagator is given by
\beq\label{eq:D}
D^\kappa(Q)=\frac{1}{Q_\kappa\cdot \bar Q_\kappa}\,,
\eeq
where we have introduced the notation $X\cdot Y\equiv X_\mu Y_\mu$. As for the tree-level gluon propagator $G^\kappa_{\mu\nu}(Q)$, it is obtained by inverting the quadratic form (\ref{eq:quad}) in the $A-h$ sector and restricting the so-obtained inverse to the $A$ sector, see App.~\ref{app:Schur} for details. One arrives at
\beq
G^\kappa_{\mu\nu}(Q) & = & G^\kappa_T(Q)P^T_{\mu\nu}(\bar Q_\kappa)+G^\kappa_L(Q)P^L_{\mu\nu}(\bar Q_\kappa)\,,\label{eq:GG}
\eeq
with
\beq
G^\kappa_T(Q) & = & \frac{1}{Q^2_\kappa+m^2}\,,\label{eq:PT}\\
G^\kappa_L(Q) & = & \frac{\bar Q^2_\kappa}{(Q_\kappa\cdot\bar Q_\kappa)^2+m^2\bar Q^2_\kappa}\,,\label{eq:PL}
\eeq
and where
\beq
P^T_{\mu\nu}(\bar Q_\kappa) & \equiv & (1-\delta_{\mu 0})(1-\delta_{\nu 0})\left(\delta_{\mu\nu}-\frac{\bar Q^\kappa_\mu \bar Q^\kappa_\nu}{q^2}\right),\label{eq:projT2}
\eeq
and
\beq
P^L_{\mu\nu}(\bar Q_\kappa) & \equiv & P^\perp_{\mu\nu}(\bar Q_\kappa)-P^T_{\mu\nu}(\bar Q_\kappa)\label{eq:projL}
\eeq
are the usual transverse projectors at finite temperature, with respect to $\bar Q_\kappa$. By construction 
\beq\label{eq:p1}
P^T_{\mu\nu}(\bar Q_\kappa)\bar Q^\kappa_\nu=P^L_{\mu\nu}(\bar Q_\kappa)\bar Q^\kappa_\nu=P^\perp_{\mu\nu}(\bar Q_\kappa)\bar Q^\kappa_\nu=0\,,
\eeq 
and then $G^\kappa_{\mu\nu}(Q_\kappa)\bar Q^\kappa_\nu=0$, in agreement with the gauge-fixing condition (\ref{eq:cond}). Very often, the property (\ref{eq:p1}) will be used as follows: if $Q$, $K$, $L$ are three momenta constrained by $\smash{Q+K+L=0}$ and $\kappa$, $\lambda$, $\tau$ are three charges constrained by $\smash{\kappa+\lambda+\tau=0}$, then
\beq
P^X_{\mu\nu}(\bar Q_\kappa)\bar K^\lambda_\nu=-P^X_{\mu\nu}(\bar Q_\kappa)\bar L^\tau_\nu\,,\label{eq:pp}
\eeq 
for $\smash{X=T,\,L}$ or $\perp$.

Let us mention that the notation $P^T_{\mu\nu}(\bar Q_\kappa)$, which was originally adopted in Ref.~\cite{Reinosa:2016iml}, is not the optimal one as it hides the fact that this projector does not depend on the frequency. In what follows, even though we continue using the old notation when referring un-specifically to the projectors, as in Eq.~(\ref{eq:pp}), we shall adopt an updated notation and an equivalent expression for this projector [latin indices are implicitly summed over]
\beq
P^T_{\mu\nu}(q) & \equiv & \delta_{\mu i}\delta_{\nu j}P^\perp_{ij}(q)\,.\label{eq:projT}
\eeq
The benefit of (\ref{eq:projT}) is that it makes explicit the fact that $P^T_{\mu\nu}(q)$ projects the frequency to $0$ and thus acts equally on momenta sharing the same vector part. In particular, for any momentum, denoted generically $\hat Q$, whose vector part is collinear to $q$:\footnote{Correspondingly, $\smash{P^L_{\mu\nu}(q)\hat Q_\nu=P^\perp_{\mu\nu}(q)\hat Q_\nu}$.}
\beq
P^T_{\mu\nu}(q)\hat Q_\nu=0\,.\label{eq:pp3}
\eeq 
This applies to $\smash{\hat Q=(0,q)}$ of course but also to $\smash{\hat Q=\bar Q_\kappa}$ and $\smash{\hat Q=Q_\kappa}$. For the same reason, the identity (\ref{eq:pp}) for $\smash{X=T}$ extends to any triplet of momenta $\hat Q$, $\hat K$ and $\hat L$ whose vector parts $q$, $k$ and $\ell$ are constrained by $\smash{q+k+\ell=0}$:
\beq
P^T_{\mu\nu}(q)\hat K_\nu=-P^T_{\mu\nu}(q)\hat L_\nu\,.\label{eq:pp2}
\eeq 
As for the projector $P^L_{\mu\nu}(\bar Q^\kappa)$, it will be sometimes convenient to rewrite it as
\beq\label{eq:df}
P^L_{\mu\nu}(\bar Q^\kappa)=\frac{\tilde Q^\kappa_\mu \tilde Q^\kappa_\nu}{\tilde Q^2_\kappa}=\frac{\tilde Q^\kappa_\mu \tilde Q^\kappa_\nu}{q^2\bar Q^2_\kappa}\,,
\label{pl}
\eeq
with
\beq
\tilde Q^\kappa_\mu\equiv (q^2,-\bar \omega^\kappa_q q)\,.
\eeq
Since the vector part of $\tilde Q_\kappa$ is collinear to $q$, the identity (\ref{eq:pp3}) applies also for $\smash{\hat Q=\tilde Q_\kappa}$. Moreover, since $P^T_{\mu\nu}(q)$ projects the frequency to zero, we have
\beq
P^T_{\mu\nu}(q)\tilde K^\lambda_\nu=-\bar\omega^\lambda_k P^T_{\mu\nu}(q) K^\lambda_\nu\,,\label{eq:pp5}
\eeq
an identity that we shall also use below.

\subsection{Interaction vertices}
As for the vertices, we find the ghost-antighost-gluon vertex
\beq
gf_{\kappa\lambda\tau}\bar Q_{\nu}^{\kappa}\,,
\eeq
where $Q$ denotes the outgoing anti-ghost momentum, $\kappa$, $\lambda$ and $\tau$ the outgoing color charges of the anti-ghost, gluon and ghost respectively and $f_{\kappa\lambda\tau}$ the corresponding structure constant. Similarly, the three-gluon vertex reads
\beq
\frac{g}{6}f_{\kappa\lambda\tau}\Big[(K_{\rho}^{\lambda}-Q_{\rho}^{\kappa})\delta_{\mu\nu}\!+\!(L_{\mu}^{\tau}-K_{\mu}^{\lambda})\delta_{\rho\nu}\!+\!(Q_{\nu}^{\kappa}-L_{\nu}^{\tau})\delta_{\mu\rho}\Big],\nonumber\\
\eeq
and the four-gluon vertex
\beq
& & \frac{g^{2}}{24}\sum_{\eta}\Big[f_{\kappa\lambda\eta}f_{\tau\xi(-\eta)}(\delta_{\mu\rho}\delta_{\nu\sigma}-\delta_{\mu\sigma}\delta_{\nu\rho})\nonumber\\
&& \hspace{1.5cm}+\,f_{\kappa\tau\eta}f_{\lambda\xi(-\eta)}(\delta_{\mu\nu}\delta_{\rho\sigma}-\delta_{\mu\sigma}\delta_{\nu\rho})\nonumber\\
&& \hspace{1.5cm}+\,f_{\kappa\xi\eta}f_{\tau\lambda(-\eta)}(\delta_{\mu\rho}\delta_{\nu\sigma}-\delta_{\mu\nu}\delta_{\sigma\rho})\Big].
\eeq

\subsection{Additional remarks}
As a sanity check, we mention that the Feynman rules that we have just derived boil down to those in the self-consistent background approach \cite{Reinosa:2014zta} when setting $\smash{r=\bar r}$ and to those in the Landau gauge when further setting $\smash{\bar r=0}$. In the present gauge with $\smash{r\neq\bar r}$, there are important differences to be emphasized.

First, the free gluon propagator has two independent transverse tensor structures at finite temperature, as compared to the corresponding propagator  in the self-consistent background approach. The transversal part is similar to the propagator in the self-consistent background approach. On the contrary, the longitudinal part is quite different with a denominator that depends quartically on the frequency. This will lead us below to reconsider how sum-integrals need to be evaluated. Second, at the level of the vertices, we note that the momentum entering the ghost-antighost-gluon vertex involves $\bar r$ while those entering the three-gluon vertex involve $r$. This will complicate the calculations somehow as compared to the self-consistent Landau gauge.

Exploiting the identity (\ref{eq:cov}) in the case of periodic transformations $\smash{U_0\in {\cal G}_0}$, it can be argued that $\bar r$ can be restricted to certain regions which are physically equivalent to each other. These are referred to as {\it Weyl chambers.} Each of them contains one particular center-symmetric instance of $\bar r$, denoted $\bar r_c$. In the SU(2) case for instance, we shall be working over the Weyl chamber $\smash{\bar r\in [0,2\pi]}$ whose center-symmetric point is $\smash{\bar r_c=\pi}$. In the SU(3) case, Weyl chambers are equilateral triangles and we shall consider the one defined by the edges $(0,0)$, $2\pi(1,1/\sqrt{3})$ and $2\pi(1,-1/\sqrt{3})$ whose center-symmetric point is located at $(4\pi/3,0)$. It is generally true that the values of $\bar r$ within a given Weyl chamber are such that $\kappa\cdot\bar r\in[2\pi k,2\pi (k+1)]$, for each $\kappa$ and for a certain $k$ that depends on $\kappa$ \cite{Reinosa:2015gxn,Reinosa:2016iml}.

As for the value of $r$, it is obtained from the minimization of the corresponding potential $V_{\bar r}(r)$ obtained from $\Gamma_{\bar A}[A]$ by restricting $\bar A$ and $A$ to (\ref{eq:bg_generic}) and (\ref{eq:1pt}) respectively, see below for the one-loop expression. For $\bar r$ in a given Weyl chamber, we have no strict mathematical argument that requires $r$ to be in the same Weyl chamber. However, for the case of interest here, namely $\bar r=\bar r_c$, we know that $r=\bar r_c$ in the low temperature phase. Moreover, the numerical minimization of the potential shows that, even though $r$ moves eventually away from $\bar r_c$, it remains in the same Weyl chamber over the range of temperatures relevant for the present study. This is good news because otherwise, the ghost propagator (\ref{eq:D}) would develop annoying poles. 

To illustrate this, consider the SU(2) case and the charged mode $\smash{\kappa=-1}$. The denominator of the ghost propagator reads $\omega_q^-\bar\omega_q^-+q^2$ with
\beq
\omega_q^-\bar\omega_q^-=(\omega_q-rT)(\omega_q-\bar r T)\,,\label{eq:q}
\eeq 
which can become negative a priori as $q$ explores $\mathds{Z}$. More precisely, strictly negative values are obtained for $\smash{\omega_q\,\in\,]\,{\rm Min}\,(r,\bar r)T,{\rm Max}\,(r,\bar r)T\,[}\,$. Clearly, the only way to avoid this situation is to take $r$ and $\bar r$ in the same Weyl chamber $[2\pi k,2\pi (k+1)]$. Indeed, in this case, $]\,{\rm Min}\,(r,\bar r)T,{\rm Max}\,(r,\bar r)T\,[\,\,\subset [2\pi k,2\pi (k+1)]$ and no Matsubara sum can reach the forbidden interval.\footnote{Still, it can be that $\omega_q^-\bar\omega_q^-$ vanishes (when $r$ or $\bar r$ lie at the border of the Weyl chamber). Then the ghost propagator has a standard pole at vanishing momentum $q$ but this is compensated in the integrals by the integration measure.} These considerations extend to SU(N).

\section{The gluon propagator}\label{sec:prop}
\begin{center}
\begin{figure}[t]
    \includegraphics[width=.9\linewidth]{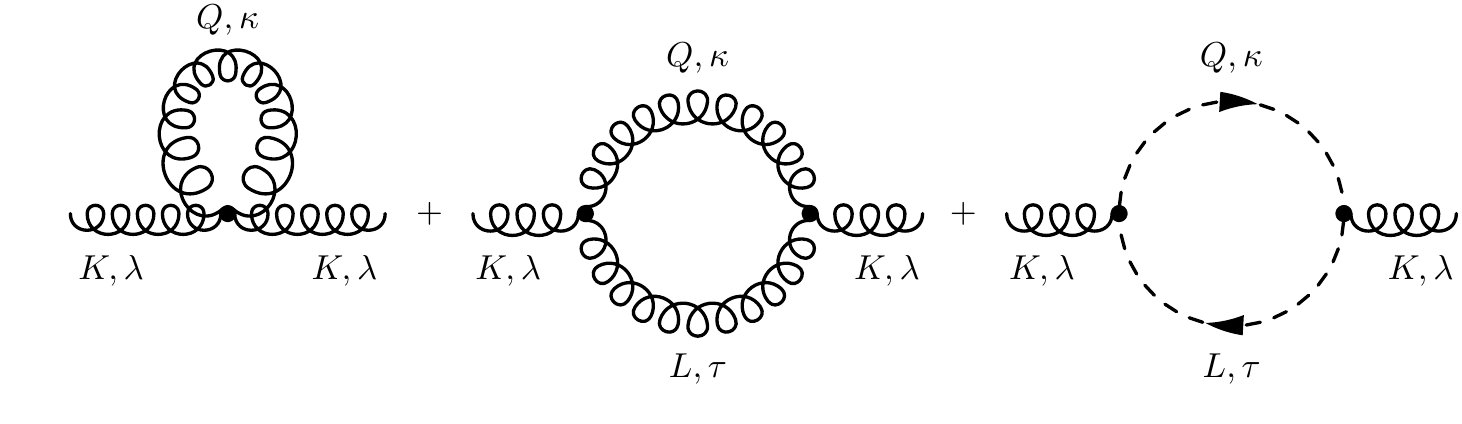}
  \caption{One-loop diagrams contributing to the gluon self-energy. These are referred to in the main text as {\it the tadpole,} {\it the gluon loop} and {\it the ghost loop,} respectively.}
  \label{fig_AA}
\end{figure}
\end{center}

In this paper, we evaluate the gluon propagator
\beq
{\cal G}_{\mu\nu}^{\kappa\lambda}(x,y)=\langle \tilde a_\mu^\kappa(x)\tilde a_\nu^\lambda(y)\rangle\,,\label{eq:prop}
\eeq
whose Fourier transform will be denoted ${\cal G}_{\mu\nu}^{\kappa\lambda}(K)$. As explained above, for the SU(2) and SU(3) cases, we can assume that
\beq
\label{eq:propertyiy}
{\cal G}_{\mu\nu}^{\kappa\lambda}(K)=\delta^{-\kappa,\lambda}{\cal G}_{\mu\nu}^\lambda(K)\,.
\eeq
Moreover, from the gauge-fixing condition (\ref{eq:cond}), the propagator is transverse with respect to $\bar K^\lambda$. In other words, it admits a similar decomposition as the free propagator above:
\beq
{\cal G}^\lambda_{\mu\nu}(K) & = & {\cal G}^\lambda_T(K)P^T_{\mu\nu}(\bar K_\lambda)+{\cal G}^\lambda_L(K)P^L_{\mu\nu}(\bar K_\lambda)\,.\label{eq:GTL}
\eeq
In fact, the Feynman rules do not give access directly to the propagator but rather to the two-point vertex function 
\beq
\label{eq:propertyiy}
\Gamma^{(2)\kappa\lambda}_{\mu\nu}(K)=\delta^{-\kappa,\lambda}\Gamma^{(2)\lambda}_{\mu\nu}(K)\,.
\eeq
The relation between ${\cal G}_{\mu\nu}^\lambda(K)$ and $\Gamma^{(2)\lambda}_{\mu\nu}(K)$ is derived in App.~\ref{app:Schur}. Here, we just quote the final result:
\beq
{\cal G}^\lambda_T(K) & = & \frac{1}{K^2_\lambda+m^2+\Pi^\lambda_T(K)}
\eeq
and
\beq
{\cal G}^\lambda_L(K) & = & \frac{1}{\frac{(K_\lambda\cdot\bar K_\lambda)^2}{\bar K^2_\lambda}+m^2+\Pi^\lambda_L(K)}\,,\label{eq:den_L2}
\eeq
with
\beq
\Pi^\lambda_T(K) & \equiv &  \frac{\Pi^\lambda_{\mu\nu}(K)P^T_{\nu\mu}(k)}{d-2}\,,\\
\Pi^\lambda_L(K) & \equiv &  \Pi^\lambda_{\mu\nu}(K)P^L_{\nu\mu}(\bar K_\lambda)\,,
\eeq
and where the self-energy is
\beq
\Pi^\lambda_{\mu\nu}(K)\equiv \Gamma^{(2)\lambda}_{\mu\nu}(K)-(K_\lambda^2+m^2)\delta_{\mu\nu}+K^\lambda_\mu K^\lambda_\nu\,.
\eeq
Below, we derive general expressions for $\Pi^\lambda_T(K)$ and $\Pi^\lambda_L(K)$ at one-loop order corresponding to the diagrams depicted in Fig.~\ref{fig_AA}. For convenience, a factor $g^2$ will be left implicit in all the self-energy contributions. Also, the latter will always appear as
\beq
\Pi^\lambda_{T/L,diag}(K)=\sum_{\kappa,\tau}{\cal C}^{\kappa\lambda\tau}\Pi^{\kappa\lambda\tau}_{T/L,diag}(K)\,,\label{eq:klt}
\eeq
where $\smash{{\cal C}^{\kappa\lambda\tau}\equiv |f^{\kappa\lambda\tau}|^2}$ and $\Pi^{\kappa\lambda\tau}_{T/L,diag}(K)$ is some expression that depends on the considered diagram. For simplicity, we shall provide this latter expression but it is implicitly assumed that it needs to be contracted with ${\cal C}^{\kappa\lambda\tau}$ and multiplied by a factor $g^2$.

Once the general expressions have been derived, we shall concentrate our analysis on the longitudinal component (\ref{eq:den_L2}) of the propagator along the neutral color modes.\footnote{For these modes, the denominator in Eq.~(\ref{eq:den_L2}) becomes more simply $K^2+m^2+\Pi^{0^{(j)}}_L(K)$.}  This is because, the transition occurs in this sector \cite{vanEgmond:2021jyx}. The other sectors (transversal component of the propagator and/or charged modes) will be studied in a separate work.

\subsection{Ghost loop}
With the Feynman rules given above, the ghost loop contribution to the gluon self-energy reads
\beq
\Pi^\lambda_{\mu\nu,gh}(K) & = & -(-1)\sum_{\kappa,\tau} f^{\kappa\lambda\tau} f^{(-\tau)(-\lambda)(-\kappa)}\nonumber\\
& & \times\,\int_Q  \bar Q^\mu_\kappa (-\bar L)^\nu_{-\tau}\,D(Q_\kappa)D(L_\tau)\,.
\eeq
Upon using Eq.~(\ref{eq:fs}), we verify that this contribution is of the form (\ref{eq:klt}). Moreover, the orientations of momenta and color charges have been chosen such that $\smash{Q+K+L=0}$ and $\smash{\kappa+\lambda+\tau=0}$ which allows us to make use of Eq.~(\ref{eq:pp}) upon projecting along $P^T_{\mu\nu}(k)$ and $P^L_{\mu\nu}(\bar K^\lambda)$. We obtain
\beq
\Pi^{\kappa\lambda\tau}_{T,gh}(K) & = & \int_Q \bar Q^\kappa\cdot \frac{P^T(k)}{d-2}\cdot \bar Q^\kappa\,\,D(Q_\kappa)D(L_\tau)\,,\\
\Pi^{\kappa\lambda\tau}_{L,gh}(K) & = & \int_Q \bar Q^\kappa\cdot P^L(\bar K^\lambda)\cdot \bar Q^\kappa\,\,D(Q_\kappa)D(L_\tau)\,.
\eeq
We have
\beq
\bar Q^\kappa\cdot P^T(k)\cdot \bar Q^\kappa & = & \frac{\hat\sigma(k,q,\ell)}{k^2}\,,
\eeq
where $\smash{\hat\sigma(k,q,\ell)\equiv k^2q^2-(k\cdot q)^2}$. This function is easily seen to be invariant under any permutation of its arguments and also under any shift of one of its arguments by a linear combination of the other two. We find similarly 
\beq
\bar Q^\kappa\cdot P^\perp(\bar K^\lambda)\cdot \bar Q^\kappa=\frac{\hat\sigma(\bar K^\lambda,\bar Q^\kappa,\bar L^\tau)}{\bar K_\lambda^2}\,,
\eeq
 from which one deduces, after a straightforward calculation, that
\beq
\bar Q^\kappa\cdot P^L(\bar K^\lambda)\cdot \bar Q^\kappa & = & \frac{\tilde\sigma(k,q,\ell)+\hat\sigma(k,q,\ell)}{\bar K_\lambda^2}-\frac{\hat\sigma(k,q,\ell)}{k^2}\nonumber\\
& = & \frac{\tilde\sigma(k,q,\ell)}{\bar K_\lambda^2}-\frac{(\bar\omega^\lambda_k)^2}{\bar K_\lambda^2}\frac{\hat\sigma(k,q,\ell)}{k^2}\,,
\eeq
where $\smash{\tilde\sigma(k,q,\ell)\equiv(\bar\omega^\lambda_k q-\bar\omega^\kappa_q k)^2}$. This function is also invariant under any permutation of its arguments. 

We have then arrived at
\beq
\Pi^{\kappa\lambda\tau}_{T,gh}(K) & = & \int_Q \frac{\hat\sigma/(d-2)}{k^2}D(Q_\kappa)D(L_\tau)\,,\\
\Pi^{\kappa\lambda\tau}_{L,gh}(K) & = & \int_Q \!\left[\frac{\tilde\sigma}{\bar K_\lambda^2}-\frac{(\bar \omega^\lambda_k)^2}{\bar K_\lambda^2}\frac{\hat\sigma}{k^2}\right] D(Q_\kappa)D(L_\tau)\,,
\eeq
where, for simplicity, we have omitted the arguments of both $\hat\sigma$ and $\tilde\sigma$ which are always $q$, $k$ and $\ell$, in any order.

\subsection{Gluon tadpole}
The gluon tadpole gives
\beq
\Pi^{\kappa\lambda\tau}_{\mu\nu, tad}=\int_Q\Big\{{\rm tr}\,G^\kappa(Q)\delta_{\mu\nu}-G^\kappa_{\mu\nu}(Q)\Big\},
\eeq
where ${\rm tr}$ refers to the trace with respect to the Euclidean spacetime indices. To evalute the corresponding projections, we use $\smash{{\rm tr}\,P^L(\bar Q^\kappa)=1}$ and $\smash{{\rm tr}\,P^T(q)=d-2}$, as well as
\beq
{\rm tr}\,P^T(q)P^T(k) & = & d-2-\frac{\hat\sigma(k,q,\ell)}{q^2k^2}\,,\\
{\rm tr}\,P^T(q)P^L(\bar K^\lambda) & = & \frac{(\bar \omega_k^\lambda)^2}{\bar K^2_\lambda}\frac{\hat\sigma(k,q,\ell)}{q^2k^2}\,,\\
{\rm tr}\,P^L(\bar Q^\kappa)P^T(k) & = & \frac{(\bar \omega_q^\kappa)^2}{\bar Q^2_\kappa}\frac{\hat\sigma(k,q,\ell)}{q^2k^2}\,,\\
{\rm tr}\,P^L(\bar Q^\kappa)P^L(\bar K^\lambda) & = & 1-\frac{\tilde\sigma(k,q,\ell)}{\bar Q^2_\kappa \bar K^2_\lambda}\nonumber\\
& - & \frac{(\bar \omega^\kappa_q)^2(\bar \omega^\lambda_k)^2}{\bar Q^2_\kappa \bar K^2_\lambda}\frac{\hat\sigma(k,q,\ell)}{q^2k^2}\,,
\eeq
which are easily derived from similar relations between $P^T$ and $P^\perp$. We eventually arrive at
\beq
\Pi^{\kappa\lambda\tau}_{T,tad} & = &  \int_Q\left[1-\frac{(\bar \omega^\kappa_q)^2}{\bar Q^2_\kappa}\frac{\hat\sigma/(d-2)}{q^2k^2}\right]G_L^\kappa(Q)\nonumber\\
& + & \int_Q\left[d-3+\frac{\hat\sigma/(d-2)}{q^2k^2}\right]G_T^\kappa(Q)\,,
\eeq
and
\beq
\Pi^{\kappa\lambda\tau}_{L,tad} & = & \int_Q\left[\frac{\tilde\sigma}{\bar Q^2_\kappa \bar K^2_\lambda}+\frac{(\bar\omega^\kappa_q)^2(\bar\omega^\lambda_k)^2}{\bar Q^2_\kappa \bar K^2_\lambda}\frac{\hat\sigma}{q^2k^2}\right]G^\kappa_L(Q)\nonumber\\
& + & \int_Q\left[d-2-\frac{(\bar \omega^\lambda_k)^2}{\bar K^2_\lambda}\frac{\hat\sigma}{q^2k^2}\right]G_T^\kappa(Q)\,.
\eeq

\subsection{Gluon loop}
The gluon loop contribution reads
\beq\label{eq:pig}
\Pi^{\kappa\lambda\tau}_{\mu\nu,gl}(K) & = & -\frac{1}{2} \int_Q V_{\mu\rho\sigma}^{\lambda\kappa\tau}(K,Q,L)G_{\rho\rho'}^\kappa(Q)\nonumber\\
& & \times\,G_{\sigma\sigma'}^\tau(L)V_{\nu\rho'\sigma'}^{\lambda\kappa\tau}(K,Q,L)\,,
\eeq
with
\beq
V_{\mu\rho\sigma}^{\lambda\kappa\tau}(K,Q,L) & \equiv & (Q^\kappa-L^\tau)_\mu\delta_{\rho\sigma}\nonumber\\
& + & (L^\tau-K^\lambda)_\rho\delta_{\mu\sigma}\nonumber\\
& + & (K^\lambda-Q^\kappa)_\sigma \delta_{\mu\rho}\,,
\eeq
and where the orientations of momenta and color charges have again been chosen such that $\smash{Q+K+L=0}$ and $\smash{\kappa+\lambda+\tau=0}$. Upon projecting along $P^T_{\mu\nu}(k)$ and $P^L_{\mu\nu}(\bar K^\lambda)$, we find
\beq
\Pi^{\kappa\lambda\tau}_{T,gl}(K) & = & \Pi^{\kappa\lambda\tau}_{gl,TTT}(K)+\Pi^{\kappa\lambda\tau}_{gl,TTL}(K)\nonumber\\
& + & \Pi^{\kappa\lambda\tau}_{gl,TLT}(K)+\Pi^{\kappa\lambda\tau}_{gl,TLL}(K)\,,\label{eq:60}
\eeq
and
\beq
\Pi^{\kappa\lambda\tau}_{L,gl}(K) & = & \Pi^{\kappa\lambda\tau}_{gl,LTT}(K)+\Pi^{\kappa\lambda\tau}_{gl,LLT}(K)\nonumber\\
& + & \Pi^{\kappa\lambda\tau}_{gl,LTL}(K)+\Pi^{\kappa\lambda\tau}_{gl,LLL}(K)\,,\label{eq:61}
\eeq
with
\beq
\Pi^{\kappa\lambda\tau}_{gl,XYZ}(K) & \equiv &  -\frac{1}{2} \int_Q W_{XYZ}(K^\lambda,Q^\kappa,L^\tau)G_Y^\kappa(Q)G_Z^\tau(L)\nonumber\\
\eeq
and
\beq
W_{XYZ}(K^\lambda,Q^\kappa,L^\tau) & \equiv & V_{\mu\rho\sigma}^{\lambda\kappa\tau}(K,Q,L)P^X_{\mu\mu'}(\bar K^\lambda)P^Y_{\rho\rho'}(\bar Q^\kappa)\nonumber\\
& & \times\,P^Z_{\sigma\sigma'}(\bar L^\tau)V_{\mu'\rho'\sigma'}^{\lambda\kappa\tau}(K,Q,L)\,.
\eeq
The function $W_{XYZ}(K^\lambda,Q^\kappa,L^\tau)$ is invariant under simultaneous permutations of $(X,Y,Z)$, $(K,Q,L)$ and $(\lambda,\kappa,\tau)$, so that we need only to determine $W_{TTT}(K^\lambda,Q^\kappa,L^\tau)$, $W_{LTT}(K^\lambda,Q^\kappa,L^\tau)$, $W_{TLL}(K^\lambda,Q^\kappa,L^\tau)$ and $W_{LLL}(K^\lambda,Q^\kappa,L^\tau)$ denoted respectively $W_{TTT}$, $W_{LTT}$, $W_{TLL}$ and $W_{LLL}$ in what follows for more readability.

\subsubsection{$W_{TTT}$}

Using Eqs.~\eqref{eq:projT} and (\ref{eq:pp2}), we obtain
\beq
W_{TTT}& = & 4\Big\{q_h\delta_{ij}+\ell_i\delta_{jh}+k_j \delta_{hi}\Big\}\nonumber\\
& & \times\,P^\perp_{hh'}(k)P^\perp_{ii'}(q)P^\perp_{jj'}(\ell)\nonumber\\
& & \times\,\Big\{q_{h'} \delta_{i'j'}+\ell_{i'}\delta_{j'h'}+k_{j'} \delta_{h'i'}\Big\}.
\eeq
This is nothing but the standard expression (for the same diagram) in the absence of background and in $d-1$ dimensions. Using Eq.~(\ref{eq:pp2}) once more, this rewrites
\beq\label{eq:vv}
& & W_{TTT}=\,4\Big(q\cdot P^\perp(k)\cdot q\,{\rm tr}\,P^\perp(q)P^\perp(\ell)\nonumber\\
& & \hspace{1.9cm}-\,2q\cdot P^\perp(k)P^\perp(q)P^\perp(\ell)\cdot q\nonumber\\
& & \hspace{1.9cm}+\,\mbox{cycl. perms.}\Big).
\eeq
Then, noting that
\beq
q\cdot P^\perp(k)\cdot q\,{\rm tr}\,P^\perp(q)P^\perp(\ell)=\frac{\hat\sigma}{k^2}\left[d-3+\frac{(q\cdot\ell)^2}{q^2\ell^2}\right]\nonumber\\
\eeq
and
\beq
& & q\cdot P^\perp(k)P^\perp(q)P^\perp(\ell)\cdot q\nonumber\\
& & \hspace{0.5cm}=\,\left(q-\frac{q\cdot k}{k^2}k\right)P^\perp(q)\left(q-\ell\frac{\ell\cdot q}{\ell^2}\right)\nonumber\\
& & \hspace{0.5cm}=\,\frac{(q\cdot k)(\ell\cdot q)}{k^2\ell^2}\,k\cdot P^\perp(q)\cdot\ell\nonumber\\
& & \hspace{0.5cm}=\,-\frac{(q\cdot k)(\ell\cdot q)}{k^2q^2\ell^2}\,\hat\sigma(k,q,\ell)\,,
\eeq
Eq.~(\ref{eq:vv}) becomes
\beq\label{eq:vv2}
& & W_{TTT}=4\hat\sigma\left\{(d-3)\left[\frac{1}{k^2}+\frac{1}{q^2}+\frac{1}{\ell^2}\right]\right.\nonumber\\
& & \hspace{2.5cm} \left.+\frac{(q\cdot\ell+\ell\cdot k+k\cdot q)^2}{k^2q^2\ell^2}\right\},
\eeq
where we have identified a perfect square in the second term. Under the constraint $\smash{k+q+\ell=0}$, we have
\beq
0 & = & (k+q+\ell)^2\nonumber\\
& = & k^2+q^2+\ell^2+2(q\cdot\ell+\ell\cdot k+k\cdot q)\,,
\eeq
and thus Eq.~(\ref{eq:vv2}) leads eventually to
\beq
& & W_{TTT}=\,4\hat\sigma\Bigg\{\left(d-\frac{5}{2}\right)\left[\frac{1}{k^2}+\frac{1}{q^2}+\frac{1}{\ell^2}\right]\nonumber\\
& & \hspace{2.0cm}+\,\left[\frac{k^2}{4q^2\ell^2}+\frac{q^2}{4\ell^2k^2}+\frac{\ell^2}{4k^2q^2}\right]\Bigg\}.
\eeq
The benefit of this expression as compared to Eq.~(\ref{eq:vv}) or Eq.~(\ref{eq:vv2}) is that we have gotten rid of scalar products in the numerator.

\subsubsection{$W_{LTT}$}
Using Eqs.~(\ref{eq:pp2}) and (\ref{eq:df}), we find
\beq
W_{LTT} & = & \Big\{(Q^\kappa-L^\tau)\cdot \tilde K^\lambda\delta_{\rho\sigma}-2K^\lambda_\rho\tilde K^\lambda_\sigma+2K^\lambda_\sigma \tilde K^\lambda_\rho\Big\}\nonumber\\
& & \hspace{1.5cm}\times\,\frac{1}{k^2\bar K^2_\lambda} P^T_{\rho\rho'}(Q^\kappa)P^T_{\sigma\sigma'}(L^\tau)\nonumber\\
& & \Big\{(Q^\kappa-L^\tau)\cdot\tilde K^\lambda\delta_{\rho'\sigma'}-2K^\lambda_{\rho'}\tilde K^\lambda_{\sigma'}+2K^\lambda_{\sigma'}\tilde K^\lambda_{\rho'}\Big\}.\nonumber\\
\eeq
Next, using Eq.~(\ref{eq:pp5}), we observe that the last two terms inside each of the curly brackets cancel. We then arrive at
\beq
&& W_{LTT} = \frac{\big((Q^\kappa-L^\tau)\cdot \tilde K^\lambda\big)^2}{k^2\bar K^2_\lambda}\,{\rm tr}\,P^\perp(q)P^\perp(\ell),
\eeq
which, upon evaluating the scalar product and using $\smash{k=-q-\ell}$, amounts to
\beq
& & W_{LTT} =   \frac{\big((\omega^\kappa_q-\omega^\tau_\ell)k^2+\bar \omega^\lambda_k(q^2-\ell^2)\big)^2}{k^2\bar K^2_\lambda}\left[d-2-\frac{\hat\sigma}{q^2\ell^2}\right].\label{eq:wltt}\nonumber\\
\eeq

\subsubsection{$W_{TLL}$}
Using the same strategy, we arrive at
\beq
W_{TLL} &= & \Big(2\tilde Q^\kappa\cdot\tilde L^\tau+\bar \omega^\tau_\ell(L^\tau-K^\lambda)\cdot \tilde Q^\kappa \nonumber\\
&& \hspace{0.3cm}-\,\bar \omega^\kappa_q(K^\lambda-Q^\kappa)\cdot\tilde L^\tau\Big)^2\,\frac{q\cdot P^\perp(k)\cdot q}{q^2\ell^2 \bar Q^2_\kappa \bar L^2_\tau}
\eeq
which, using $k+q+\ell=0$, finally gives
\beq
W_{TLL}&=& \,\Big(\bar \omega^\tau_\ell(\omega^\tau_\ell-\omega^\lambda_k)q^2+\bar \omega^\kappa_q(\omega^\kappa_q- \omega^\lambda_k)\ell^2\nonumber\\
&& \hspace{0.5cm}-\,\bar \omega^\kappa_q \bar \omega^\tau_\ell k^2+2q^2\ell^2\Big)^2\,\frac{\hat\sigma}{k^2q^2\ell^2 \bar Q^2_\kappa \bar L^2_\tau}\,.\label{eq:wtll}
\eeq

\subsubsection{$W_{LLL}$}
In the same fashion, we find
\beq
W_{LLL}&=& \Big(( Q^\kappa-L^\tau)\cdot\tilde K^\lambda\,\tilde Q^\kappa\cdot\tilde L^\tau+\mbox{cycl. perms.}\Big)^2\nonumber\\
& & \times\,\frac{1}{k^2q^2\ell^2 \bar K_\lambda^2 \bar Q_\kappa^2 \bar L_\tau^2}
\eeq
which gives
\beq
& & W_{LLL}\!=\!\Big(\big((\omega^\kappa_q-\omega^\tau_\ell)k^2+\bar\omega^\lambda_k(q^2-\ell^2)\big)\big(q^2\ell^2+\bar\omega^\kappa_q\bar\omega^\tau_\ell q\cdot\ell\big)\nonumber\\
& & \hspace{1.5cm}+\,\mbox{cycl. perms.}\Big)^2\frac{1}{k^2q^2\ell^2 \bar K_\lambda^2 \bar Q_\kappa^2 \bar L_\tau^2}\,. 
\eeq
A little algebra allows one to rewrite the expression inside the square as
\beq
& & \frac{1}{2}(\omega^\kappa_q-\omega^\tau_\ell)\bar\omega^\kappa_q\bar\omega^\tau_\ell k^2 (k^2-q^2-\ell^2)+\bar\omega^\lambda_k(q^2-\ell^2)q^2\ell^2\nonumber\\
& & +\,\mbox{cycl. perms.}
\eeq

\section{Longitudinal component\\ in the neutral color sector}\label{sec:long}
As already mentioned, the most interesting contribution to the propagator is its longitudinal component along the neutral color modes. This is because, in the center-symmetric Landau gauge (corresponding to $\smash{\bar r=\bar r_c}$), the expectation value of the temporal gluon field along the neutral color directions (which, up to some trivial factors corresponds to $r$), is an order parameter for the confinement-deconfinement transition \cite{vanEgmond:2021jyx}. Thus, in this work, we restrict to this particular sector. 

For later use, it is convenient to recall the expression of the potential $V_{\bar r}(r)$ whose minimization with respect to $r$ in the gauge $\smash{\bar r=\bar r_c}$ allows one to identify the confinement-deconfinement transition as the departure of $r_{\rm min}$ from the center-symmetric value $\smash{r=\bar r_c}$. At one-loop order, this potential rewrites in terms of the tree-level propagators as \cite{vanEgmond:2021jyx}
\beq
V_{\bar r}(r) & = & (r_j-\bar r_j)^2\frac{m^2T^2}{2g^2}+\frac{1}{2}\int_Q \ln D^2(Q_\kappa)\nonumber\\
& - & \frac{d-2}{2}\int_Q \ln G_T^\kappa(Q)-\frac{1}{2}\int_Q \ln \frac{G_L^\kappa(Q)}{\bar Q_\kappa^2}\,,\label{eq:V}
\eeq
where summations over $j$ and $\kappa$ are implied. In the SU(2) case, concomittantly to the departure of $r_{\rm min}$ from $\bar r_c$, the curvature mass\footnote{The factor $g^2/T^2$ originates in the factor $g/T$ that exists between $r$ and the gauge field, see Eq.~(\ref{eq:1pt}).}
\beq
\frac{g^2}{T^2}\left.\frac{\partial^2 V_{\bar r_c}(r)}{\partial r^2}\right|_{r=\bar r_c}\label{eq:curv}
\eeq
vanishes, thus leading to a singular behavior of the zero-momentum longitudinal propagator in the neutral color sector. In the SU(3) case, although the mass never really vanishes, the propagator in this limit is enhanced \cite{vanEgmond:2021jyx}.

\subsection{Zero-frequency limit}
For simplicity, we shall also restrict ourselves to a vanishing Matsubara frequency.\footnote{Higher Matsubara frequencies are expected to be less sensitive to thermal effects.} In this limit, the ghost loop writes 
\beq
\Pi^{\kappa0^{(j)}\tau}_{L,gh}(k)=\int_Q\,(\bar\omega_q^\kappa)^2\,D(Q_\kappa)D(L_\tau)\,,\label{eq:l1}
\eeq
and the tadpole gives
\beq
\Pi^{\kappa0^{(j)}\tau}_{L,tad}=\int_Q\Bigg\{(d-2) G_T^\kappa(Q)+(\bar\omega_q^\kappa)^2\frac{G^\kappa_L(Q)}{\bar Q^2_\kappa}\Bigg\}\,.\label{eq:l2}
\label{tdpl}
\eeq
The relevant components of the gluon loop become
\beq
& & W_{LTT}(K^\lambda,Q^\kappa,L^\tau)\to(\omega^\kappa_q)^2\left(d-2-\frac{\hat{\sigma}}{q^2\ell^2}\right)\,,\label{eq:l3}\\
& & W_{LTL}(K^\lambda,Q^\kappa,L^\tau)\to 4\hat\sigma\frac{(L_{\tau}\cdot \bar L_{\tau})^2}{q^2\ell^2\bar L^2_\tau}\,,\label{eq:l33}\\\nonumber\\
& & W_{LLT}(K^\lambda,Q^\kappa,L^\tau)\to 4\hat{\sigma}\frac{(Q_{\kappa}\cdot \bar Q_{\kappa})^2}{q^2\ell^2\bar{Q}_{\kappa}^2}\,,\label{eq:l44}\\
& & W_{LLL}(K^\lambda,Q^\kappa,L^\tau)\nonumber\\
& & \hspace{0.5cm}\to\,\frac{(\bar\omega^\kappa_q)^2}{q^2 \ell^2 \bar{Q}^2_\kappa\bar{L}^2_\tau}\Big( \ell^2(L_{\tau}\cdot \bar L_{\tau})+q^2 (Q_{\kappa}\cdot \bar Q_{\kappa})\nonumber\\
&& \hspace{2.9cm}-\,k^2 (\ell^2+q^2+\omega_q^{\kappa}\bar \omega_q^{\kappa})\Big)^2\,.\label{eq:l4}
\eeq
\begin{widetext}
The total contribution to the longitudinal component of the self-energy in the zero-frequency limit is then
\beq\label{eq:zero}
\Pi^{\kappa0^{(j)}\tau}_{L}(k) & = & \int_Q\,(\bar \omega_q^\kappa)^2\Bigg[D^\kappa(Q)D^\tau(L)+\frac{G^\kappa_L(Q)}{\bar Q^2_\kappa}-\frac{\Big(\ell^2(L_{\tau}\cdot \bar L_{\tau})+q^2 (Q_{\kappa}\cdot \bar Q_{\kappa})+k^2 (\ell^2+q^2+\omega_q^{\kappa}\bar \omega_q^{\kappa})\Big)^2}{2q^2 \ell^2 }\frac{G^\kappa_L(Q)}{\bar{Q}^2_\kappa}\frac{G^\tau_L(L)}{\bar{L}^2_\tau}\Bigg]\nonumber\\
&+& 2\int_Q\,\Bigg[( \omega_q^\kappa)^2 G_T^\kappa(Q)G_T^{\tau}(L)-\frac{ (L_{\tau}\cdot \bar L_{\tau})^2}{\ell^2}\frac{G_L^{\tau}(L)}{\bar L_\tau^2}\frac{G_T^\kappa(Q)}{q^2}- \frac{ (Q_{\kappa}\cdot \bar Q_{\kappa})^2}{q^2}\frac{G_L^\kappa(Q)}{\bar{Q}_{\kappa}^2 }\frac{G_T^{\tau}(L)}{\ell^2}\Bigg]\hat \sigma\nonumber\\
& + & (d-2)\int_Q \,\Bigg[G_T^\kappa(Q)-2(\omega^\kappa_q)^2 G_T^\kappa(Q)G_T^{\tau}(L)\Bigg].
\eeq
As already explained above, this expression needs to be multiplied by $\smash{{\cal C}^{\kappa0^{(j)}\kappa}\equiv |f^{\kappa 0^{(j)}\tau}|^2}$ and contracted w.r.t. the color indices $\kappa$ and $\tau$. Now, since $\smash{|f^{\kappa 0^{(j)}\tau}|^2=\kappa_j^2\delta^{\tau,(-\kappa)}}$, see Ref.~\cite{Reinosa:2015gxn}, this boils down to multiplying the right-hand side of Eq.~(\ref{eq:zero}) by $\kappa_j^2$ while summing over $\kappa$. We recall that the expression needs also to be multiplied by a factor $g^2$.\\
\end{widetext}

\subsection{Retrieving the curvature mass}
In the zero-momentum limit, the previous expression should yield the one-loop contribution to the curvature mass (\ref{eq:curv}).\footnote{We note that for $\omega^\lambda_k=0$, we have $\smash{P^L_{\mu\nu}(\bar K^\lambda)=\delta_{\mu0}\delta_{\nu0}}$.} Let us use this as a cross-check of our propagator calculation. We find
\beq\label{eq:zerozero}
\Pi^{\kappa0^{(j)}\tau}_{L} & = & \int_Q\,(\bar \omega_q^\kappa)^2\left[D^2(Q_\kappa)+\frac{G^\kappa_L(Q)}{\bar Q^2_\kappa}\right.\nonumber\\
& & \hspace{1.5cm} \left. -\,2\frac{\big(Q_\kappa\cdot\bar Q_\kappa\big)^2}{\bar Q_\kappa^4}\,(G^\kappa_L(Q))^2\right]\nonumber\\
& + & (d-2)\int_Q \,\Bigg[G_T^\kappa(Q)-2(\omega^\kappa_q)^2(G_T^\kappa(Q))^2\Bigg].\nonumber\\
\eeq
Now, upon using Eq.~(\ref{eq:V}) as well as
\beq
\frac{\partial}{\partial r_j} G_T(Q_\kappa) & = & -2T\kappa_j\omega_q^\kappa G_T^2(Q_\kappa)\,,\nonumber\\
\frac{\partial}{\partial r_j} G_L(Q_\kappa) & = & -2T\kappa_j\bar\omega_q^\kappa (Q_\kappa\cdot\bar Q_\kappa) \frac{G_L^2(Q_\kappa)}{\bar Q_\kappa^2}\,,\\
\frac{\partial}{\partial r_j} D(Q_\kappa) & = & -T\kappa_j\bar\omega_q^\kappa  D^2(Q_\kappa)\,,\nonumber
\eeq
it is readily checked that
\beq
\frac{g^2}{T^2}\frac{\partial^2 V}{\partial r_j^2}=m^2+g^2\kappa_j^2\Pi^{\kappa0^{(j)}\tau}_{L}\,,
\eeq
as announced.

\section{Analytical evaluation\\ of the momentum integrals}\label{sec:mom}
The next step is to determine the sum-integrals in Eq.~(\ref{eq:zero}). The standard path would be to determine the Matsubara sums analytically and the resulting momentum integrals numerically (after prior extraction of the UV divergences). However, the fact that the denominator of $G_L$ involves a quartic polynomial in the frequencies, see Eq.~(\ref{eq:PL}), makes the implementation of this approach quite cumbersome. Here, we put forward an alternative strategy based, instead, on the analytical evaluation of the momentum integrals followed by the numerical evaluation of the resulting Matsubara sums. In this section, we concentrate on the analytical determination of the momentum integrals and leave the discussion of the resulting Matsubara sums for the next section.

Rather than presenting the strategy in full glory for the (zero-frequency) self-energy (\ref{eq:zero}), we first consider the simpler case of its zero-momentum limit (\ref{eq:zerozero}) which is already general enough to illustrate the subtleties of the approach. The corresponding result for non-vanishing momentum is quoted at the end of the section and the details are given in App.~\ref{app:reduction}.

\subsection{Vanishing momentum}
Looking at the various terms in Eq.~(\ref{eq:zerozero}), we observe that they can all be expressed in terms of $(d-1)$-dimensional scalar propagators
\beq
G_M(q)\equiv \frac{1}{q^2+M^2}\,.
\eeq
Indeed, it is easily seen that
\beq
 D^\kappa(Q) & = & G_{M_{0,q}^\kappa}(q)\,,\label{eq:M0}\\
G^\kappa_T(Q) & = & G_{M^\kappa_{T,q}}(q)\,,\\
\frac{G^\kappa_L(Q)}{\bar Q^2_\kappa} & = & G_{M^\kappa_{L+,q}}(q)G_{M^\kappa_{L-,q}}(q)\nonumber\\
& = & -\frac{G_{M^\kappa_{L+,q}}(q)-G_{M^\kappa_{L-,q}}(q)}{(M^\kappa_{L+,q})^2-(M^\kappa_{L-,q})^2}\,,\label{eq:GLsQ}
\eeq
with [the index $q$ accompagning the various here-defined masses is the Matsubara index $q\in\mathds{Z}$]
\beq
(M^\kappa_{0,q})^2 & = & \omega_q^\kappa\bar \omega_q^\kappa\,,\\
(M^\kappa_{T,q})^2 & = & (\omega^\kappa_q)^2+m^2\,,\\
(M^\kappa_{L\pm,q})^2 & = & \omega_q^\kappa\bar \omega_q^\kappa+\frac{m^2}{2}\nonumber\\
& \pm & \sqrt{\left(\omega_q^\kappa\bar \omega_q^\kappa+\frac{m^2}{2}\right)^2-(\bar\omega^\kappa_q)^2((\omega^\kappa_q)^2+m^2)}\nonumber\\
& = & \omega_q^\kappa\bar \omega_q^\kappa+\frac{m^2}{2}\pm\frac{m^2}{2}\sqrt{1+4\frac{\bar \omega_q^\kappa(\omega_q^\kappa-\bar \omega_q^\kappa)}{m^2}}\,.\nonumber\\
\eeq
We note that the ``square'' masses $(M^\kappa_{L\pm,q})^2$  are either both real or complex conjugate of each other. In any case, one has ${\rm Re}\,(M^\kappa_{L\pm,q})^2\geq 0$. We shall also need to consider the square roots of these square masses which we denote as $M^\kappa_{L\pm,q}$. 

The square masses $(M^\kappa_{L\pm,q})^2$ are the roots of the quadratic polynomial
\beq
& & X^2-(2\omega^\kappa_q\bar\omega^\kappa_q+m^2) X+(\bar\omega^\kappa_q)^2((\omega^\kappa_q)^2+m^2)\nonumber\\
& & \hspace{1.0cm}=\,(X-\omega^\kappa_q\bar\omega^\kappa_q)^2-m^2(X-(\bar\omega^\kappa_q)^2)\,.\label{eq:pol}
\eeq
Thus, in particular,
\beq
(M^\kappa_{L+,q})^2(M^\kappa_{L-,q})^2 & \!=\! & (\bar \omega^\kappa_q)^2((\omega^\kappa_q)^2+m^2)\,,\label{eq:rrr}\\
(M^\kappa_{L+,q})^2+(M^\kappa_{L-,q})^2 & \!=\! & 2\bar \omega^\kappa_q\omega^\kappa_q+m^2\,,\label{eq:rr1}\\
(M^\kappa_{L+,q})^2-(M^\kappa_{L-,q})^2 & \!=\! & m^2\sqrt{1+4\frac{\bar \omega_q^\kappa(\omega_q^\kappa-\bar \omega_q^\kappa)}{m^2}}\,.\nonumber\\
\eeq
From the second form of the polynomial (\ref{eq:pol}), we also find the identity
\beq
\big((M^\kappa_{L\pm,q})^2-\omega^\kappa_q\bar\omega^\kappa_q\big)^2=m^2\big((M^\kappa_{L\pm,q})^2-(\bar\omega^\kappa_q)^2\big)\,,\label{eq:rr2}\nonumber\\
\eeq
which will allow us to perform some simplifications below.

Thanks to Eqs.~(\ref{eq:M0})-(\ref{eq:GLsQ}), we can re-express the momentum integrals appearing in Eq.~(\ref{eq:zerozero}) in terms of the basic integral
\beq
J_M=\int_q G_M(q)\,,\label{eq:JM}
\eeq
and possibly of its $M^2$-derivative which we denote as $J'_M$. This is already pretty clear for those contributions involving $D$ or $G_T$, so let us concentrate on those involving $G_L$. Introducing the simplifying notation $\smash{M_\pm\equiv M^\kappa_{L\pm,q}}$. We have
\beq
& & \frac{\big(Q_\kappa\cdot\bar Q_\kappa\big)^2}{\bar Q_\kappa^4}\,(G^\kappa_L(Q))^2\nonumber\\
& & \hspace{0.5cm}=\,\left(\frac{(\omega^\kappa_q\bar\omega^\kappa_q+q^2)G_{M_+}(q)-(\omega^\kappa_q\bar\omega^\kappa_q+q^2)G_{M_-}(q)}{M_+^2-M_-^2}\right)^2\nonumber\\
& & \hspace{0.5cm}=\,\left(\frac{(\omega^\kappa_q\bar\omega^\kappa_q-M_+^2)G_{M_+}(q)-(\omega^\kappa_q\bar\omega^\kappa_q-M_-^2)G_{M_-}(q)}{M_+^2-M_-^2}\right)^2\nonumber\\
& & \hspace{0.5cm}=\,\frac{(\omega^\kappa_q\bar\omega^\kappa_q-M_+^2)^2G^2_{M_+}(q)+(\omega^\kappa_q\bar\omega^\kappa_q-M_-^2)^2G^2_{M_-}(q)}{(M_+^2-M_-^2)^2}\nonumber\\
& & \hspace{0.5cm}+\,\frac{2(\omega^\kappa_q\bar\omega^\kappa_q-M_+^2)(\omega^\kappa_q\bar\omega^\kappa_q-M_-^2)\big(G_{M_+}(q)-G_{M_-}(q)\big)}{(M_+^2-M_-^2)^3}\,.\nonumber\\
\eeq
When combining this with Eq.~(\ref{eq:GLsQ}), we find
\beq
& & \frac{G^\kappa_L(Q)}{\bar Q^2_\kappa}-2\frac{\big(Q_\kappa\cdot\bar Q_\kappa\big)^2}{\bar Q_\kappa^4}\,(G^\kappa_L(Q))^2\nonumber\\
& & \hspace{0.5cm}=\,-\,2\frac{(\omega^\kappa_q\bar\omega^\kappa_q-M_+^2)^2G^2_+(q)+(\omega^\kappa_q\bar\omega^\kappa_q-M_-^2)^2G^2_-(q)}{(M_+^2-M_-^2)^2}\nonumber\\
& & \hspace{0.9cm}-\,\frac{(2\omega^\kappa_q\bar\omega^\kappa_q-M_+^2-M_-^2)^2\big(G_{M_+}(q)-G_{M_-}(q)\big)}{(M_+^2-M_-^2)^3}\,.\nonumber\\
\eeq
We eventually arrive at\footnote{Some of the prefactors within the first square bracket could be further simplified using Eqs.~(\ref{eq:rr1}) and (\ref{eq:rr2}). However, we will refrain from doing so for the moment.}
\beq
& & \Pi^{\kappa0^{(j)}\tau}_{L}=T\sum_{q\in\mathds{Z}} (\bar \omega_q^\kappa)^2\Bigg[-J'_{M_0}\nonumber\\
& & \hspace{1.7cm}-\,\frac{(2\omega^\kappa_q\bar\omega^\kappa_q-M_+^2-M_-^2)^2\big(J_{M_+}-J_{M_-}\big)}{(M_+^2-M_-^2)^3}\nonumber\\
& & \hspace{1.7cm}+\,2\frac{(\omega^\kappa_q\bar\omega^\kappa_q-M_+^2)^2J'_{M_+}+(\omega^\kappa_q\bar\omega^\kappa_q-M_-^2)^2J'_{M_-}}{(M_+^2-M_-^2)^2}\Bigg]\nonumber\\
& & \hspace{1.2cm}+\,(d-2)T\sum_{q\in\mathds{Z}} \,\Big[J_{M_T}+2(\omega^\kappa_q)^2J'_{M_T}\Big]\,,\label{eq:fsum}
\eeq
\vglue1mm
\noindent{where, for simplicity, we have also set $\smash{M_T\equiv M^\kappa_{T,q}}$, $\smash{M_0\equiv M^\kappa_{0,q}}$. The function $J_M$ is easily determined within dimensional regularization,}
\beq
J_M=\frac{(M^2)^{d/2-3/2}}{(4\pi)^{d/2-1/2}}\Gamma(3/2-d/2)\,,
\eeq
\vglue0.5mm
\noindent{see App.~\ref{app:basic}, thus turning Eq.~(\ref{eq:fsum}) into an explicit Matsubara sum which still depends, however, on the dimensional regularization parameter $\smash{\epsilon\equiv(4-d)/2}$. We explain how to evaluate this type of sums in Sec.~\ref{sec:sums}.}

\subsection{Non-vanishing momentum}
Before doing so, let us briefly summarize how the same strategy allows one to rewrite the self-energy (\ref{eq:zero}) as a Matsubara sum involving the scalar integrals $J_M$ and
\beq
I_{M_1M_2}(k) & \equiv & \int_q G_{M_1}(q)G_{M_2}(\ell)\,.\label{eq:IMM}
\eeq
Details are provided in App.~\ref{app:reduction}. In order to arrive at compact formulas, it is convenient to introduce the finite difference operators
\beq
d_M f(M)& \equiv & \frac{f(M)-f(0)}{M^2}\,,\label{dm}\\
d_{M^\kappa_{L\pm,q}}f(M^\kappa_{L+,q})& \equiv & \frac{f(M^\kappa_{L+,q})-f(M^\kappa_{L-,q})}{(M^\kappa_{L+,q})^2-(M^\kappa_{L-,q})^2}\,,\label{dm2}
\eeq
as well as the auxiliary function
\beq
 \hat I_{M_1M_2}(k) & \equiv & \int_q \hat\sigma\,G_{M_1}(q)G_{M_2}(\ell)\,,\label{eq:Ih}
\eeq
which is related to the basic integrals in the following way, see App.~\ref{app:reduction} for more details,
\beq
\hat I_{M_1M_2}(k) & = & \frac{k^2-M^2_1+M^2_2}{4}J_{M_1}\nonumber\\\nonumber\\
& + & \frac{k^2+M^2_1-M^2_2}{4}J_{M_2}\nonumber\\
& - & \frac{\Delta(-k^2,M^2_1,M^2_2)}{4}I_{M_1M_2}\,,
\eeq
with
\beq
& & \Delta(-k^2,M^2_1,M^2_2)=k^4+M^4_1+M^4_2\nonumber\\
& & \hspace{1.5cm}+\,2k^2M^2_1+2k^2M^2_2-2M^2_1M^2_2\,.
\eeq 
With these auxiliary notations, the ghost loop (\ref{eq:l1}) reads
\beq
\Pi^{\kappa0^{(j)}\tau}_{L,gh}(k)=T\sum_{q\in\mathds{Z}}(\bar \omega_q^\kappa)^2I_{M_{0,q}^\kappa M_{0,\ell}^\tau}(k)\,,\label{eq:s0}
\eeq
while the tadpole (\ref{eq:l2}) rewrites
\beq
\Pi^{\kappa0^{(j)}\tau}_{L,tad}(k)=T\sum_{q\in\mathds{Z}}\Big[(d-2) J_{M_{T,q}^\kappa}-(\bar \omega_q^\kappa)^2d_{M_{L+,q}^\kappa}J_{M_{L+,q}^\kappa}\Big]\,.\label{eq:s00}\nonumber\\
\eeq
\begin{widetext}
As for the relevant components of the gluon loop, see Eqs.~(\ref{eq:l3})-(\ref{eq:l4}), they read
\beq
\Pi_{gl,LTT}^{\kappa0^{(j)}\tau}(k) & = & -2\,T\sum_{q\in\mathds{Z}}(\omega^\kappa_q)^2\,\Big[(d-2)I_{M_{T,q}^\kappa M_{T,\ell}^\tau}(k)-d_{M_{T,q}^\kappa}d_{M_{T,\ell}^\tau}\hat I_{M_{T,q}^\kappa M_{T,\ell}^\tau}(k)\Big]\,,\label{eq:s1}\\
\Pi^{\kappa0^{(j)}\tau}_{LTL}(k) & = & -2\,T\sum_{q\in\mathds{Z}}d_{M_{T,q}^\kappa}\left[\frac{(\omega_q^\kappa)^2}{(\omega_q^\kappa)^2+m^2}\hat I_{M_{T,q}^\kappa0}(k)-m^2d_{M_{L\pm,\ell}^\tau}\left(\frac{(\bar \omega_q^\kappa)^2}{(M_{L+,\ell}^\tau)^2}-1\right)\hat I_{M_{T,q}^\kappa M_{L+,\ell}^\tau}(k)\right],\label{eq:s2}\\
\Pi^{\kappa0^{(j)}\tau}_{LLT}(k) & = & -2\,T\sum_{q\in\mathds{Z}}d_{M_{T,\ell}^\tau}\left[\frac{(\omega_q^\kappa)^2}{(\omega_q^\kappa)^2+m^2}\hat I_{M_{T,\ell}^\tau0}(k)-m^2d_{M_{L\pm,q}^\kappa}\left(\frac{(\bar \omega_q^\kappa)^2}{(M_{L+,q}^\kappa)^2}-1\right)\hat I_{M_{T,\ell}^\tau M_{L+,q}^\kappa}(k)\right],\label{eq:s3}
\eeq
and
\beq
\Pi^{\kappa0^{(j)}\tau}_{LLL}(k) & = & -\frac{1}{2}T\sum_{q\in\mathds{Z}}(\bar \omega^\kappa_q)^2\,\Bigg[d_{M_{L\pm,q}^\kappa}d_{M_{L+,q}^\kappa}d_{M_{L\pm,\ell}^\tau}d_{M_{L+,\ell}^\tau}\nonumber\\
& & \hspace{3.0cm}\times\,\Big(\bar \omega^\kappa_q\omega^\kappa_qk^2+((M_{L+,q}^\kappa)^2+(M_{L+,\ell}^\tau)^2)(\bar \omega^\kappa_q \omega^\kappa_q-k^2)\nonumber\\
& & \hspace{7.5cm}-\,((M_{L+,q}^\kappa)^4+(M_{L+,\ell}^\tau)^4)\Big)^2 I_{M_{L+,q}^\kappa M_{L+,\ell}^\tau}(k)\label{eq:s44}\nonumber\\
& & \hspace{2.5cm}-\,2d_{M_{L\pm,q}^\kappa}d_{M_{L+,q}^\kappa}(k^2+m^2+(M^\kappa_{L+,q})^2)J_{M^\kappa_{L+,q}}\Bigg].\label{eq:s4}
\eeq
\end{widetext}
It should be noted that, even though $\smash{M_{L+,q}^\kappa=M_{L+,\ell}^\tau}$ in the case of a vanishing external Matsubara frequency and a neutral mode, they have to be distinguished in order for the operators \eqref{dm} and \eqref{dm2} to apply according to the labels $q$ and $\ell$.

As it was already the case for $J_M$, the basic integral $I_{M_1M_2}(k)$, and in turn $\hat I_{M_1M_2}(k)$, can be determined within dimensional regularization, see App.~\ref{app:basic}, thus turning (\ref{eq:s0})-(\ref{eq:s4}) into explicit Matsubara sums that we now explain how to evaluate.

\section{Numerical evaluation\\ of the Matsubara sums}\label{sec:sums}

As we have just illustrated, after performing the $(d-1)$-dimensional momentum integrals analytically, one is in general left with the evaluation of Matsubara sums $\sum_{q\in\mathds{Z}}f(q,\epsilon)$, where the summand $f(q,\epsilon)$ depends on the dimensional regularization parameter $\smash{\epsilon\equiv(4-d)/2}$. Before considering the numerical evaluation of these sums, we need to investigate their behavior as $\smash{\epsilon\to 0}$. 

Here, we observe that, because the UV counting of the original sum-integrals involves only even powers of the momentum, the UV counting of the integrals that appear in $f(q,\epsilon)$ involves odd powers of the momentum which do not lead to divergences in dimensional regularization. Thus, the summand $f(q,\epsilon)$ admits a finite $\smash{\epsilon\to 0}$ limit, as can be checked in particular for the summand in Eq.~(\ref{eq:zerozero}) using that
\beq
J_M\to -\frac{(M^2)^{1/2}}{4\pi}\,.
\eeq
Similar conclusions apply to the summands of Eqs.~(\ref{eq:s0})-(\ref{eq:s4}). From these considerations, one may wrongly conclude that it is always licit to reverse the order between the summation over $q\in\mathds{Z}$ and the expansion in $\epsilon$. 

This view is of course too na\"\i ve for it misses the presence of potential UV divergences in the original sum-integrals that should manifest themselves as poles in $1/\epsilon$.\footnote{We will see below that the situation is even more delicate since even when the inversion might seem licit for it leads to a convergent sum, finite, order $\epsilon^0$ contributions might be missed.} What happens in the presence of these divergences is that, despite the $\smash{\epsilon\to 0}$ limit of the summand being well defined, the corresponding sum is not convergent as $f(q,0)$ does not decrease fast enough for $|q|\to\infty$. This makes it clear that $\epsilon$ plays the role of a regulator for the Matsubara sum in those cases and that the $\epsilon$-expansion cannot be done prior to the sum.\footnote{In the standard approach where the original Matsubara sums are performed analytically, leading to effective $(d-1)$-dimensional momentum integrals, the fact that the regulator $\epsilon$ is inevitably connected to the very definition of $(d-1)$-dimensional integration makes it clear that the $\epsilon$-expansion and the momentum integration cannot be interchanged.} 

To add to these remarks, we recall that the $\epsilon$-dependence should be seen, in principle, as resulting from an analytical continuation which, a priori, requires knowing the sums analytically for those integer values of $d$ that make the sums convergent. Of course, it is not always possible to perform the sums analytically. In order to seek some guidance regarding the strategy that we should adopt, we then start by considering a simple example.

\subsection{Guiding example}
Consider the sums 
\beq
\sum_{q>0} \frac{1}{q^{a+b\epsilon}}\,,\label{eq:ex}\nonumber
\eeq 
to which our generic situation $\sum_{q\in\mathds{Z}}f(q,\epsilon)$ would resemble in the UV.\footnote{Here, we have in mind the region $q\to\infty$ but the same discussion applies to $q\to-\infty$.} As already stated above, these sums need to be understood as resulting from an analytical continuation in $\epsilon$. In that case, they can be expressed in terms of the Riemann zeta function as $\zeta(a+b\epsilon)$, and,  in the limit $\epsilon\to 0$, one has
\beq\label{eq:122}
\lim_{\epsilon\to 0}\sum_{q>0}1/q^{a+b\epsilon}=\left\{
\begin{array}{l}
\zeta(a)<\infty, \quad \mbox{if $a<1$,}\\
\sim 1/(b\epsilon), \quad \;\;\mbox{if $a=1$,}\\
\zeta(a)<\infty, \quad \mbox{if $a>1$.}
\end{array}
\right.
\eeq
On the other hand, were we to assume that the $\epsilon$-expansion can be done before the sum we would arrive at
\beq\label{eq:123}
\sum_{q>0} \lim_{\epsilon\to 0} 1/q^{a+b\epsilon}=\left\{
\begin{array}{l}
\infty, \quad\quad\quad\quad\, \mbox{if $a< 1$,}\\
\infty, \quad\quad\quad\quad\, \mbox{if $a=1$,}\\
\zeta(a)<\infty, \quad \mbox{if $a>1$.}
\end{array}
\right.
\eeq
This illustrates that the question of the interchangeability of the $q$-sum and the $\epsilon$-expansion is a subtle one. In particular, by comparing the first line of Eq.~(\ref{eq:122}) and the first line of Eq.~(\ref{eq:123}) we see that the fact that the original sum admits a well defined $\smash{\epsilon\to 0}$ limit does not necessarily mean that the sum and the $\epsilon$-expansion can be interchanged. 

One criterion that does seem to work though is when the sum that is obtained after setting $\epsilon\to 0$ in the summand is convergent, last line of Eq.~(\ref{eq:122}) and last line of Eq.~(\ref{eq:123}). Let us now build on this remark to devise our strategy.

\subsection{Strategy}
 Coming back to the generic case of a sum $\sum_{q\in\mathds{Z}}f(q,\epsilon)$, the idea is then to rewrite it as
\beq
\sum_{q\in\mathds{Z}}g(q,\epsilon)+\sum_{q\in\mathds{Z}}\Big[f(q,\epsilon)-g(q,\epsilon)\Big]\,.
\eeq
The function $g(q,\epsilon)$ should satisfy two requirements. First, the associated sum $\sum_{q\in\mathds{Z}}g(q,\epsilon)$ should be simpler than the original sum. By this, we mean that it should be possible to evaluate it analytically or to treat it with the usual method, that is rewriting it as a sum-integral and performing the original Matsubara sum analytically. Second, the subtracted $q$-sum $\sum_{q\in\mathds{Z}}\lim_{\epsilon\to 0}\big[f(q,\epsilon)-g(q,\epsilon)\big]$ should be convergent. Below, we will see how to construct functions $g(q,\epsilon)$ meeting these requirements. 

Once such a function $g(q,\epsilon)$ has been identified, not only do we know that the poles in $1/\epsilon$ of the original sum $\sum_{q\in\mathds{Z}}f(q,\epsilon)$ are entirely contained in $\sum_{q\in\mathds{Z}}g(q,\epsilon)$ but, also, interchanging the $q$-sum and the $\epsilon\to 0$ limit becomes licit for the subtracted sum:
\beq\label{eq:inversion}
& & \lim_{\epsilon\to 0} \sum_{q\in\mathds{Z}}\big[f(q,\epsilon)-g(q,\epsilon)\big]\nonumber\\
& & \hspace{0.7cm}=\,\sum_{q\in\mathds{Z}}\lim_{\epsilon\to 0}\big[f(q,\epsilon)-g(q,\epsilon)\big].
\eeq
In fact, this reasoning is correct except for one particular case which we have overlooked  because we did not take into consideration possible prefactors that could multiply Eqs.~(\ref{eq:122}) and (\ref{eq:123}). Indeed, suppose that $f(q,\epsilon)-g(q,\epsilon)$ contains certain terms of the form $\sim\epsilon\times \sum_{q>0}1/q^{a+b\epsilon}$. A priori, this does not affect the fact that the $q$-sum $\sum_{q\in\mathds{Z}}\lim_{\epsilon\to 0}\big[f(q,\epsilon)-g(q,\epsilon)\big]$ is convergent. However, re-analyzing Eqs.~(\ref{eq:122}) and (\ref{eq:123}) including the prefactor $\epsilon$, we find
\beq\label{eq:126}
\lim_{\epsilon\to 0}\epsilon\sum_{q>0}1/q^{a+b\epsilon}=\left\{
\begin{array}{l}
0 \quad \mbox{if $a<1$,}\\
\frac{1}{b} \quad \mbox{if $a=1$,}\\
0 \quad \mbox{if $a>1$,}
\end{array}
\right.
\eeq
while
\beq\label{eq:127}
\sum_{q>0} \lim_{\epsilon\to 0} \epsilon/q^{a+b\epsilon}=\left\{
\begin{array}{l}
0 \quad \mbox{if $a< 1$,}\\
0 \quad \mbox{if $a=1$,}\\
0 \quad \mbox{if $a>1$.}
\end{array}
\right.
\eeq
We then find new cases for which the sum obtained by setting $\smash{\epsilon=0}$ is convergent, and, out of these, there is one, corresponding to $\smash{a=1}$, where interchanging the $q$-sum and the $\epsilon$-expansion is not licit. This seems to invalidate the rule that we inferred above.

There are two possible solutions. One possibility is to update $g(q,\epsilon)$ in order to absorb these terms and make the interchange (\ref{eq:inversion}) licit again. In those approaches that build $g(q,\epsilon)$ out of terms of the form $1/q^{a+b\epsilon}$, see below, this is easily done and in fact quite natural since the terms $1/q^{1+b\epsilon}$ are also those that lead to the divergences in $1/\epsilon$. In other approaches, it might be cumbersome to incorporate the terms $\epsilon\times 1/q^{1+b\epsilon}$ in $g(q,\epsilon)$. In this case, a simpler method is to identify these terms and correct the formula (\ref{eq:inversion}) accordingly. More precisely, if the $q$-sum $\sum_{q\in\mathds{Z}}\lim_{\epsilon\to 0}\big[f(q,\epsilon)-g(q,\epsilon)\big]$ is convergent but $f(q,\epsilon)-g(q,\epsilon)$ contains a term $C_+\epsilon/q^{1+\gamma_+\epsilon}$ for $q\to\infty$ and a term $C_-\epsilon/(-q)^{1+\gamma_-\epsilon}$ for $q\to-\infty$, then
\beq\label{eq:inversion_2}
& & \lim_{\epsilon\to 0} \sum_{q\in\mathds{Z}}\big[f(q,\epsilon)-g(q,\epsilon)\big]\nonumber\\
& & \hspace{0.5cm}=\,\sum_{q\in\mathds{Z}}\lim_{\epsilon\to 0}\big[f(q,\epsilon)-g(q,\epsilon)\big]+\frac{C_+}{\gamma_+}+\frac{C_-}{\gamma_-}\,.
\eeq
which generalizes formula (\ref{eq:inversion}).

In what follows, we make the discussion more explicit by describing two possible methods for constructing $g(q,\epsilon)$ and by illustrating the use of Eq.~(\ref{eq:inversion_2}) on the particular cases of Eq.~(\ref{eq:fsum}) and Eqs.~(\ref{eq:s0})-(\ref{eq:s4}).

\subsection{Expansions in $q$ or $q+\bar r/2\pi$}
The most direct choice for $g(q,\epsilon)$ is to take the first orders in the expansion of $f(q,\epsilon)$ at large $q$ (by this we mean the regions $q\to\infty$ and $q\to-\infty$). The associated sum $\sum_{q\in\mathds{Z}}g(q,\epsilon)$ is similar to (\ref{eq:ex}) and can be evaluated analytically for any $\epsilon$ in terms of the Riemann $\zeta$ function. In particular, possible divergences come from terms of the form $\sum_{q>0} 1/q^{1+a\epsilon}=\zeta(1+a\epsilon)\sim 1/(a\epsilon)$. One should then include these terms in $g(q,\epsilon)$. Since it is easy to include these terms in $g(q,\epsilon)$ irrespectively of their prefactor, the subtlety discussed above can always be avoided and one can rely on (\ref{eq:inversion}).

Actually, in the present context, the integer $q$ appears always shifted by $q+\bar r/2\pi$ or $q+r/2\pi$. It might then be convenient to construct $g(q,\epsilon)$ from the expansion at large $q+\bar r/2\pi$. In this case, the associated sum  $\sum_{q\in\mathds{Z}}g(q,\epsilon)$ can be evaluated analytically for any $\epsilon$ in terms of the Hurwitz or generalized $\zeta$ function. In App.~\ref{app:Matsubara}, we illustrate this strategy using various examples.

\subsection{Expansion in $r-\bar r$}
Yet another possibility for constructing $g(q,\epsilon)$ is to take the first orders in the expansion of $f(q,\epsilon)$ around $\smash{r=\bar r}$. This strategy is quite natural since, in the background field formalism, the divergences are organized in powers of $r-\bar r$ \cite{vlongue}. This is particularly useful when the divergences are contained in the first term of the expansion. In this case indeed, it is enough to choose $g(q,\epsilon)=f(q,\epsilon)|_{r=\bar r}$ and to compute the sum as
\beq
\sum_{q\in\mathds{Z}}f(q,\epsilon)=\sum_{q\in\mathds{Z}}f(q,\epsilon)|_{r=\bar r}+\sum_{q\in\mathds{Z}}\Delta f(q,\epsilon)\,,\label{eq:sub}
\eeq
with
\beq
\Delta F(r,\bar r)\equiv F(r,\bar r)-F(\bar r,\bar r)\,.
\eeq
The first term in Eq.~(\ref{eq:sub}) can be computed via the standard method after rewriting it as a sum-integral while the second term is finite in the limit $\epsilon\to 0$ limit. However, the evaluation of this limit might be subtle due to the presence of terms of the form $\epsilon\times \sum_{q\in\mathds{Z}} 1/q^{1+b\epsilon}$ which requires the identification of the corresponding coefficients $C_\pm$ and the use of Eq.~(\ref{eq:inversion_2}).\footnote{Another way to tackle the difficulty would be to add and subtract the next term in the Taylor expansion in powers of $r-\bar r$. This can be cumbersome, however, for it implies pushing the Taylor expansion to the next order.}

\subsection{Back to the self-energy}
The gluon self-energy in the neutral color sector enters the category of functions that can be treated using Eq.~(\ref{eq:sub}). We shall then compute it as\footnote{For the gluon self-energy with $\lambda\neq 0$, we would need to subtract an extra term in the expansion. This is related to the presence of coupling type divergences which are not needed in the case $\lambda=0$ because the associated structure constants vanish in this case.}
\beq
\Pi^{0^{(j)}}_L(K)=\Pi^{0^{(j)}}_L(K)|_{r=\bar r}+\Delta\Pi^{0^{(j)}}_L(K)\,,\label{eq:sub2}
\eeq
with
\beq
\Delta\Pi^{0^{(j)}}_L(K)\equiv\Pi^{0^{(j)}}_L(K)-\Pi^{0^{(j)}}_L(K)|_{r=\bar r}.\label{eq:sub3}
\eeq
The first term in Eq.~(\ref{eq:sub2}) corresponds to our old result for the self-energy in the presence of a self-consistent background \cite{Reinosa:2016iml}, while the second term is finite and can be computed by first performing the momentum integrals analytically and then the resultant Matsubara sums numerically after the $\smash{\epsilon\to 0}$ limit has been taken using Eq.~(\ref{eq:inversion_2}).

\subsubsection{Vanishing momentum}
To see more precisely how this works, let us consider the case of the mass (\ref{eq:fsum}). In this case, a careful analysis reveals that the correction terms in Eq.~(\ref{eq:inversion_2}) stem from those terms involving the propagator $G_T$, so let us concentrate on these terms for now on and more precisely on their contribution to the subtracted piece (\ref{eq:sub3}). Were we to set $\smash{\epsilon=0}$, we would find that the summand of the subtracted sum behaves as
\beq
\Delta\left[T^2\frac{\kappa\cdot r}{\pi}-\frac{3m^4\kappa\cdot r}{128\pi^5T^2q^4}+\frac{3(\kappa\cdot r)^2m^4T^2}{128\pi^6T^4q^5}\dots\right]\label{eq:143}\nonumber\\
\eeq
as $q\to\infty$, and
\beq
\Delta\left[-T^2\frac{\kappa\cdot r}{\pi}+\frac{3m^4\kappa\cdot r}{128\pi^5T^2q^4}-\frac{3(\kappa\cdot r)^2m^4T^2}{128\pi^6T^4q^5}\dots\right]\label{eq:144}\nonumber\\
\eeq
as $q\to -\infty$. When adding these contributions (renaming $q$ as $-q$ when appropriate), we would find that the first two terms cancel while the last ones add up leading to a rather convergent subtracted sum which seems (na\"\i vely) to comfort us in our choice of setting $\smash{\epsilon=0}$ from the beginning. 

However, redoing the analysis for $\epsilon\neq 0$, in addition to the terms that lead to the expansions (\ref{eq:143}) and (\ref{eq:144}), we find the terms
\beq
-\epsilon\times\frac{\Delta(\kappa\cdot r)^2T^2}{2\pi^2q^{1+2\epsilon}}
\eeq
for $q\to\infty$ and 
\beq
-\epsilon\times\frac{\Delta(\kappa\cdot r)^2T^2}{2\pi^2(-q)^{1+2\epsilon}}
\eeq
for $q\to -\infty$. These terms do not contribute if we take the $\epsilon\to 0$ limit prior to the sum over $q$. However, if we take the sum over $q$ first, they contribute together as
\beq
-\frac{\Delta(\kappa\cdot r)^2T^2}{2\pi^2}\,,
\eeq
where we used that $\smash{\epsilon\,\zeta(1+2\epsilon)\to 1/2}$.  The final contribution to $\Delta\Pi^{\kappa 0^{(j)}\tau}_L(K)$ from the last line of Eq.~(\ref{eq:fsum}) is then
\beq
-\frac{T}{2\pi}\Delta\left\{\sum_{q\in\mathds{Z}}\Big[(M^2_{T})^{1/2}+(\omega^\kappa_q)^2(M^2_{T})^{-1/2}\Big]-\frac{(\kappa\cdot r)^2T}{\pi}\right\}.\label{eq:ccc}\nonumber\\
\eeq
We have checked numerically that this formula coincides with the one obtained by applying the usual strategy of performing the Matsubara sums analytically on the last line of Eq.~(\ref{eq:fsum}). In the case of a vanishing gluon mass, because the result of this last line is known exactly, we can even check analytically the relevance of the extra term Eq.~(\ref{eq:ccc}), see App.~\ref{app:Matsubara}.

As already mentionned, the other contributions to Eq.~(\ref{eq:fsum}) do not lead to correction terms in Eq.~(\ref{eq:inversion_2}).\footnote{For the sake of completeness, we should mention that in \cite{vanEgmond:2021jyx} the curvature mass was computed by using our new computational strategy only on the part that depends on $D$ and $G_L$ while the contribution involving $G_T$ was computed using the usual strategy of performing the Matsubara sums analytically. In this mixed approach, we did not have to deal with the subtleties discussed in the present work, although we did not realize this immediately.} Thus, $\epsilon$ can be safely set equal to $0$ in the corresponding contribution to $\Delta\Pi^{\kappa 0^{(j)}\tau}_L(K)$. We find
\beq\label{eq:interm}
&  &\frac{T}{4\pi}\sum_{q\in\mathds{Z}}(\bar\omega_q^\kappa)^2\Delta\Bigg\{\frac{1}{2M_0}+\frac{1}{(M_+^2-M_-^2)^2}\nonumber\\
& & \hspace{0.5cm}\times\,\Bigg[-\frac{(\omega^\kappa_q\bar\omega^\kappa_q-M_+^2)^2}{M_+}-\frac{(\omega^\kappa_q\bar\omega^\kappa_q-M_-^2)^2}{M_-}\nonumber\\
& & \hspace{1.0cm}+\,\frac{(2\omega^\kappa_q\bar\omega^\kappa_q-M_+^2-M_-^2)^2}{M_++M_-}\Bigg]\Bigg\},
\eeq
which becomes, after some algebra,
\beq
&  & \frac{T}{4\pi}\sum_{q\in\mathds{Z}}(\bar\omega_q^\kappa)^2\Delta\Bigg\{\frac{1}{2M_0}-\frac{(\omega^\kappa_q\bar\omega^\kappa_q+M_+M_-)^2}{M_+M_-(M_++M_-)^3}\Bigg\}.\nonumber\\
\eeq
Alternatively, one could use Eqs.~(\ref{eq:rr1}) and (\ref{eq:rr2}) to arrive at
\beq
&  & \frac{T}{4\pi}\sum_{q\in\mathds{Z}}(\bar\omega_q^\kappa)^2\Delta\Bigg\{\frac{1}{2M_0}+\frac{m^2}{(M_+^2-M_-^2)^2}\nonumber\\
& & \times\Bigg[\left(\frac{(\bar\omega^\kappa_q)^2}{M_+M_-}-1\right)(M_++M_-)+\frac{m^2}{M_++M_-}\Bigg]\Bigg\}.\nonumber\\
\eeq
These formulas can be further simplified using Eqs.~(\ref{eq:rrr}) and (\ref{eq:rr1}) which lead to
\beq
M_+M_- & = & |\bar\omega_q^\kappa|\sqrt{(\omega^\kappa_q)^2+m^2}\,,\label{eq:eq11}\\
M_++M_- & = & \sqrt{|\omega_q^\kappa|+\sqrt{(\omega_q^\kappa)^2+m^2}}\nonumber\\
& \times & \sqrt{\sqrt{(\omega_q^\kappa)^2+m^2}-|\omega_q^\kappa|+2|\bar\omega_q^\kappa|}\,,\label{eq:eq22}
\eeq 
and also to (recall that $\smash{\omega_q^\kappa\bar\omega_q^\kappa\geq 0}$ and thus $\smash{\omega_q^\kappa\bar\omega_q^\kappa=|\omega_q^\kappa||\bar\omega_q^\kappa|}$)
\beq
\omega_q^\kappa\bar\omega_q^\kappa+M_+M_-=|\bar\omega_q^\kappa|\left(|\omega_q^\kappa|\!+\!\sqrt{(\omega^\kappa_q)^2\!+m^2}\right)\!.
\eeq

\subsubsection{Non-vanishing momentum}
As for the non-vanishing momentum case, the same strategy can be applied. For the case of interest, it can be checked that the correction term in Eq.~(\ref{eq:inversion_2}) does not depend on $k$, so we should use the same correction term as in Eq.~(\ref{eq:ccc}).

\section{Results}\label{sec:res}

Let us now present our results for the longitudinal SU(2) and SU(3) propagators at vanishing frequency and along the neutral color modes.

\begin{figure}[t]
	\includegraphics[width=\linewidth]{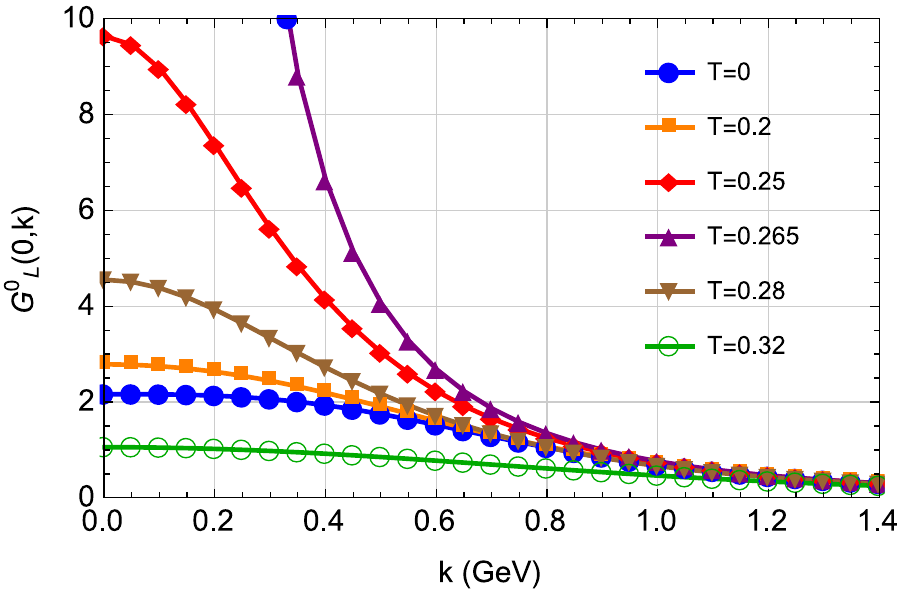}\\\includegraphics[width=\linewidth]{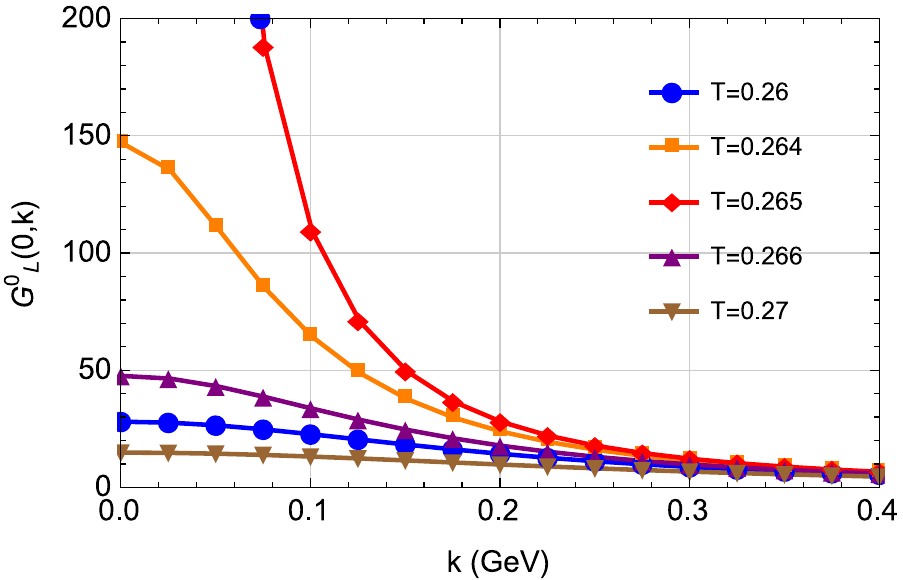}
	\caption{The one-loop SU(2) longitudinal propagator in the neutral sector at vanishing frequency as a function of the spatial momentum $k$ for various temperatures (top) and for temperatures around $T_c$ (bottom). Temperature units are in GeV.}
	\label{fig:su2}
\end{figure}

\begin{figure}[t]
	\includegraphics[width=\linewidth]{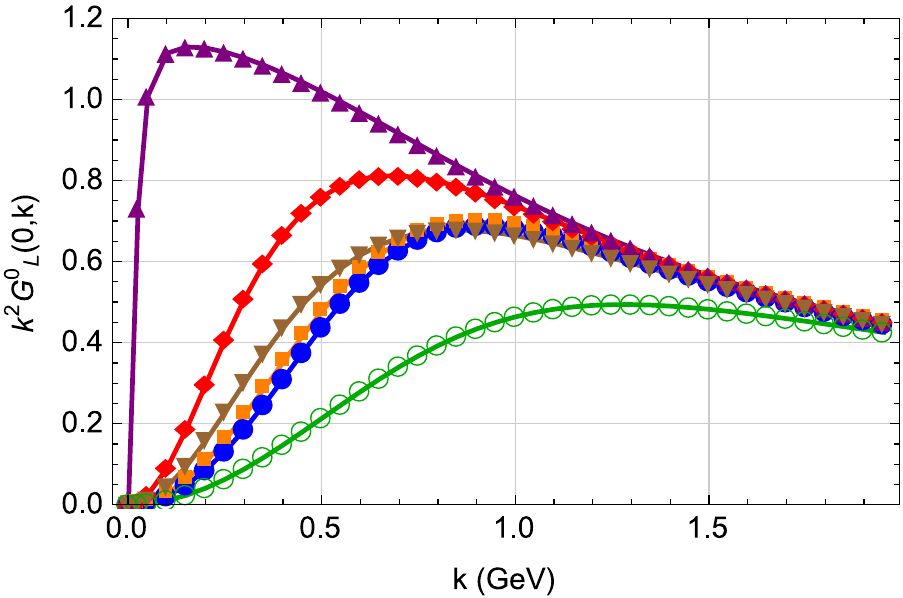}\\ \includegraphics[width=\linewidth]{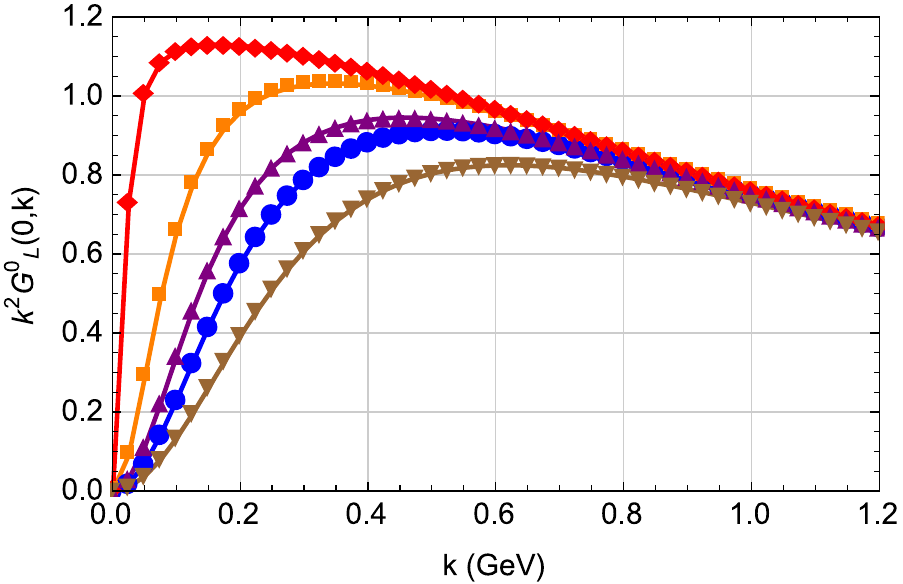}
	\caption{The one-loop SU(2) longitudinal dressing functions at vanishing frequency as functions of the spatial momentum $k$ for various temperatures (top) and for temperatures around $T_c$ (bottom). The color codes are those of the corresponding figures in Fig.~\ref{fig:su2}.}
	\label{fig:dressing}
\end{figure}

\subsection{SU(2)}

For the SU(2) group, there is one neutral mode. Therefore, the vector $r^j$ has only one component, which we denote $r$. We use the parameter values $\smash{m=680}$ MeV and $\smash{g=7.5}$ that were obtained in Ref.~\cite{Tissier:2011ey} by fitting one-loop zero-temperature CF propagators to lattice data in the Landau gauge, using the same vanishing momentum (VM) renormalization scheme as in the present work. The values of $r$ are found by minimizing the potential $V_{\bar r_c}(r)$, as was done in Ref.~\cite{vanEgmond:2021jyx}. For SU(2), a center-symmetric state corresponds to $\smash{r_{\text{min}}=\bar r_c=\pi}$, and it was found in \cite{vanEgmond:2021jyx} that $r_{\text{min}}$ deviates from this value at the transition temperature $T_c \simeq 265$ MeV. 

Our results for the propagator are presented in Fig.~\ref{fig:su2} while the corresponding dressing function is shown in Fig.~\ref{fig:dressing} for the same temperatures, below, at, and above $T_c$. It should be stressed that, for $\smash{T=0}$, the self-energy features a $k^2\ln k^2/\mu^2$ contribution at small $k$ \cite{Tissier:2011ey} which in turn implies a non-monotonous behavior in this region (not visible in the plots). The origin of this term is the loop of ghosts,\footnote{It can be argued that the one-loop ghost diagram dominates in the deep infrared \cite{Tissier:2011ey}.} see the second diagram of Fig.~\ref{fig_AA}, which are massless at zero temperature. In contrast, for any strictly positive temperature, the color charged ghosts become effectively massive in the presence of the background.\footnote{By this, we mean that the corresponding propagators are not singular in the limit of vanishing Matsubara frequency and momentum.} Since these are the only ghosts that are coupled in the loop to the neutral gluons, the $k^2\ln k^2/\mu^2$ term is absent from the infrared expansion of the $\smash{T>0}$ expressions. Nevertheless, the propagators remain non-monotonic in a range of momenta below $k\sim 60$ MeV for temperatures below $T\sim 30$ MeV. This is because, the $k^2\ln k^2/\mu^2$ logarithmic behavior at $T=0$  leaves place, at $T>0$, to $k^2$ terms in the infrared expansion whose coefficient comes with the opposite sign as compared to the standard situation.\footnote{A simple toy model to understand these features is provided by the family of functions $1/(1+x\ln(x+a))$ parametrized by $a\geq 0$.} Above these temperatures, the propagators are smooth, decreasing functions of $k$ that become almost independent of temperature for momenta larger than $800$ MeV.

We find various signatures of the transition from the propagator/dressing function. In particular, in the considered momentum range, the propagator at any given value of the momentum $k$ increases with increasing $T$ until $T_c$ and then decreases. Obviously the same behavior applies to the dressing function. These signatures are however not easy to connect to the breaking of center-symmetry.

As we have already argued above, a clear-cut signature of center-symmetry breaking is the electric susceptibility or zero-momentum value of the propagator. As a function of $T$, it shows the same non-monotonic behaviour as the propagator at any other value of $k$, with the added property that it diverges at $T_c$. This divergence is directly connected to the continous breaking of symmetry of the center-symmetric potential $V_{\bar r_c}(r)$ \cite{vanEgmond:2021jyx} and thus provides a clear signature of center-symmetry breaking. As we discuss below, this signature is absent in other choices of gauge (that is other choices of $\bar r$).

The divergence of the susceptibility at $T_c$ is paramount to the vanishing of the zero-momentum mass. Another relevant scale that shows a similar behavior is given by the momentum $k_{\rm max}(T)$ at which the dressing function is maximal.\footnote{We note that this scale is scheme-independent by construction as opposed to the zero-momentum or curvature mass.} We find that it decreases as one increases the temperature in the confining phase until $\smash{T=T_c}$ at which $\smash{k_{\rm max}(T_c)=0}$. The reason why the scale vanishes at $T_c$ is that the dressing functions for  $\smash{T<T_c}$ all vanish at $\smash{k=0}$, whereas the dressing function at $\smash{T=T_c}$ reaches instead a maximal non-zero value at $\smash{k=0}$. We stress that this behavior is specific of our computational one-loop set-up which does not provide an accurate description of the vicinity of the critical region. In particular, as discussed above, the low momentum behavior of the inverse propagator for the neutral mode at $\smash{T=T_c}$ involves a pure $k^2$ contribution with no logarithm and thus no trace of anomalous behavior. In contrast, in the presence of a (strictly positive) anomalous dimension $\eta$, the dressing function at $T_c$ would also vanish at $k=0$ and therefore the temperature dependence of $k_{\rm max}(T)$ would need to be revisited. Since $\eta$ is tiny, however,\footnote{The value of $\eta$ should be that of the Ising model in $\smash{d=3}$ dimensions. This can be seen by noticing that, at $T_c$, the effective theory for the massless modes $\tilde a_0^{0^{(3)}}$ and $c^{0^{(3)}}$ involves no coupling between gluons and ghosts (due to the anti-ghost shift symmetry of the original action), and, in the gluon sector, amounts to a scalar theory with $Z_2$ (center) symmetry. Moreover, only the zero Matsubara mode can become critical explaining why $\smash{d=3}$.} we expect the present conclusions to be modified only slightly, with a turning point value $k_{\rm max}(T_c)$ that remains close to $0$.

\subsection{SU(3)}

\begin{figure}[t]
	\includegraphics[width=\linewidth]{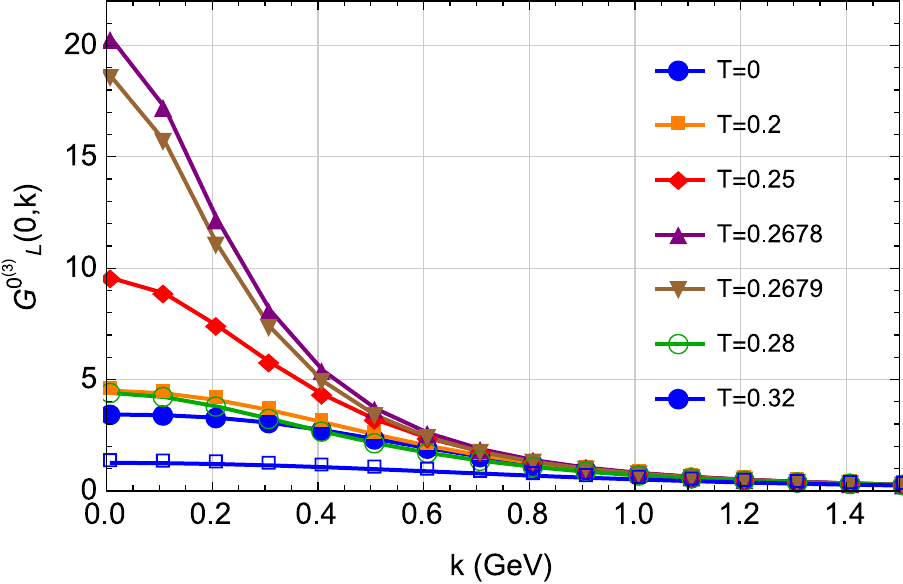}\\	\includegraphics[width=\linewidth]{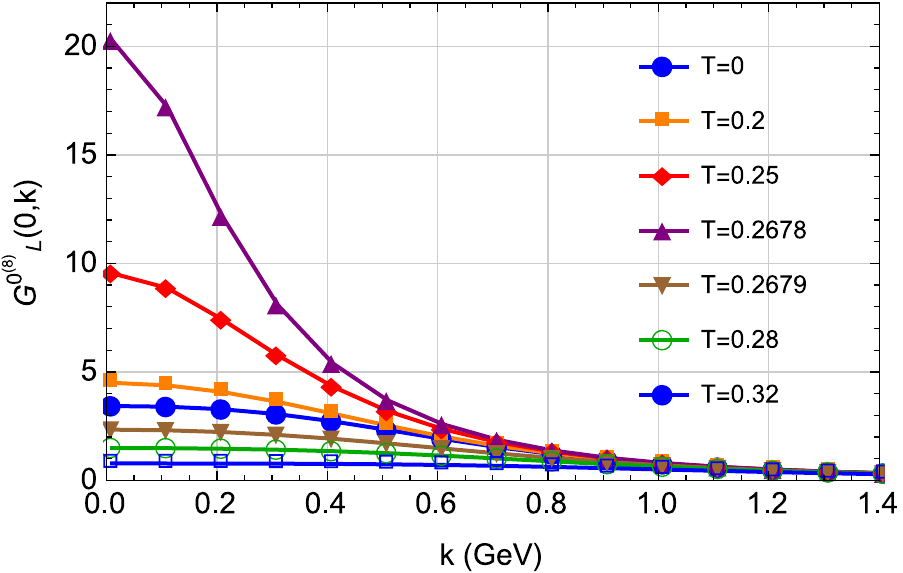}
	\caption{The one-loop SU(3) longitudinal propagator at vanishing frequency as functions of the spatial momentum $k$ for various temperatures for the neutral mode $0^{(3)}$ (top) and the neutral mode  $0^{(8)}$ (bottom). The temperature $T=0.2679$ corresponds to a temperature slightly above the transition.}
	\label{fig:su3}
\end{figure}

\begin{figure}[t]
	\includegraphics[width=\linewidth]{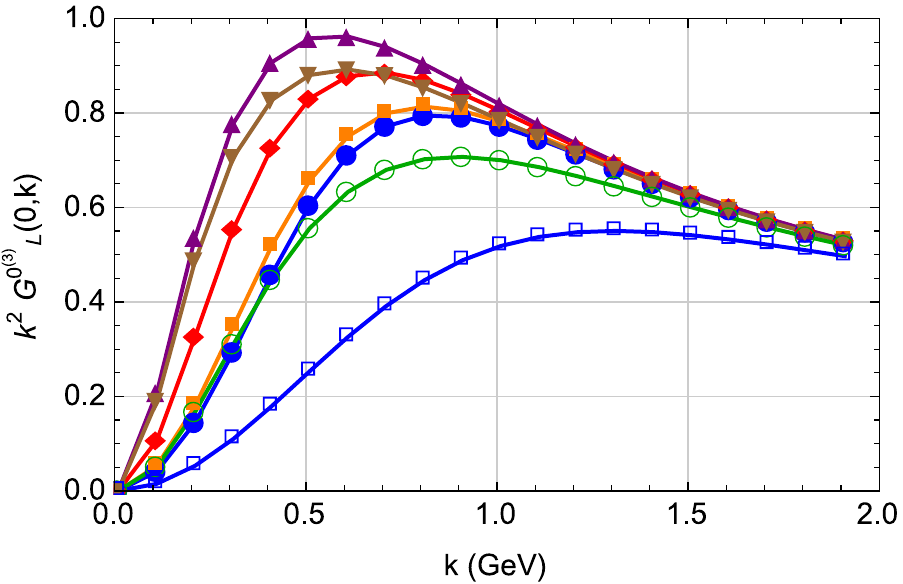}\\ \includegraphics[width=\linewidth]{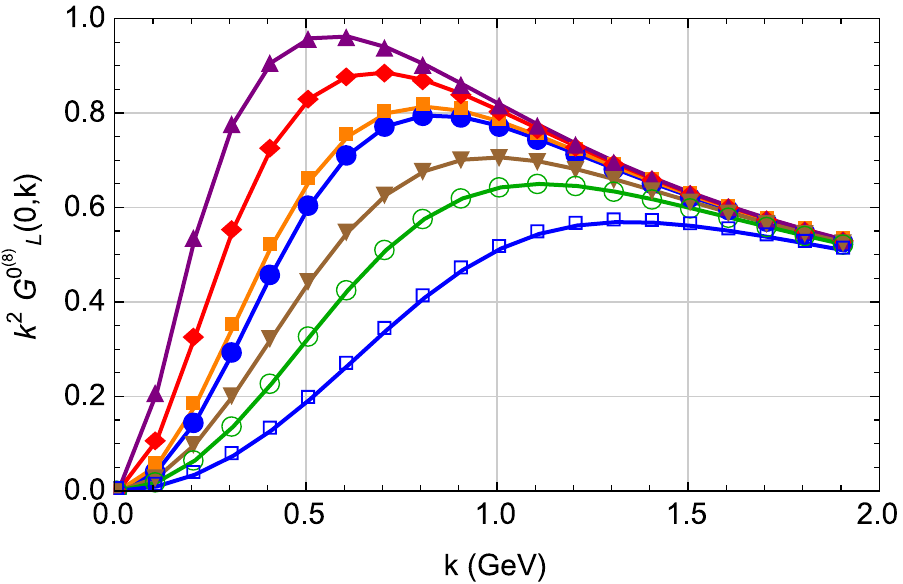}
	\caption{The one-loop SU(3) longitudinal propagator at vanishing frequency as functions of the spatial momentum $k$ for various temperatures for the neutral mode $0^{(3)}$ (top) and the neutral mode  $0^{(8)}$ (bottom). The color codes are those of the corresponding figures in Fig.~\ref{fig:su3}.}
	\label{fig:dressing3}
\end{figure}

In the SU(3) case there are two neutral modes, $0^{(3)}$ and $0^{(8)}$. The vector $r^j$ is thus two-dimensional $\smash{r=(r_3, r_8)}$, and the center-symmetric background configuration that defines the center-symmetric Landau gauge is $\smash{\bar r_c=(4 \pi/3,0)}$. Because YM theory is charge-conjugation invariant, we can, without loss of generality, restrict our analysis to\footnote{This is true as long as we also choose $\smash{\bar r_8=0}$.} $\smash{r_8 = 0}$ \cite{Reinosa:2015gxn,Reinosa:2020mnx}. Yet, we can investigate the propagator in both neutral color directions $0^{(3)}$ and $0^{(8)}$. The corresponding self-energies can be both conveniently expressed in terms of the SU(2) one-loop self-energy $\Pi^{0}_{L, SU(2)} (r,4\pi/3)$ in the gauge $\smash{\bar r=4\pi/3}$.\footnote{We stress that this is not the center-symmetric Landau gauge in the SU(2) case.} We find
\beq
\Pi^{0^{(3)}}_{L, SU(3)} (r_3) & = & \Pi^{0}_{L, SU(2)}\left(r_3,  \frac{4 \pi}{3}\right)\nonumber\\
& + & \frac{1}{2}\Pi^{0}_{L, SU(2)}\left(\frac{r_3}{2}, \frac{2 \pi}{3}\right),
\eeq
for the neutral mode $0^{(3)}$, while
\beq
\Pi^{0^{(8)}}_{L, SU(3)} (r_3)=\frac{3}{2}\Pi^{0}_{L, SU(2)}\left(\frac{r_3}{2}, \frac{2 \pi}{3}\right),
\eeq
for the neutral mode $0^{(8)}$. We use the parameters $\smash{m=540}$ MeV and $\smash{g=4.9}$ that were obtained in the same way as for the SU(2) case, by fitting to the zero-temperature lattice data in Landau gauge. With these parameters, it was found in Ref.~\cite{vanEgmond:2021jyx} that a first-order phase transition occurs at $\smash{T_c \simeq 267}$ MeV.

The propagators for both neutral modes are shown in Fig.~\ref{fig:su3} and the corresponding dressing functions in Fig.~\ref{fig:dressing3}. Some of the signatures of the transition are common to the SU(2) case, in particular the fact that the propagators/dressing functions increase with $T$ until $T_c$ and then decrease. Also the scale $k_{\rm max}(T)$ has its turning point at $T_c$. However, the fact that the transition is first-order means that we do not expect the electric susceptibility to diverge at $T_c$, and this is indeed what we see in Fig. \ref{fig:su3}. Similarly, the scale $k_{\rm max}(T)$ does not vanish at $T_c$.

\begin{figure}[t]
	\includegraphics[width=\linewidth]{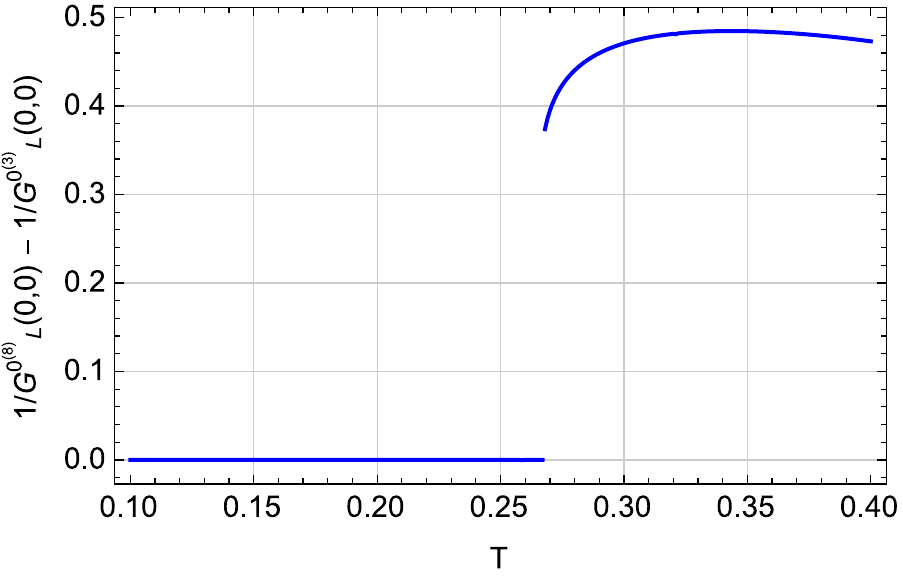}\\ \includegraphics[width=\linewidth]{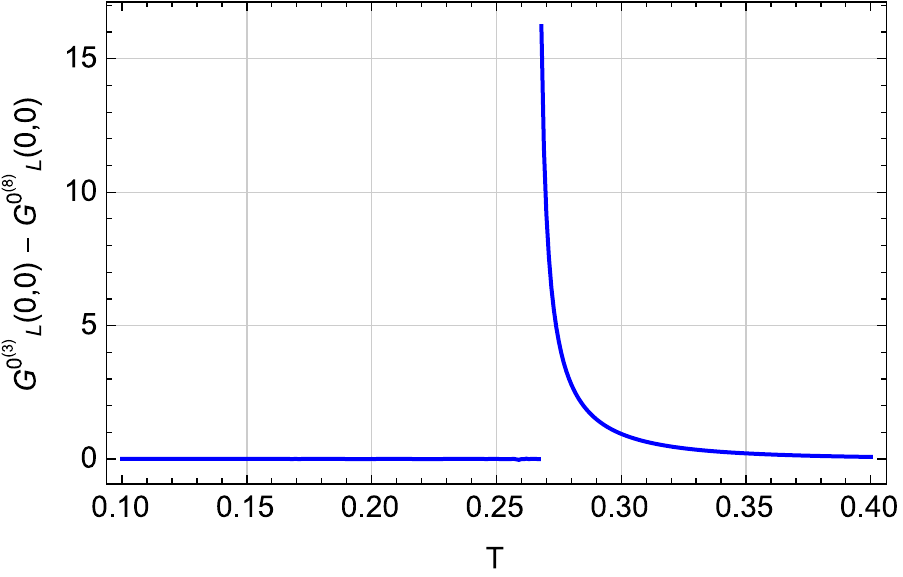}
	\caption{Order parameters for the SU(3) transition constructed from the gluon propagator in the center-symmetric Landau gauge. Top: difference of the zero-momentum square masses along the color directions $3$ and $8$. Bottom: difference of the zero-momentum propagators along the color directions $3$ and $8$.}
	\label{fig:order}
\end{figure}

The question is then whether, in the SU(3) case, we can find clear-cut signatures for center-symmetry breaking as we did for the SU(2) case. For instance, even though the zero-momentum propagator does not diverge at the transition, we do see an enhancement of the propagator towards $T_c$ where it reaches its maximal value. This observation is in line with the findings based on the effective potential and the curvature mass in Ref.~\cite{vanEgmond:2021jyx}. However, it is not obvious how to connect it to the breaking of center-symmetry. We also mention that the propagator jumps at the transition (due to the fact that the order parameter jumps). The jump is, however, not that dramatic along the color direction $3$ and therefore should not be easily detectable in lattice simulations. We find that the jump in the color direction $8$ is larger but we have no first principle argument of why this should be so generically. Moreover, it is again not obvious how to connect this jump to the breaking of symmetry. 

In contrast we can construct a clear-cut probe for the transition as follows. We first note that
\beq
& & \Pi^{0^{(8)}}_{L, SU(3)} (r_3)-\Pi^{0^{(3)}}_{L, SU(3)} (r_3)\nonumber\\
& & \hspace{0.2cm}=\,\Pi^{0}_{L, SU(2)}\left(\frac{r_3}{2}, \frac{2 \pi}{3}\right)-\Pi^{0}_{L, SU(2)}\left(r_3, \frac{4 \pi}{3}\right).
\eeq
Then, owing to the fact that $\smash{r_3=4\pi/3}$ in the low temperature phase and that $\Pi^{0}_{L, SU(2)}\left(r,\bar r\right)$ is invariant under the center-symmetry transformation $\smash{(r,\bar r)\to(2\pi-r,2\pi-\bar r)}$, it follows that the propagators for both neutral modes coincide in the confining phase. This is clearly visible in Fig.~\ref{fig:su3} and identifies yet another possible practical probe of the transition. In fact, what we have found is that the difference of the propagators along the $3$ and $8$ directions in the center-symmetric Landau gauge is an order parameter for the SU(3) transition. We illustrate this in Fig.~\ref{fig:order} by showing this difference at vanishing momentum, as a function of the temperature. As we argue below, this signature is absent in other choices of gauge.

\begin{figure}[t]
	\includegraphics[width=\linewidth]{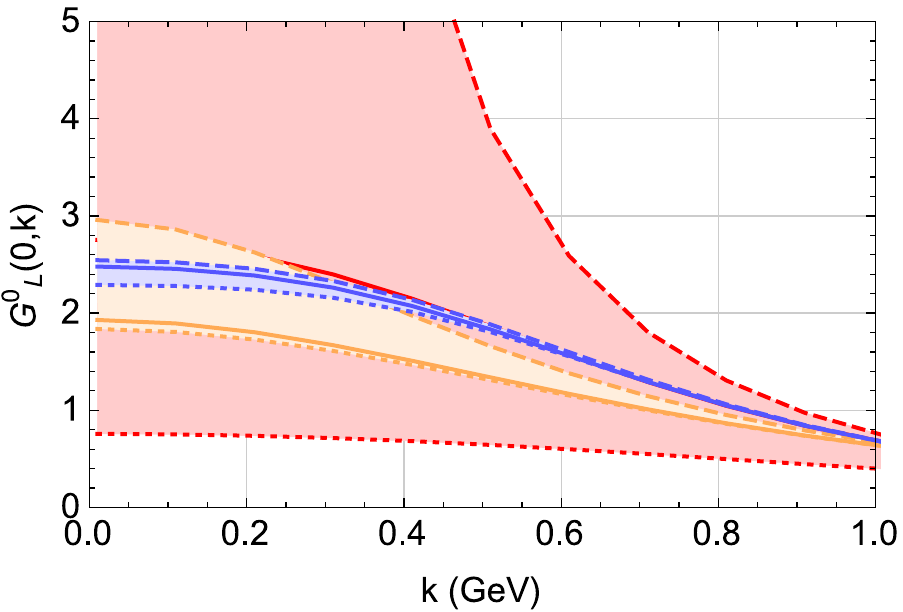}
	\caption{The one-loop SU(2) longitudinal propagator in various gauges: center-symmetric Landau gauge (red), self-consistent Landau gauge (orange), standard Landau gauge (blue). In each gauge we show the propagator at the temperature $T_{\rm peak}$ at which it reaches is maximum height (dashed curves), as well as the temperatures $0.75T_{\rm peak}$ (plain curves) and $1.25T_{\rm peak}$ (dotted curves).}
	\label{fig:comparison}
\end{figure}

\subsection{Discussion}
It is interesting to confront the above results to those obtained in other gauges\footnote{Here, we restrict to those works that rely on the Curci-Ferrari model. In the case of the standard Landau gauge, there exist of course many approaches to the evaluation of the propagators at finite temperature \cite{Fister:2011uw, Fischer:2012vc, Fukushima:2013xsa, Huber:2012kd, Quandt:2015aaa}.} such as the standard Landau gauge \cite{Reinosa:2013twa} or the self-consistent background Landau gauge \cite{Reinosa:2016iml}. The former is obtained by setting $\smash{\bar r=0}$ in our expressions, as well as $\smash{r=0}$ since it turns out that the minimum of $V_{\bar r=0}(r)$ is always located at $\smash{r=0}$.\footnote{This is guaranteed by the fact that there is no preferred color direction in the Landau gauge and the gluon field should then average to $0$. For high enough temperatures we find a non-trivial local minimum though but it never becomes the absolute one.} As for the propagator in the self-consistent Landau gauge, it is obtained by setting $\smash{r=\bar r}$ (we have checked that this gives back the result of Ref.~\cite{Reinosa:2016iml}) and determining $\bar r$ from the minimization of the {\it background effective potential} $\smash{\tilde V(\bar r)=V_{\bar r}(r=\bar r)}$. The latter is center-symmetric, just as $\smash{V_c(r)\equiv V_{\bar r_c}(r)}$, and thus offers an alternative way to study the deconfinement transition \cite{vanEgmond:2021jyx,vlongue}.

Here, it is important to stress that, even though the center-symmetric Landau gauge and the self-consistent Landau gauge allow one to capture the breaking of center-symmetry, they lead to different one-loop predictions for the transition temperatures. In the center-symmetric Landau gauge, we find a transition temperature $\smash{T_c\simeq 265\,}$MeV while in the self-consistent Landau gauge, for the same zero-temperature parameters, we find $\smash{\bar T_c\simeq 227\,}$ MeV, smaller than $T_c$. As we have pointed out in Ref.~\cite{vanEgmond:2021jyx}, we attribute these differences to the fact that the self-consistent background approach relies on an identity which is not exactly fulfilled when approximations or modelling are considered. On the other hand, we notice that both propagators are maximal at their respective transition temperatures. For more readability, we shall therefore compare the propagators in terms of a rescaled temperature such that the maximal values are all obtained when this rescaled temperature is equal to $1$. This is even more necessary in the Landau gauge, where we do not have a good grasp on the order parameter, and, therefore we cannot access the value of the transition temperature.

The comparison for the SU(2) case is presented in Fig.~\ref{fig:comparison}. We stress that the propagator being a gauge-variant quantity, we do not expect its realizations in various gauges to coincide. The goal of Fig.~\ref{fig:comparison} is then less to illustrate the quantitative differences between the various propagators but more their qualitative differences, in particular in regard to their ability to identify the deconfinement transition. It is clear that the propagator in the center-symmetric Landau gauge is the more prone to allow for such an identification.

It is maybe surprising at first sight that the propagator in the self-consistent Landau gauge does not diverge at the transition. Indeed, since the self-consistent background equals $\bar r_c$ in the low temperature phase, it should coincide in this phase with the propagator in the center-symmetric Landau gauge. However, because the corresponding transition temperatures $\bar T_c$ and $T_c$ do not coincide for the reasons mentioned above, and even though the two propagators indeed coincide for $T<\bar T_c$, the one in the self-consistent Landau gauge departs from the one in the center-symmetric Landau gauge above $\bar T_c$ and it does not have the possibility to become critical at $T_c$. Only in the absence of spurious effects related to the use of approximations or modelling, would the two temperatures coincide and the propagator in the self-consistent Landau gauge would become critical at the transition as well.\footnote{Note that $\tilde V^{(2)}(\bar r=\bar r_c)$ always vanishes at $\bar T_c$ by definition, but this is not the zero-momentum mass obtained from the propagator.}

\begin{figure}[t]
	\includegraphics[width=\linewidth]{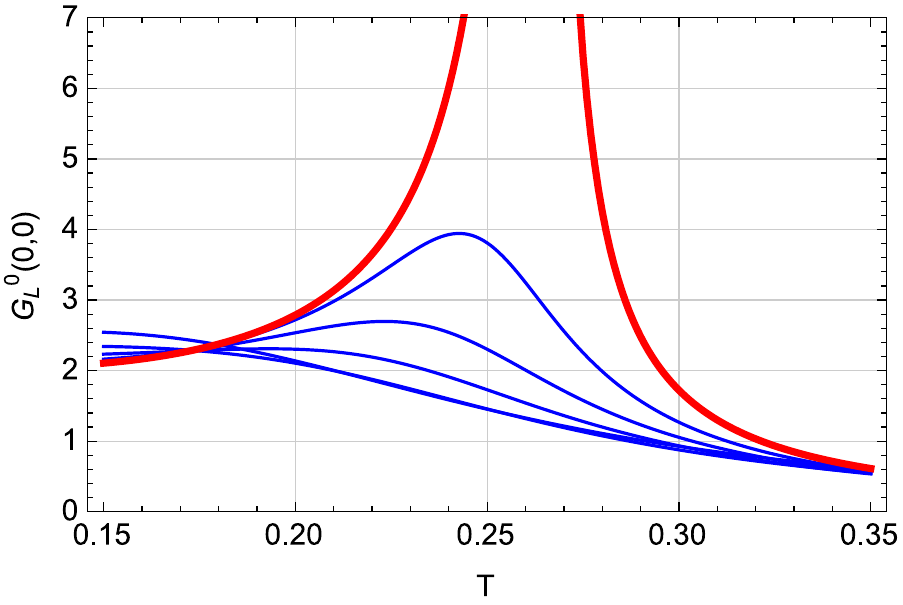}
	\caption{The electric SU(2) susceptibility as a function $T$, as computed within the CF model and for various choices of $\bar r$ ($\bar r/\pi=0$, $0.2$, $0.4$, $0.6$ and $0.8$). The red curve corresponds to the center-symmetric value $\bar r=\pi$.}
	\label{fig:gauge}
\end{figure}

It is also worth wondering why the Landau gauge propagator does not become critical at the transition. After all, in the exact theory, one should obtain the same physics from minimizing $V_{\bar r}(r)$ for any value of $\bar r$ including $\smash{\bar r=0}$. However, we have already emphasized that this is usually not enough to account for the symmetry because what matters is that the potential is center-symmetric with respect to $r$ and this occurs only in the gauge $\bar r=\bar r_c$. We illustrate this in Fig.~\ref{fig:gauge} where we plot the zero-momentum propagator as a function of $T$ in various gauges obtained by varying $\bar r$ in the interval $[0,\pi]$, the Landau gauge corresponding to $\bar r=0$. Only in the gauge where the potential is center-symmetric, does the susceptibility diverges at the transition.

To demistify the question even further, we can consider the following model. Suppose that the potential writes ($x=r-\bar r_c$, $y=\bar r-\bar r_c$):\footnote{Of course, this is a very simplistic model since in general, one is not able, without a priori knowledge, to separate so easily the physical information $-p(T)$ contained in the potential  from the gauge degrees of freedom.}
\beq
V(x,y)=-p(T)+W(x,y)-W(x_{\rm min}(y),y)
\eeq
with
\beq
W(x,y)=-a(T)\frac{x^2}{2}+\frac{x^4}{4}+xy\,,
\eeq
and where $x_{\rm min}(y)$ is any of the absolute minima of $V(x,y)$ for a given $y$ (representing a choice of gauge). By construction, we have $V(x_{\rm min}(y),y)=-p(T)$ which does not depend on $y$, so the same physics is captured in any gauge. On the other hand, the propagator mass at the transition (corresponding to $a(T_c)=0$) is
\beq
\left.\frac{\partial^2V}{\partial x^2}\right|_{x_{\rm min}(y),y;T_c}=3x^2_{\rm min}(y)\,,
\eeq
with
\beq
0=x^3_{\rm min}(y)+y\,.
\eeq
It follows that the propagator mass vanishes only in the case $y=0$, which is the only case where the symmetry $x\to -x$ is explicit in the potential $V(x,y=0)$. Of course, this is just a result within a very simplistic toy model. So, it could still be that the Landau gauge propagator becomes critical at the transition. Let us here list some arguments in favor of this scenario. These are of course speculative and deserve a further study that goes beyond the scope of the present work. 

The first one relies on the similarities between the perturbative expressions for the propagators along the neutral color directions in the confined phase. Indeed, the Feynman integrals that enter these quantities in the center-symmetric Landau gauge are essentially the same as those in the Landau gauge, with the only difference that, in the former case, the internal frequencies are shifted by a quantity proportional to the background field \cite{Reinosa:2015gxn,Reinosa:2020mnx}. This pretty much resembles the relation between the pressure in the Landau and in the center-symmetric Landau gauge. For this latter quantity, since the pressure is gauge-independent, there should exist a mechanism explaining how one can move from an infinite set of diagrams with shifted internal frequencies, to the same diagrams with unshifted frequencies. If the same mechanism holds for the propagators along the neutral color modes, this could open a connection with the same quantity computed in the Landau gauge.\footnote{This would not mean that this quantity is gauge-independent, just that it coincides in two different gauges.} A second argument in favor of a connection between the propagators in the various gauges is that the SU(2) propagator in the center-symmetric Landau gauge has a pole at vanishing momentum for $\smash{T=T_c}$. Poles of YM correlations functions are expected to be gauge-independent, so the Landau gauge propagator should also feature a pole at vanishing momentum for $\smash{T=T_c}$. This assumes, however, that the gauge-independence of poles extends to background gauges, which is not clear {\it a priori} \cite{Kobes:1990dc}. Further investigations in this direction would require studying the background field Nielsen identities.\footnote{In principle, this is beyond the scope of the CF in the presence of fixed mass. But one could extend the model by allowing the mass to depend on the background (after all, the CF mass serves as an effective description of the gauge-fixing and, as such, could depend on the background) such that certain observables remain background invariant. This strategy is currently under investigation.}

\begin{figure}[t]
	\includegraphics[width=\linewidth]{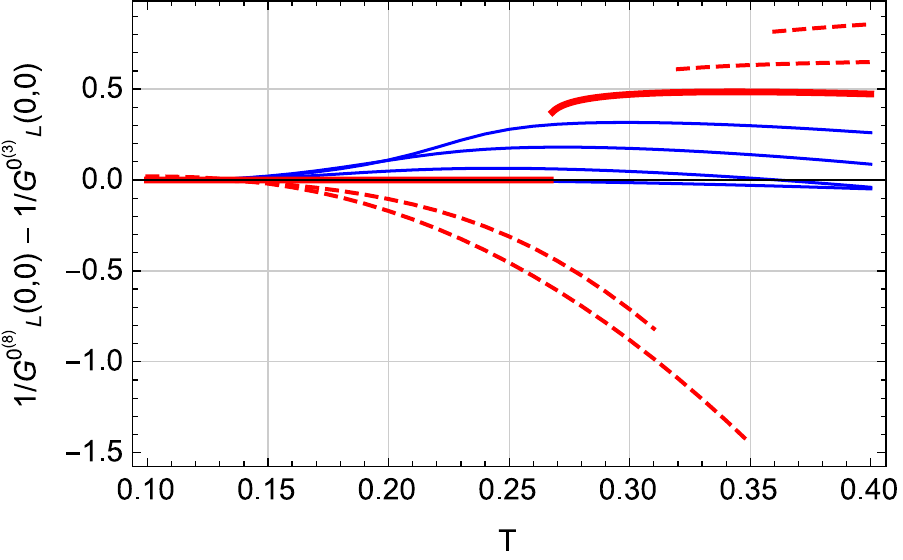}
	\caption{The difference of square masses across the transition and for different values of $\bar r_3$ (we set $\bar r_8=0$). The plain red curve corresponds to the center-symmetric Landau gauge $\bar r_3=4\pi/3$ . The blue curves correspond to $\bar r_3<4\pi/3$ ($\bar r_3/\pi=0.1$, $0.4$, $0.7$, $1.0$), while the red dashed curves to $4\pi/3<\bar r_3<2\pi$ ($\bar r_3/\pi=1.6$, $1.9$).}
	\label{fig:gauge_2}
\end{figure}

The discussion in the case of SU(3) is a bit more subtle. Indeed even if one considers a choice of $\bar r$ for which the potential is not center-symmetric, the minimum of the potential could still jump at some temperature impacting the propagators accordingly. To illustrate this, in Fig.~\ref{fig:gauge_2}, we investigate the fate of the order parameter identified in the first plot of Fig.~\ref{fig:order}, as $\bar r_3$ is chosen away from the center-symmetric value $4\pi/3$. For simplicity, we keep $\bar r_8=0$, which allows us to restrict the minimization of the potential to the subspace $r_8=0$. We find that slightly below $\bar r_3=4\pi/3$, the first order transition of the potential in the gauge $\bar r_3=4\pi/3$ turns into a smooth crossover which prevents a clear identification of the transition, in particular in the Landau gauge $\bar r_3=0$. Interestingly enough, for values of $\bar r_3$ above $4\pi/3$ (and below $2\pi$), the potential still undergoes a first order transition which leads to the dashed curves in Fig.~\ref{fig:gauge_2}. However, this jump does not occur from a center-symmetric minimum to a center-breaking minimum and accordingly the would be order parameter is not constant in any phase. Thus, as before, the only case in which the behavior of the quantity plotted in Fig.~\ref{fig:gauge_2} can be unambiguously linked with center-symmetry breaking is the center-symmetric choice of background $\bar r_3=4\pi/3$.

\section{Conclusions and Outlook}

We have investigated the gluon propagator at finite temperature in the recently introduced center-symmetric Landau gauge. We use the background field CF model as an effective description of the gauge-fixed Lagrangian in the infrared which has shown quite promising results both in the vacuum and at finite temperature, see Ref.~\cite{Pelaez:2021tpq}. In its applications to YM theories, the model is known to be perturbative which allows one to access correlation functions within a systematic scheme and with a good control on the error. For this reason, we have restricted to a one-loop determination of the gluon propagator.

In the center-symmetric Landau gauge, the gluon propagator has two components known has transversal (with respect to the 3d momentum) and longitudinal. It has also various color components referred to as neutral and charged components. Even though we have derived one-loop expressions for all these quantities, we have then restricted our focus to the longitudinal component along the neutral color directions. The reason for this choice is that the deconfinement transition occurs in this sector and, therefore, the corresponding components of the gluon propagator are the more prone to reflect the phase transition. A detailed study of the other components is left to a future work.

For simplicity, we have also restricted our analysis to the propagator in the zero-frequency limit since this is the one we expect to be more sensitive to variations of the temperature. Our results confirm and extend what was already anticipated in Ref.~\cite{vanEgmond:2021jyx}. The gluon propagator in the center-symmetric Landau gauge can be used as a clear-cut probe of the deconfinement transition. In the SU(2) case for instance, it becomes critical at $T_c$. In the SU(3) case, although the propagator does not reach a critical form (since the transition is first order), it is much enhanced and features a slight discontinuity at the transition. More importantly, it becomes non-degenerate along the two diagonal color directions in the deconfined phase, allowing one to define a new order parameter. These results are in sharp contrast with those obtained in other gauges such as the standard Landau gauge or the self-consistent Landau gauge \cite{Fister:2011uw,Reinosa:2016iml}. In the latter case, we have argued that the exact SU(2) propagator should be critical in principle but it does not in practical implementations due to the necessary use of approximations which generically violate nontrivial identities. In the case of the standard Landau gauge, we have explained that the standard argument based on symmetries does not impose a priori any constraint on the behavior of the propagator at the transition. We have nevertheless listed some arguments in favor of a possible crirical behavior of the propagator. Those arguments remain, however, possible scenarios that further investigations would need to clarify.

It would now be interesting to test the present results in actual lattice simulations. As it was argued in Ref.~\cite{vanEgmond:2021jyx}, the center-symmetric Landau gauge in the space of periodic (in imaginary time) gauge field configurations can be mapped onto the standard Landau gauge in the space of twisted gauge field configurations. It should therefore be possible to use the standard minimization algorithms of the lattice Landau gauge in order to simulate the center-symmetric Landau gauge. Work in this direction is in progress \cite{Orlando}. The aim of this study would be two-fold. First, confirm that correlation functions in the center-symmetric Landau gauge are good probes for the deconfinement transition.\footnote{We mention that part of our results, in particular in the SU(2) case are based on the analyticity of the effective action. Since no such notion exists on the lattice in the presence of Gribov copies \cite{Maas:2013sca}, this might represent an obstacle to checking some of our results on the lattice.} Second, test the predictions of the perturbative CF model in this particular gauge. In the same vein, it would be interesting to investigate possible connections between the present proposal and the lattice ``sectorized'' propagators of Ref.~\cite{Silva:2016onh}.

Let us finally mention that, in the SU(2) case, a more accurate description of the vicinity of the critical region requires a renormalization group analysis. This is also work in progress.

\acknowledgements{We would like to thank O. Oliveira, J. Serreau, P. Silva and M. Tissier for enlightening discussions at various stages of this work, as well as for critical comments on the manuscript.}

\appendix

\section{Quadratic form inversion}\label{app:Schur}
The free gluon propagator is obtained from the inversion of the quadratic form (\ref{eq:quad}) in the $A-h$ sector. In matrix form, the latter writes
\beq\label{eq:M}
M_{\mu\nu}=\left(\begin{array}{cc}
(Q^2+m^2)\delta_{\mu\nu}-Q_\mu Q_\nu & -\bar Q_\mu\\
\bar Q_\nu & 0 
\end{array}\right).
\eeq
For simplicity, we have omitted the color structure since it is diagonal in the chosen Cartan-Weyl basis. A similar inversion is needed when relating the full gluon propagator (\ref{eq:prop}) to the one-particle-irreducible (1PI) diagrams that enter the two-point vertex function $\Gamma^{(2)}_{\mu\nu}(Q)$, and, because the $A-h$ sector is not renormalized, the matrix to be inverted in this case has a similar structure:
\beq\label{eq:calM}
{\cal M}_{\mu\nu}=\left(\begin{array}{cc}
\Gamma^{(2)}_{\mu\nu} & -\bar Q_\mu\\
\bar Q_\nu & 0 
\end{array}\right).
\eeq
In what follows, we thus kill two birds with one stone by inverting (\ref{eq:calM}) which automatically gives us access to the inverse of (\ref{eq:M}).

There are various strategies to invert (\ref{eq:calM}). One possibility is to use the Schur complement just as it was done in \cite{vanEgmond:2021jyx} for evaluating the determinant. Here, we follow a more direct approach by writing the inverse as
\beq
{\cal M}^{-1}_{\mu\nu}=\left(\begin{array}{cc}
{\cal G}_{\mu\nu} & {\cal B}_\mu\\
-{\cal B}_\nu & {\cal C} 
\end{array}\right).
\eeq
Expressing that $\smash{{\cal M}{\cal M}^{-1}=\mathds{1}}$, this leads to the set of equations
\beq
\Gamma^{(2)}_{\mu\rho}{\cal G}_{\rho\nu}+\bar Q_\mu {\cal B}_\nu & = & \delta_{\mu\nu}\,,\label{eq:one}\\
\Gamma^{(2)}_{\mu\rho}{\cal B}_{\rho}-\bar Q_\mu {\cal C} & = & 0\,,\label{eq:two}\\
\bar Q_\rho{\cal G}_{\rho\nu} & = & 0\,,\\
\bar Q_\rho{\cal B}_\rho & = & 1\,.
\eeq
The third equation is a direct consequence of the gauge-fixing condition (\ref{eq:cond}), which, in turn, tells us that the propagator decomposes along the two transverse projectors (\ref{eq:projT2})-(\ref{eq:projL}) at finite temperature, see Eq.~(\ref{eq:GTL}). The two components ${\cal G}_T$ and ${\cal G}_L$ can then be obtained by contracting (\ref{eq:one}) with $P_T(\bar Q)$ and $P_L(\bar Q)$ respectively. One finds
\beq
{\cal G}_T(Q) & = & \frac{d-2}{\Gamma^{(2)}_{\mu\rho}(Q)P^T_{\rho\mu}(\bar Q)}\,,\\
{\cal G}_L(Q) & = & \frac{1}{\Gamma^{(2)}_{\mu\rho}(Q)P^L_{\rho\mu}(\bar Q)}\,.
\eeq
It is convenient to split $\Gamma^{(2)}_{\mu\nu}(Q)$ into a tree-level part, as given by the first block of the matrix (\ref{eq:M}), and the loop corrections which we gather into a self-energy $\Pi_{\mu\nu}(Q)$:
\beq
\Gamma^{(2)}_{\mu\nu}(Q)=(Q^2+m^2)\delta_{\mu\nu}-Q_\mu Q_\nu+\Pi_{\mu\nu}(Q)\,.
\eeq
Then, a trivial calculation using Eqs.~(\ref{eq:projL}) and (\ref{eq:pp3}) leads to
\beq
{\cal G}_T(Q) & = & \frac{1}{Q^2+m^2+\Pi_T(Q)}\,,\\
{\cal G}_L(Q) & = & \frac{1}{\frac{(Q\cdot\bar Q)^2}{\bar Q^2}+m^2+\Pi_L(Q)}\,.\label{eq:den_L}
\eeq
with
\beq
\Pi_T(Q) & \equiv &  \frac{\Pi_{\mu\rho}(Q)P^T_{\rho\mu}(\bar Q)}{d-2}\,,\\
\Pi_L(Q) & \equiv &  \Pi_{\mu\rho}(Q)P^L_{\rho\mu}(\bar Q)\,.
\eeq
In particular, by switching off the loop corrections, we arrive at the free propagators (\ref{eq:PT}) and (\ref{eq:PL}).

We have introduced the notations $\Pi_T$ and $\Pi_L$ for convenience. However, this does not mean that they corresponds to the components of $\Pi_{\mu\nu}$ along $P^T_{\mu\nu}(\bar Q)$ and $P^L_{\mu\nu}(\bar Q)$. In fact, we have a priori
\beq
\Pi_{\mu\nu} & = & \alpha_T P^T_{\mu\nu}(\bar Q)+\alpha_L P^L_{\mu\nu}(\bar Q)\nonumber\\
& + & \alpha_\parallel P^\parallel_{\mu\nu}(\bar Q)+\hat\alpha_\parallel P^\parallel_{\mu\nu}(Q)\,,\label{eq:decomp}
\eeq
with $\{P^T(\bar Q),P^L(\bar Q),P^\parallel(\bar Q)\}$ forming an orthogonal set of projectors, but not $\{P^T(\bar Q),P^L(\bar Q),P^\parallel(\bar Q),P^\parallel(Q)\}$. Owing to Eq.~(\ref{eq:pp3}), we find indeed that $\smash{\Pi_T=\alpha_T}$. However, using Eq.~(\ref{eq:projL}), we also find that 
\beq
\Pi_L=\alpha_L+\hat\alpha_\parallel \frac{Q^2\bar Q^2-(Q\cdot\bar Q)^2}{Q^2\bar Q^2}\,.\label{eq:conv}
\eeq
We have decided to define this combination as $\Pi_L$ because this is the one that eventually enters ${\cal G}_L$.\footnote{Had we decided instead to write $\alpha_X\to\Pi_X$ in Eq.~(\ref{eq:decomp}), the denominator in Eq.~(\ref{eq:den_L}) would have written
\beq
Q^2+m^2+\Pi_L-(Q^2-\hat\Pi_\parallel)\frac{Q^2\bar Q^2-(Q\cdot\bar Q)^2}{Q^2\bar Q^2}\,.
\eeq}
The other components enter the determination of ${\cal B}_\mu$ and ${\cal C}$ which are not needed for our analysis.

\section{Reduction of momentum integrals}\label{app:reduction}
As we have stated in the main text, all the required momentum integrals can be reduced in terms of the two basic integrals $J_M$ and $I_{M_1M_2}(k)$ defined in Eqs.~(\ref{eq:JM}) and (\ref{eq:IMM}). In this section, we provide more details on this reduction while in the next section, we evaluate the basic integrals.

\subsection{Reduction of $\hat I_{M_1M_2}(k)$}
To reduce $\hat I_{M_1M_2}(k)$, defined in Eq.~(\ref{eq:Ih}), we use
\beq
q^2G_M(q)=1-M^2G_M(q)\,,\label{eq:trick0}
\eeq 
as well as
\beq
q\cdot\ell & = & \frac{1}{2}\big(k^2-q^2-\ell^2\big)\nonumber\\
& = & \frac{1}{2}\big(k^2+M^2_1+M^2_2\nonumber\\
& & \hspace{1.0cm}-\,q^2-M^2_1-\ell^2-M^2_2\big)\,.
\eeq
This leads to
\beq\label{eq:43}
& & q^2\ell^2 G_{M_1}(q)G_{M_2}(\ell)\nonumber\\
& & \hspace{0.5cm}=\,1-M^2_1G_{M_1}(q)-M^2_2G_{M_2}(\ell)\nonumber\\
& & \hspace{0.9cm}+\,M^2_1M^2_2G_{M_1}(q)G_{M_2}(\ell)\,,
\eeq
and
\beq
& & (q\cdot\ell)^2 G_{M_1}(q)G_{M_2}(\ell)\nonumber\\
& & \hspace{0.5cm}=\,\frac{(k^2+M^2_1+M^2_2)^2}{4}G_{M_1}(q)G_{M_2}(\ell)\nonumber\\
& & \hspace{0.5cm}-\,\frac{k^2+M^2_1+M^2_2}{4}\big(G_{M_1}(q)+G_{M_2}(\ell)\big)\nonumber\\
& & \hspace{0.5cm}-\,\frac{q\cdot\ell}{2}\big(G_{M_1}(q)+G_{M_2}(\ell)\big)\,,
\eeq
from which one easily deduces that
\beq
\hat I_{M_1M_2}(k) & = & \frac{k^2-M^2_1+M^2_2}{4}J_{M_1}\nonumber\\
& + & \frac{k^2+M^2_1-M^2_2}{4}J_{M_2}\nonumber\\
& - & \frac{\Delta(-k^2,M^2_1,M^2_2)}{4}I_{M_1M_2}(k)\,,
\eeq
with
\beq
& & \Delta(-k^2,M^2_1,M^2_2)=k^4+M^4_1+M^4_2\nonumber\\
& & \hspace{1.0cm}+\,2k^2M^2_1+2k^2M^2_2-2M^2_1M^2_2\,.\label{eq:Delta}
\eeq

\subsection{Reduction of the self-energy}
To reduce the self-energy (\ref{eq:zero}), we use Eq.~(\ref{eq:trick0}) in the form
\beq
\frac{G_M}{q^2} =-d_M G_M\,,\label{eq:trick00}
\eeq
where $d_M$ denotes the finite difference operator
\beq
d_{M}f(M)\equiv \frac{f(M)-f(0)}{M^2}\,.
\eeq 
Similarly, we rewrite Eq.~(\ref{eq:GLsQ}) as
\beq
\frac{G^\kappa_L(Q)}{Q^2_\kappa}=-d_{M^\kappa_{L\pm,q}}G_{M^\kappa_{L+,q}}(q)\,,\label{eq:52}
\eeq
where $d_{M^\kappa_{L\pm,q}}$ denotes the finite difference operator 
\beq
d_{M^\kappa_{L\pm,q}}f(M^\kappa_{L+,q})\equiv \frac{f(M^\kappa_{L+,q})-f(M^\kappa_{L-,q})}{(M^\kappa_{L+,q})^2-(M^\kappa_{L-,q})^2}\,.
\eeq 
The formulas (\ref{eq:s0})-(\ref{eq:s1}) are quite easily obtained. In order to obtain Eq.~(\ref{eq:s3}) from Eq.~(\ref{eq:l44}), we use Eq.~(\ref{eq:trick00}) as well as
\beq\label{eq:trick2}
\frac{(Q_\kappa\cdot\bar Q_\kappa)^2}{\bar Q^2_\kappa}G^\kappa_L(Q)=1-m^2G^\kappa_L(Q)\,,
\eeq
and
\beq\label{eq:GL}
G^\kappa_L(Q) & = & -d_{M^\kappa_{L\pm,q}}\Big[\bar Q^2_\kappa-(\bar\omega_q^\kappa)^2+(M_{L+,q}^\kappa)^2\nonumber\\
& & \hspace{1.5cm}+\,(\bar\omega_q^\kappa)^2-(M_{L+,q}^\kappa)^2\Big]G_{M^\kappa_{L+,q}}(q)\nonumber\\
& = & -d_{M^\kappa_{L\pm,q}}\Big[((\bar\omega^\kappa_q)^2-(M^\kappa_{L+,q})^2)G_{M^\kappa_{L+,q}}(q)\Big].\nonumber\\
\eeq
which leads to
\beq
\frac{G^\kappa_L(Q)}{q^2} & = & d_{M^\kappa_{L\pm,q}}\big(((\bar\omega^\kappa_q)^2-(M^\kappa_{L+,q})^2)d_{M^\kappa_{L+,q}}G_{M^\kappa_{L+,q}}(q)\big)\nonumber\\
& = & \frac{(\bar\omega_q^\kappa)^2G_0(q)}{(M^\kappa_{L+,q})^2(M^\kappa_{L-,q})^2}\nonumber\\
&  & +\,d_{M^\kappa_{L\pm,q}}\frac{((\bar\omega_q^\kappa)^2-(M_{L+,q}^\kappa)^2)G_{M^\kappa_{L+,q}}(q)}{(M^\kappa_{L+,q})^2}\nonumber\\
& = & \frac{G_0(q)}{(\omega_q^\kappa)^2+m^2}\nonumber\\
& & +\,d_{M^\kappa_{L\pm,q}}\frac{((\bar\omega_q^\kappa)^2-(M_{L+,q}^\kappa)^2)G_{M^\kappa_{L+,q}}(q)}{(M^\kappa_{L+,q})^2}\,.\label{eq:GLsq}\nonumber\\
\eeq
In the last step, we have used Eq.~(\ref{eq:rrr}). The sames steps lead from Eq.~(\ref{eq:l33}) to Eq.~(\ref{eq:s44}).

Finally, to obtain Eq.~(\ref{eq:s4}) from Eq.~(\ref{eq:l4}), we rewrite
\beq
\Big( \ell^2(L_{\tau}\cdot \bar L_{\tau})+q^2 (Q_{\kappa}\cdot \bar Q_{\kappa})-k^2 (\ell^2+q^2+\omega_q^{\kappa}\bar \omega_q^{\kappa})\Big)^2\nonumber\\
\eeq
as
\beq
\Big(\omega^\kappa_q\bar\omega^\kappa_qk^2+(q^2+\ell^2)(k^2-\omega^\kappa_q\bar\omega^\kappa_q)-(q^4+\ell^4)\Big)^2\,.\label{eq:b15}\nonumber\\
\eeq
Then, we notice that ($n\geq 0$)
\beq
& & d_{M^\kappa_{L+,q}} (M^\kappa_{L+,q})^{2n}q^2G_{M^\kappa_{L+,q}}(q)\nonumber\\
& & \hspace{0.5cm}=\,d_{M^\kappa_{L+,q}}(M^\kappa_{L+,q})^{2n} [1-(M_{L+,q}^\kappa)^2G_{M^\kappa_{L+,q}}(q)]\nonumber\\
& & \hspace{0.5cm}=\,(M^\kappa_{L+,q})^{2n-2}-d_{M^\kappa_{L+,q}}(M_{L+,q}^\kappa)^2G_{M^\kappa_{L+,q}}(q)\,,\nonumber\\
\eeq
and
\beq
& & d_{M^\kappa_{L+,q}} (M^\kappa_{L+,q})^{2n}q^4G_{M^\kappa_{L+,q}}(q)\nonumber\\
& & \hspace{0.5cm}=\,d_{M^\kappa_{L+,q}} (M^\kappa_{L+,q})^{2n}[q^2-(M^\kappa_{L+,q})^2\nonumber\\
& & \hspace{4.0cm}+\,(M^\kappa_{L+,q})^4G_{M^\kappa_{L+,q}}(q)]\nonumber\\
& & \hspace{0.5cm}=\,(M^\kappa_{L+,q})^{2n-2}q^2-(M^\kappa_{L+,q})^{2n}\nonumber\\
& & \hspace{0.7cm}+\,d_{M^\kappa_{L+,q}} (M^\kappa_{L+,q})^{2n+4}G_{M^\kappa_{L+,q}}(q)\,,
\eeq
with $(M^\kappa_{L+,q})^{2n-2}=0$ if $n=0$. More precisely, we will need the case where the prefactor is a quadratic polynomial $\alpha+\beta (M_{L+,q}^\kappa)^2+\gamma (M_{L+,q}^\kappa)^4$ and where we act in addition with $d_{M_{L\pm,q}}$. We find
\beq
& & d_{M^\kappa_{L\pm,q}}d_{M^\kappa_{L+,q}} (\alpha+\beta (M_{L+,q}^\kappa)^2+\gamma (M_{L+,q}^\kappa)^4)q^2G_{M^\kappa_{L+,q}}(q)\nonumber\\
& & \hspace{0.5cm}=\gamma-d_{M^\kappa_{L\pm,q}}d_{M^\kappa_{L+,q}} (\alpha+\beta (M_{L+,q}^\kappa)^2+\gamma (M_{L+,q}^\kappa)^4)\nonumber\\
& & \hspace{3.5cm}\times\,(M^\kappa_{L+,q})^2G_{M^\kappa_{L+,q}}(q)\,,
\eeq
and
\beq
& & d_{M^\kappa_{L\pm,q}}d_{M^\kappa_{L+,q}} (\alpha+\beta (M_{L+,q}^\kappa)^2+\gamma (M_{L+,q}^\kappa)^4)q^4G_{M^\kappa_{L+,q}}(q)\nonumber\\
& & \hspace{0.5cm}=-\beta+\gamma(q^2-M_{L+,q}^2-M_{L-,q}^2)\nonumber\\
& & \hspace{0.7cm}+\,d_{M^\kappa_{L\pm,q}}d_{M^\kappa_{L+,q}} (\alpha+\beta (M_{L+,q}^\kappa)^2+\gamma (M_{L+,q}^\kappa)^4)\nonumber\\
& & \hspace{3.7cm}\times\,(M^\kappa_{L+,q})^4G_{M^\kappa_{L+,q}}(q)\,.
\eeq
We note that in the first term, we can replace $(M^\kappa_{L+,q})^2+(M^\kappa_{L-,q})^2$ by $2\omega^\kappa_q\bar\omega^\kappa_q+m^2$.

In a first step, we can apply these identities with $\beta=\gamma=0$ and $\alpha=1$ to one instance of the factor that is squared in (\ref{eq:b15}). We find
\beq
 & & \big(\omega^\kappa_q\bar\omega^\kappa_qk^2+((M_{L+,q}^\kappa)^2+(M_{L+,\ell}^\kappa)^2)(\omega^\kappa_q\bar\omega^\kappa_q-k^2)\nonumber\\
 & & \hspace{4.5cm}-\,((M_{L+,q}^\kappa)^4+(M_{L+,\ell}^\kappa)^4)\big)\nonumber\\
 & & \big(\omega^\kappa_q\bar\omega^\kappa_qk^2+(q^2+\ell^2)(k^2-\omega^\kappa_q\bar\omega^\kappa_q)-(q^4+\ell^4)\big)\,.
\eeq
The new expression is now suited for the application of the same identities with appropriate choices of $\alpha$, $\beta=\omega^\kappa_q\bar\omega^\kappa_q-k^2$ and $\gamma=-1$. We arrive at
\begin{widetext}
\beq
& & d_{M^\kappa_{L\pm,q}}d_{M^\kappa_{L+,q}}d_{M^\tau_{L\pm,\ell}}d_{M^\tau_{L+,\ell}}\Big(\omega^\kappa_q\bar\omega^\kappa_qk^2+(q^2+\ell^2)(k^2-\omega^\kappa_q\bar\omega^\kappa_q)-(q^4+\ell^4)\Big)^2G_{M^\kappa_{L+,q}}(q)G_{M^\tau_{L+,\ell}}(\ell)\nonumber\\
& & \hspace{0.5cm}=\,d_{M^\kappa_{L\pm,q}}d_{M^\kappa_{L+,q}}d_{M^\tau_{L\pm,\ell}}d_{M^\tau_{L+,\ell}}\nonumber\\
& & \hspace{0.7cm}\big(\omega^\kappa_q\hat\omega^\kappa_qk^2+((M_{L+,q}^\kappa)^2+(M_{L+,\ell}^\tau)^2)(\omega^\kappa_q\bar\omega^\kappa_q-k^2)-((M_{L+,q}^\kappa)^4+(M_{L+,\ell}^\tau)^4)\big)^2G_{M^\kappa_{L+,q}}(q)G_{M^\tau_{L+,\ell}}(\ell)\nonumber\\
& & \hspace{0.5cm}+\,(2(\omega^\kappa_q\bar\omega^\kappa_q-k^2)+\ell^2-2\omega_q^\kappa\bar\omega_q^\kappa-m^2)d_{M^\kappa_{L\pm,q}}d_{M_{L+,q}}G_{M^\kappa_{L+,q}}(q)\nonumber\\
& & \hspace{0.5cm}+\,(2(\omega^\kappa_q\bar\omega^\kappa_q-k^2)+q^2-2\omega_q^\kappa\bar\omega_q^\kappa-m^2)d_{M^\tau_{L\pm,\ell}}d_{M^\tau_{L+,\ell}}G_{M^\tau_{L+,\ell}}(\ell)\,,
\eeq
where we have used that $\omega_\ell^\tau=\omega_q^\kappa$. We should not use yet that  $M_{L\pm,q}=M_{L\pm,\ell}$.
After some simplifications, this leads eventually to
\beq
\Pi^{\kappa0^{(j)}\tau}_{LLL}(k) & = & -\frac{1}{2}T\sum_{q\in\mathds{Z}}(\bar \omega^\kappa_q)^2\,\Bigg[d_{M_{L\pm,q}^\kappa}d_{M_{L+,q}^\kappa}d_{M_{L\pm,\ell}^\tau}d_{M_{L+,\ell}^\tau}\nonumber\\
& & \hspace{3.0cm}\times\,\Big(\bar \omega^\kappa_q\omega^\kappa_qk^2+((M_{L+,q}^\kappa)^2+(M_{L+,\ell}^\tau)^2)(\bar \omega^\kappa_q \omega^\kappa_q-k^2)\nonumber\\
& & \hspace{7.5cm}-\,((M_{L+,q}^\kappa)^4+(M_{L+,\ell}^\tau)^4)\Big)^2 I_{M_{L+,q}^\kappa M_{L+,\ell}^\tau}(k)\label{eq:s44}\nonumber\\
& & \hspace{2.5cm}-\,2d_{M_{L\pm,q}^\kappa}d_{M_{L+,q}^\kappa}(k^2+m^2+(M^\kappa_{L+,q})^2)J_{M^\kappa_{L+,q}}\Bigg],
\eeq
which is the formula (\ref{eq:s4}) quoted in the main text.
\end{widetext}

\section{Basic momentum integrals}\label{app:basic}

\subsection{Integral $J_M$}
In dimensional regularization, we have the general formula
\beq
\int_q \frac{1}{(q^2+M^2)^n}=\frac{(M^2)^{d/2-1/2-n}}{(4\pi)^{d/2-1/2}}\frac{\Gamma(n+1/2-d/2)}{\Gamma(n)}\,,\nonumber\\
\eeq
where we took into account the fact that the integral is a $(d-1)$-dimensional one. In particular
\beq
J_M & \equiv & \int_q \frac{1}{q^2+M^2}\nonumber\\
& = & \frac{(M^2)^{d/2-3/2}}{(4\pi)^{d/2-1/2}}\Gamma(3/2-d/2)\,,
\eeq
as well as
\beq
J'_M & \equiv & -\int_q \frac{1}{(q^2+M^2)^2}\nonumber\\
& = & -\frac{(M^2)^{d/2-5/2}}{(4\pi)^{d/2-1/2}}\Gamma(5/2-d/2)\,.
\eeq
In the limit $d\to 4$ this gives
\beq
J_M & \to & -\frac{(M^2)^{1/2}}{4\pi}\,,\\
J'_M & \to & -\frac{(M^2)^{-1/2}}{8\pi}\,.
\eeq
We recall that, eventually we only need the formulae in the limit $d\to 4$. The formulae for arbitrary $d$ are needed to identify potential correction terms in the inversion formula (\ref{eq:inversion_2}).

\subsection{Integral $I_{M_1M_2}(k)$}
Next, we consider the integral [$\smash{\ell\equiv -k-q}$]
\beq
I_{M_1M_2}(k)\equiv\int_q \frac{1}{(q^2+M^2_1)(\ell^2+M^2_2)}\,.
\eeq
Using the Feynman trick, we arrive at
\beq
I_{M_1M_2}(k) & = & \int_0^1 dx \int_q \frac{1}{(q^2+M^2)^2}\nonumber\\
& = & \frac{\Gamma(5/2-d/2)}{(4\pi)^{d/2-1/2}}\int_0^1 dx\, (M^2)^{d/2-5/2}\,,
\eeq
with $M^2\equiv xM^2_1+(1-x)M^2_2+x(1-x)k^2$. For the purpose of studying the presence of correction terms in the formula (\ref{eq:inversion_2}), we need to consider this expression for an arbitrary value of $d$. Otherwise, we can take the limit $\smash{d\to 4}$ and we find
\beq
I_{M_1M_2}(k)=\frac{1}{8\pi k}\int_0^1 \frac{dx}{\,\left(x\frac{M^2_1}{k^2}+(1-x)\frac{M^2_2}{k^2}+x(1-x)\right)^{1/2}}\,.\nonumber\\
\eeq
Upon the change of variables $x=y+1/2$, this rewrites
\beq
I_{M_1M_2}(k)=\frac{1}{8\pi k}\int_{-1/2}^{1/2}\frac{dy}{\left(\frac{\Delta}{4k^4}-\left(y+\frac{M^2_2-M^2_1}{2k^2}\right)^2\right)^{1/2}}\,,\nonumber
\eeq
with $\Delta$ given in Eq.~(\ref{eq:Delta}).

The evaluation of the $y$-integral requires some care in the case where $M^2_1$ or $M^2_2$ are complex. Let us first consider the case where they are real (and positive), in which case $\Delta>0$. Using the change of variables
\beq
y=\frac{\Delta^{1/2}}{2k^2}z-\frac{M^2_2-M^2_1}{2k^2}\,,
\eeq
we obtain
\beq
I_{M_1M_2}(k) & = & \frac{1}{8\pi k}\int_{\frac{M^2_2-M^2_1-k^2}{\Delta^{1/2}}}^{\frac{M^2_2-M^2_1+k^2}{\Delta^{1/2}}}\frac{dz}{\left(1-z^2\right)^{1/2}}\nonumber\\
& = & -\frac{1}{8\pi}\int_{{\rm Arccos}\,\frac{M^2_2-M^2_1-k^2}{\Delta^{1/2}}}^{{\rm Arccos}\,\frac{M^2_2-M^2_1+k^2}{\Delta^{1/2}}}d\theta \,1\nonumber\\
 & = & \frac{1}{8\pi k}\left[{\rm Arccos}\,\frac{M^2_2-M^2_1-k^2}{\Delta^{1/2}}\right.\nonumber\\
 & & \hspace{1.0cm}\left.-\,{\rm Arccos}\,\frac{M^2_2-M^2_1+k^2}{\Delta^{1/2}}\right].
\eeq
Consider more generally the integral
\beq
\int_{-1/2}^{1/2}dy\,\frac{1}{\sqrt{a-(y+b)^2}}\,,
\eeq
with $a$ and $b$ complex and such that the argument of the square root never gets negative  in the interval $y\in [-1/2,1/2]$ (we are always in this case). The change of variables
\beq
y=\sqrt{a}z-b\,,
\eeq
leads to
\beq
\int_{-1/2}^{1/2}dy\,\frac{1}{\sqrt{a-(y+b)^2}}=\int_{(b-1/2)/\sqrt{a}}^{(b+1/2)/\sqrt{a}} dz\,\frac{\sqrt{a}}{\sqrt{a(1-z^2)}}\,,\nonumber\\
\eeq
where the integral is to be interpreted as contour integral along the segment $z(y)=(b+y)/\sqrt{a}$ in the complex plane connecting $(b-1/2)/\sqrt{a}$ to $(b+1/2)/\sqrt{a}$. To continue, we need a formula that allows us to split $\sqrt{a(1-z^2)}$ into two square-roots. In general
\begin{eqnarray}
\sqrt{z_1z_2} & = & \sqrt{|z_1z_2|}e^{i{\rm Arg}\,(z_1z_2)/2}\,\nonumber,\\
\sqrt{z_1} & = & \sqrt{|z_1|}e^{i{\rm Arg}\,(z_1)/2}\,,\\
\sqrt{z_2} & = & \sqrt{|z_2|}e^{i{\rm Arg}\,(z_2)/2}\,,\nonumber
\end{eqnarray}
with $-\pi<{\rm Arg}\,z\leq\pi$. It follows that
\beq
\sqrt{z_1z_2}=\sqrt{z_1}\sqrt{z_2}e^{i\{{\rm Arg}\,(z_1z_2)-{\rm Arg}\,(z_1)-{\rm Arg}\,(z_2)\}/2}\,.\nonumber\\
\eeq
Since the expression between brackets is a multiple of $2\pi$, the exponential factor is either $+1$ or $-1$. It follows that
\beq
& & \int_{-1/2}^{1/2}dy\,\frac{1}{\sqrt{a-(y+b)^2}}\nonumber\\
& & \hspace{0.5cm}=\,\int_{(b-1/2)/\sqrt{a}}^{(b+1/2)/\sqrt{a}} dz\,\frac{e^{i\{{\rm Arg}\,[a(1-z^2)]-{\rm Arg}\,a-{\rm Arg}\,[1-z^2]\}/2}}{\sqrt{(1-z^2)}}\,.\nonumber\\
\eeq
As one integrates along the segment $z(y)=(b+y)/\sqrt{a}$  joining $(b-1/2)/\sqrt{a}$ and $(b+1/2)/\sqrt{a}$, 
\beq
\tau(y)\equiv e^{i\{{\rm Arg}\,[a(1-z^2(y))]-{\rm Arg}\,a-{\rm Arg}\,[1-z^2(y)]\}/2},
\eeq
can jump at certain values of $y$. This is due to the fact that the combination
\beq
{\rm Arg}\,[a(1-z^2(y))]-{\rm Arg}\,a-{\rm Arg}\,[1-z^2(y)]\,,\label{eq:comb}
\eeq
which is a multiple of $2\pi$ can jump by $\pm 2\pi$ at certain values of $y\in[-1/2,1/2]$. These jumps happen when either $a(1-z^2(y))$ or $(1-z^2(y))$ lie on the negative real axis. As we have already mentionned, the first possibility is excluded so we are left with the conditions
\beq
\text{Im}\,z^2(y)=0\quad \mbox{and} \quad \text{Re}\,z^2(y) \geq 1\,,
\label{ee}
\eeq
together of course with $-1/2\leq y\leq 1/2$, that determine the possible changes of sign along the integration range. We will see below that these equations have at most one solution $y_0$. In that case, the integral rewrites
\beq
& & \int_{-1/2}^{1/2}\frac{dy}{\sqrt{a-(y+b)^2}}\nonumber\\
& & \hspace{0.2cm}=\tau(-1/2)\Bigg(\,\int_{(b-1/2)/\sqrt{a}}^{(b+y_0-\epsilon)/\sqrt{a}} \frac{dz}{\sqrt{(1-z^2)}}\nonumber\\
&& \hspace{2cm}-\int_{(b+y_0+\epsilon)/\sqrt{a}}^{(b+1/2)/\sqrt{a}} \frac{dz}{\sqrt{(1-z^2)}}\Bigg),
\eeq
where $\epsilon>0$ is needed not to evaluate the square root strictly on a negative number. The case where there is no change of sign can be formally incorporated in the same formula by choosing $\smash{y_0=1/2}$.

We are then left with the integration of $1/\sqrt{1-z^2}$ over two distinct intervals. To do so, we use the primitive
\beq
{\rm Arcsin}\,z\equiv -i\ln(iz+\sqrt{1-z^2})\,.
\eeq
We verify that
\beq
{\rm Arcsin}'\,z&=&-i\frac{i-z/\sqrt{1-z^2}}{iz+\sqrt{1-z^2}}=\frac{1}{\sqrt{1-z^2}}.
\eeq
Moreover it can be checked that its branch-cuts do not cross the considered integration ranges.\footnote{By construction, this is true for the branch-cut $1-z^2=-u$ with $u>0$. Another potential branch-cut is $iz+\sqrt{1-z^2}=-u$ with $u>0$. This implies $z=-i(1/u-u)/2$. However, $iz+\sqrt{1-z^2}=(1/u-u)/2+\sqrt{1+(1/u-u)^2/4}=\sqrt{(1/u+u)^2/4}=(1/u-u)/2+(1/u+u)/2=1/u$ and thus this is not a solution.} We then arrive at
\beq
& & \int_{-1/2}^{1/2}dy\,\frac{1}{\sqrt{a-(y+b)^2}}\nonumber\\
& & \hspace{0.5cm}=\tau(-1/2)\Big(\,{\rm Arcsin}\,z(y_0-\epsilon)-{\rm Arcsin}\,z(-1/2)\nonumber\\
& & \hspace{2.5cm}+\,{\rm Arcsin}\,z(y_0+\epsilon)-{\rm Arcsin}\,z(1/2)\Big)\,.\label{23}\nonumber\\
\eeq
We note that, in the case where $a$ is strictly negative, our convention for $\sqrt{a}$ is $i\sqrt{|a|}$. Had we chosen instead the convention $-i\sqrt{|a|}$, it is easily checked that this would lead to Eq.~(\ref{23}).

Let us now determine $y_0$ in the various cases of interest. 

\subsubsection{$M_1=M_{\pm}$ and $M_2=0$}
We have
\beq
a= \frac{(k^2+M^2_{\pm})^2}{4k^4}, \,\,\, b=-\frac{M^2_{\pm}}{2k^2}\,,
\eeq
so that 
\beq
z^2(y)=\frac{(M^2_{\pm}-2k^2 y)^2}{(k^2+M^2_{\pm})^2},
\eeq
The first condition in (\ref{ee}) occurs iff the imaginary part of the numerator equals the imaginary part of the denominator that is iff $\smash{y_0=-1/2}$. Then, the second condition in (\ref{ee}) is automatically satisfied. This case is marginal though since the change of sign occurs at the boundary of the integration domain.

\subsubsection{$M_1=M_{+}$ and $M_2=M_{-}$}
We first $M^2_{\pm}=m^2_R \pm i m^2_I$, so that
\beq
a=\frac{1}{4}-\left(\frac{m_I^2}{k^2}\right)^2+\frac{m_R^2}{k^2}, \,\,\, b=-i \frac{m_I^2}{k^2},
\eeq
and then
\beq
z^2(y)=-\frac{4 (m_I^2+i k^2 y)^2}{k^4-4 m_I^4+4 k^2 m_R^2}\,.
\eeq
The first condition in (\ref{ee}) yields $\smash{y_0=0}$, while the second condition rewrites
\beq
\frac{k^4}{4m_I^4}+\frac{k^2m_R^2}{m_I^4}\leq 1\,.
\eeq

\subsubsection{$M_1=M_{\pm}$ and $M_2=M_{\pm}$}
We have
\beq
a= \frac{1}{4}+\frac{M^2_{\pm}}{k^2}, \,\,\, b=0,
\eeq
so that
\beq
z^2(y)=\frac{4k^2 y^2}{k^2+4M_{\pm}^2},
\eeq
which is incompatible with \eqref{ee}.

\subsubsection{$M_1^2=M^2_{L,\pm}$ and $M_2^2=M^2_T$}
We have
\beq
a&=&(b-1/2)^2+\frac{M_T^2}{k^2}, \,\,\,\, b=\frac{M_T^2-M^2_{L,\pm}}{2k^2},
\eeq
so that  
\beq
z^2(y)=\frac{(b+y)^2}{(b-1/2)^2+\frac{M_T^2}{k^2}}.
\eeq
This rewrites
\beq
\frac{(y+{\rm Re}\,b)^2-({\rm Im}\, b)^2+2i(y+{\rm Re}\,b){\rm Im}\,b}{(-1/2+{\rm Re}\,b)^2-({\rm Im}\, b)^2+\frac{M^2_T}{k^2}+2i(-1/2+{\rm Re}\,b){\rm Im}\,b}\,,\nonumber\\
\eeq
or, introducing the variables,
\beq
Y & = & \frac{y+{\rm Re}\,b}{-1/2+{\rm Re}\,b}\,,\\
Z & = & \frac{{\rm Im}\,b}{-1/2+{\rm Re}\,b}\,,\\
X & = & \frac{M_T}{-1/2+{\rm Re}\,b}\,,
\eeq
this simplifies into
\beq
z^2(y)=\frac{Y^2-Z^2+2iYZ}{1-Z^2+X^2+2iZ}\,.\label{eq:case}
\eeq
The first condition in (\ref{ee}) rewrites
\beq
Y^2-Y(1-Z^2+X^2)-Z^2=0
\eeq
which admits the solutions
\beq
Y=\frac{1}{2}\Bigg(1-Z^2+X^2\pm\sqrt{(1-Z^2+X^2)^2+4Z^2}\Bigg)\,,\nonumber\\
\eeq
from which one can easily deduce the corresponding values for $y_0$. To test wether these solutions are compatible with the second condition in (\ref{ee}), we note that, given a complex number $w=(A+iB)/(C+iD)$, requiring that its imaginary part vanishes imposes $\smash{AD-BC=0}$. In that case, assuming that $\smash{C\neq 0}$, the real part of $w$ takes the simpler form
\beq
{\rm Re}\,w=\frac{AC+BD}{C^2+D^2}=\frac{AC+AD^2/C}{C^2+D^2}=\frac{A}{C}\,,
\eeq
and, similarly, assuming that $\smash{D\neq 0}$, the real part takes the simpler form $\smash{{\rm Re}\,w=B/D}$. Coming back to the case of (\ref{eq:case}), when the imaginary part vanishes, the real part is necessary equal to $Y$. Since the real part of $z^2(y)$ should be larger than $1$, this automatically excludes the solution with a $-$. The solution with a $+$ can be compatible with the constraint depending on the parameters. The condition writes
\beq
X^2-Z^2+\sqrt{(1-Z^2+X^2)^2+4Z^2}\geq 1\,,
\eeq
which is trivially satisfied if $X^2>Z^2$. In the case $X^2<Z^2$, it is equivalent to
\beq
(1-Z^2+X^2)^2+4Z^2\geq (1+Z^2-X^2)^2\,,
\eeq
that is
\beq
2(X^2+Z^2)\geq 0
\eeq
which is also always satisfied. It follows that the solution with a $+$ always fulfil the second condition in (\ref{ee}). We find a change of sign along the integration range when the corresponding $y_0$ lies in the interval $[-1/2,1/2]$.

\section{More on Matsubara sums}\label{app:Matsubara}
In the main text, we explained that there were various ways of dealing with the Matsubara sums that result from performing the momentum integrals analytically. We explored one of them in detail, based on an expansion of the corresponding summand around $r=\bar r$. Here we consider a more general strategy based on an expansion at large Matsubara frequencies. We shall consider the example of the tadpole sum-integral
\beq
{\cal J}_{M_T^\kappa}\equiv T\sum_{q\in\mathds{Z}}\int_q\frac{1}{q^2+(\omega_q^\kappa)^2+m^2}\,,
\eeq
but the strategy applies to much general situations. In a first step, we perform the momentum integral analytically in $d-1$ dimensions. This gives
\beq
{\cal J}_{M_T^\kappa}=\frac{T\Gamma(3/2-d/2)}{(4\pi)^{d/2-1/2}}\sum_{q\in\mathds{Z}}\big((\omega_q^\kappa)^2+m^2\big)^{d/2-3/2}\,.
\eeq
As we have already discussed, inverting the $\epsilon$-expansion and the $q$-summation makes no sense at this point. To circumvent this difficulty, we add and subtract to the sum the first terms of its expansion at large $|q|$. In fact, since $q$ only enters via $(\omega_q^\kappa)^2$, it makes more sense to consider an expansion in that variable, which corresponds in fact to an expansion in $m^2$. We then have
\beq
{\cal J}_{M_T^\kappa} & = & {\cal J}_{M_T^\kappa}(m=0)-m^2{\cal J}_{M_T^\kappa}'(m=0)\nonumber\\
& + & \frac{T\Gamma(3/2-d/2)}{(4\pi)^{d/2-1/2}}\sum_{q\in\mathds{Z}}\Big[\big((\omega_q^\kappa)^2+m^2\big)^{d/2-3/2}\nonumber\\
& & \hspace{1.5cm}-|\omega_q^\kappa|^{d-3}-\frac{d-3}{2}m^2|\omega_q^\kappa|^{d-5}\Big]\,.
\eeq
The subtracted bracket behaves as $\smash{|\omega_q^\kappa|^{d-7}=|\omega_q^\kappa|^{-3-2\epsilon}}$. This means that for this term reverting the $\epsilon$-expansion and the $q$-sum is licit so, up to terms of order $\epsilon$, we can write
\beq
{\cal J}_{M_T^\kappa} & = & {\cal J}_{M_T^\kappa}(m=0)-m^2{\cal J}'_{M_T^\kappa}(m=0)\nonumber\\
& - & \frac{T}{4\pi}\sum_{q\in\mathds{Z}}\left[\big((\omega_q^\kappa)^2+m^2\big)^{1/2}-|\omega_q^\kappa|-\frac{m^2}{2|\omega_q^\kappa|}\right]\,.\label{eq:decomp}\nonumber\\
\eeq
Let us now determine ${\cal J}_{M_T^\kappa}(m=0)$ and ${\cal J}'_{M_T^\kappa}(m=0)$. We have
\beq
{\cal J}_{M_T^\kappa}(m=0) & = & \frac{T\Gamma(3/2-d/2)}{(4\pi)^{d/2-1/2}}\sum_{q\in\mathds{Z}}|\omega_q^\kappa|^{d-3}\nonumber\\
& = & \frac{T^{d-2}\Gamma(3/2-d/2)}{4\pi^{5/2-d/2}}\sum_{q\in\mathds{Z}}\frac{1}{\left|q+\frac{\kappa\cdot r}{2\pi}\right|^{3-d}}\nonumber\\
\eeq
Let us then shift the integer $q$ in the sum by $-\lfloor\kappa\cdot r/(2\pi)\rfloor$, where $\lfloor x\rfloor$ denotes the floor function that gives the largest integer less or equal to $x$. Then, introducing the notation $\{x\}\equiv x-\lfloor x\rfloor \in [0,1[$, we find
\beq
\sum_{q\in\mathds{Z}}\frac{1}{\left|q+\left\{\frac{\kappa\cdot r}{2\pi}\right\}\right|^{3-d}} & = & \sum_{q=0}^\infty \frac{1}{\left(q+\left\{\frac{\kappa\cdot r}{2\pi}\right\}\right)^{3-d}}\nonumber\\
& + & \sum_{q=1}^\infty\frac{1}{\left(q-\left\{\frac{\kappa\cdot r}{2\pi}\right\}\right)^{3-d}}\,.
\eeq
where we could remove the absolute values and we considered the change of variables $\smash{q\to -q}$ in the last line. In terms of the Hurwitz zeta function
\beq
\zeta(s,a)\equiv\sum_{n=0}^\infty \frac{1}{(n+a)^s}\,,
\eeq
this rewrites
\beq
& & {\cal J}_{M_T^\kappa}(m=0)=\frac{T^{d-2}\Gamma(3/2-d/2)}{4\pi^{5/2-d/2}}\Bigg[\zeta\left(3-d,\left\{\frac{\kappa\cdot r}{2\pi}\right\}\right)\nonumber\\
& & \hspace{0.7cm}+\,\zeta\left(3-d,-\left\{\frac{\kappa\cdot r}{2\pi}\right\}\right)-\left(-\left\{\frac{\kappa\cdot r}{2\pi}\right\}\right)^{d-3}\Bigg].
\eeq
Now, $\zeta(s,a)$ is regular for any $s=-n<0$ and we have
\beq
\zeta(-n,a)=-\frac{B_{n+1}(a)}{n+1}\,,
\eeq
where $B_n$ denotes the $n^{th}$ Bernouilli polynomial. Together, with the identity $(B_2(x)+B_2(-x))/2-x=B_2(x)$, this leads to
\beq
{\cal J}_{M_T^\kappa}(m=0) & = & \frac{T^2}{2}B_2\left(\left\{\frac{\kappa\cdot r}{2\pi}\right\}\right),\label{eq:J0}
\eeq
up to terms of order $\epsilon$. We have recovered a well known result, see for instance \cite{Reinosa:2016iml}. Proceeding similarly for $J'_{M_T^\kappa}(m=0)$, we arrive at
\beq
{\cal J}'_{M_T^\kappa}(m=0) & = & -\frac{T\Gamma(5/2-d/2)}{(4\pi)^{d/2-1/2}}\sum_{q\in\mathds{Z}}|\omega_q^\kappa|^{d-5}\nonumber\\
& = & -\frac{T^{d-4}\Gamma(5/2-d/2)}{16\pi^{9/2-d/2}}\sum_{q\in\mathds{Z}}\frac{1}{\left|q+\frac{\kappa\cdot r}{2\pi}\right|^{5-d}}\nonumber\\
\eeq
and thus
\beq
& & {\cal J}'_{M_T^\kappa}(m=0)=-\frac{T^{d-4}\Gamma(5/2-d/2)}{16\pi^{9/2-d/2}}\Bigg[\zeta\left(5-d,\left\{\frac{\kappa\cdot r}{2\pi}\right\}\right)\nonumber\\
& &\hspace{0.7cm} +\,\zeta\left(5-d,-\left\{\frac{\kappa\cdot r}{2\pi}\right\}\right)-\left(-\left\{\frac{\kappa\cdot r}{2\pi}\right\}\right)^{d-5}\Bigg].
\eeq
This time, we use that 
\beq
\zeta(1+2\epsilon,x) & = & 1/(2\epsilon)-\psi(x)\,,\\
\Gamma(x+\epsilon) & = & \Gamma(x)[1+\epsilon\psi(x)]\,,
\eeq
and we find
\beq
{\cal J}'_{M_T^\kappa}(m=0) & = & -\frac{(\pi T^2)^{-\epsilon}}{16\pi^2}\Bigg[\frac{1}{\epsilon}+\psi\left(\frac{1}{2}\right)-\psi\left(\left\{\frac{\kappa\cdot r}{2\pi}\right\}\right)\nonumber\\
& & -\,\psi\left(-\left\{\frac{\kappa\cdot r}{2\pi}\right\}\right)+\left\{\frac{2\pi}{\kappa\cdot r}\right\}\Bigg].\label{eq:Jp0}
\eeq
Using Eqs.~(\ref{eq:J0}) and (\ref{eq:Jp0}) in Eq.~(\ref{eq:decomp}) we obtain an alternative formula for the massive tadpole at finite temperature, in terms of an explicit contribution that includes the divergent part and an implicit contribution in the form of a finite Matsubara sum to be determined numerically. We have checked that this coincides with the result obtained by first performing the Matsubara sum analytically and the resulting momentum integral numerically after the UV divergences have been extracted.

In view of this result, it also interesting to revisit the evaluation of the last line of Eq.~(\ref{eq:fsum}) and in particular the relevance of the correction term in Eq.~(\ref{eq:ccc}). Let us do this in the case $\smash{m=0}$ which allows for analytical calculation. In this case, the subtracted version of the last line of Eq.~(\ref{eq:fsum}) rewrites entirely in terms of ${\cal J}_{M_T^\kappa}(m=0)$. We find
\beq
(d-2)^2\Delta{\cal J}_{M_T^\kappa}(m=0)\,,
\eeq
which according to Eq.~(\ref{eq:J0}) has the finite $\epsilon\to 0$ limit
\beq
& & 2T^2\Delta B_2\left(\left\{\frac{\kappa\cdot r}{2\pi}\right\}\right)\nonumber\\
& & \hspace{0.5cm}=\,2T^2\Delta\left[\left(\left\{\frac{\kappa\cdot r}{2\pi}\right\}\right)^2-\left(\left\{\frac{\kappa\cdot r}{2\pi}\right\}\right)\right].\label{eq:res}
\eeq
Alternatively, if just perform the momentum integrals, we find
\beq
& &(d-2)^2\frac{\Gamma(3/2-d/2)}{(4\pi)^{d/2-1/2}}T\sum_{q\in\mathds{Z}} \Delta ((\omega^\kappa_q)^2)^{d/2-3/2}\,.\nonumber\\
\eeq
Since the sum is convergent (after cancellation of leading terms for $q\to\infty$ and $q\to -\infty$), it seems legitimate to set $\epsilon=0$ prior to the sum. We find
\beq
& & -\frac{T}{\pi}\sum_{q\in\mathds{Z}} \Big[|\omega^\kappa_q|-|\bar\omega^\kappa_q|\Big]\nonumber\\
& & \hspace{0.5cm}=\,-2T^2\sum_{q\in\mathds{Z}} \Bigg[\left|q+\frac{r\cdot\kappa}{2\pi}\right|-\left|q+\frac{\bar r\cdot\kappa}{2\pi}\right|\Bigg].\label{eq:fff}
\eeq
This sum corresponds to the first two terms of Eq.~(\ref{eq:ccc}). We evaluate it as follows. We have already mentioned that if $r$ and $\bar r$ lie in the same Weyl chamber, then $r\cdot\kappa$ and $\bar r\cdot\kappa$ are in the same interval $[2\pi k,2\pi (k+1)]$ for a certain $k$. By shifting $q$ under the sum by $-k$, the sum rewrites
\beq
& & -2T^2\sum_{q\in\mathds{Z}} \Big[\left|q+\frac{r\cdot\kappa}{2\pi}-k\right|-\left|q+\frac{\bar r\cdot\kappa}{2\pi}-k\right|\Big]\nonumber\\
& & \hspace{0.5cm}=\,-2T^2\sum_{q=0}^\infty \Big[q+\frac{r\cdot\kappa}{2\pi}-k-q-\frac{\bar r\cdot\kappa}{2\pi}+k\Big]\nonumber\\
& & \hspace{0.5cm}+\,2T^2\sum_{q=1}^\infty \Big[q-\frac{r\cdot\kappa}{2\pi}+k-q+\frac{\bar r\cdot\kappa}{2\pi}-k\Big]\nonumber\\
& & \hspace{0.5cm}=\,-2T^2\Delta\left\{\frac{r\cdot\kappa}{2\pi}\right\}.
\eeq
We thus see that the sum (\ref{eq:fff}) which corresponds to the first two terms of Eq.~(\ref{eq:ccc}) and which was obtained by setting $\smash{\epsilon=0}$ prior to the sum accounts for the linear term in Eq.~(\ref{eq:res}) but completely misses the quadratic term. The latter is accounted for by the correction term in (\ref{eq:ccc}). The reason why this term has been missed in the present derivation is again that, in setting $\smash{\epsilon=0}$, we miss a term of the form
\beq
& &(d-2)^2\frac{(d-3)(d-4)}{2}\frac{\Gamma(3/2-d/2)}{(4\pi)^{d/2-1/2}}T\Delta (\kappa\cdot r)^2\sum_{q\in\mathds{Z}} |\omega_q|^{d-5}\nonumber
\eeq
in the expansion at large $|q|$. The $\smash{\epsilon\to 0}$ limit of this missed contribution is precisely the correction term in Eq.~(\ref{eq:ccc}) or the missing term in Eq.~(\ref{eq:res}).

\end{document}